\documentclass{article}
\usepackage[english]{babel}
\usepackage{amsfonts}
\usepackage[letterpaper,top=2cm,bottom=2cm,left=2cm,right=2cm,marginparwidth=1.75cm]{geometry}
\usepackage{subcaption}
\usepackage{amsmath}
\usepackage{graphicx}
\usepackage{float}
\usepackage[colorlinks=true, allcolors=blue]{hyperref}
\numberwithin{equation}{section}
\usepackage{appendix}
\usepackage{authblk}
\usepackage{placeins}
\usepackage{subcaption} 
\usepackage{caption} 
\title{Oscillatory Instabilities of a One-Spot Pattern in the Schnakenberg Reaction-Diffusion System in $3$-D Domains}

\author{Siwen Deng\thanks{\raggedright School of Mathematical and Physical Sciences, Macquarie University, Sydney, Australia; \href{mailto:Siwenddeng@gmail.com}{Siwenddeng@gmail.com}.} \quad Justin C. Tzou\thanks{School of Mathematical and Physical Sciences, Macquarie University, Sydney Australia; \href{mailto:tzou.justin@gmail.com}{tzou.justin@gmail.com}.} \quad Shuangquan Xie\thanks{School of Mathematics, Hunan University, Changsha City, Hunan, China; \href{mailto:xieshuangquan2013@gmail.com}{xieshuangquan2013@gmail.com}.}}

\usepackage[parfill]{parskip}
\begin{document}

\date{}

\maketitle

\begin{abstract} 
\noindent For an activator-inhibitor reaction-diffusion system in a bounded three-dimensional domain $\Omega$ of $\mathcal{O}(1)$ volume and small $\mathcal{O}(\varepsilon^2) \ll \mathcal{O}(1)$ activator diffusivity, we employ a hybrid asymptotic-numerical method to investigate two instabilities of a localized one-spot equilibrium that result from Hopf bifurcations: an amplitude instability leading to growing oscillations in spot amplitude, and a translational instability leading to growing oscillations of the location of the spot's center $\mathbf{x}_0 \in \Omega$. Here, a one-spot equilibrium is one in which the activator concentration is exponentially small everywhere in $\Omega$ except in an $\mathcal{O}(\varepsilon)$ localized region about $\mathbf{x}_0$ where its concentration is $\mathcal{O}(1)$. We find that the translation instability is governed by a $3 \times 3$ nonlinear matrix eigenvalue problem. The entries of this matrix involve terms calculated from certain Green’s functions, which encode information about the domain’s geometry. In this nonlinear matrix eigenvalue system, the most unstable eigenvalue determines the oscillation frequency at onset, while the corresponding eigenvector determines the direction of oscillation. We demonstrate the impact of domain geometry and defects on this instability, providing analytic insights into how they select the preferred direction of oscillation. For the amplitude instability, we illustrate the intricate way in which the Hopf bifurcation threshold $\tau_H$ varies with a feed-rate parameter $A$. In particular, we show that the $\tau_H$ versus $A$ relationship possesses two saddle-nodes, with different branches scaling differently with the small parameter $\varepsilon$. All asymptotic results are confirmed by finite elements solutions of the full reaction-diffusion system. 
\end{abstract}


\section{Introduction}
Research on spot patterns in the three-dimensional ($3$D) Schnakenberg activator-inhibitor system is driven by applications in fields as diverse as biological morphogenesis (cf. \cite{bio1}, \cite{bio2}) and chemical reactions (cf. \cite{chem1}, \cite{chem2}), where understanding the conditions for stability is essential. Localized equilibrium patterns can exhibit a diverse range of complex dynamic behaviors, including spot self-replication, spot annihilation, spot amplitude temporal oscillations, and slow spot drift (cf. \cite{CHEN_WARD_2009}, \cite{Chen2D}, \cite{Kolokolnikov2009}, \cite{TZOU_BAYLISS_MATKOWSKY_VOLPERT_2011},\cite{WEI2002478}). 

There has been extensive work in analyzing slow spot drift and spot amplitude temporal oscillations. It is demonstrated in \cite{translation1D3,KOLOKOLNIKOV2005258,translational1D} that a single spike in a one-dimensional Gray-Scott model can undergo destabilizing oscillations in either its amplitude or location. {\color{blue}The type of instability that arises first is primarily determined by the feed-rate $A$, which represents the amount of substrate chemical being pumped into the system.} Typically, amplitude oscillations are triggered at a lower feed-rate than translational oscillatory instabilities. In \cite{WM_Hopf_1D}, Hopf bifurcations leading to oscillations in spike heights 
in the Gierer-Meinhardt model were studied across various ranges of the reaction-time constant. In two dimensions, a comprehensive study of the translational oscillatory instability analysis for the Schnakenberg model is presented in \cite{Tzou_2023} and \cite{Xie_2017} while the amplitude oscillatory stability is analyzed in \cite{hopf2d}. 
{\color{blue}Moreover, \cite{hopf2d} introduced an anomalous scaling of the instability threshold for destabilizing amplitude oscillations, resolving a long-standing open problem in nonlocal eigenvalue problems (NLEP).}

{\color{blue}In three-dimensional domains, competition and spot self-replication instabilities of localized spot solutions have been studied for the Schnakenberg \cite{jst-3d} and Gierer-Meinhardt models \cite{Gomez2020}. The former is a global instability that results in the annihilation of one or more spots due to an  ``over-crowding'', while the latter is a local instability that results in one or more spots self-replicating into two.}  
However, Hopf bifurcations are more intricate in $3$D and detailed insights into the translational oscillatory instability and amplitude oscillatory instabilities in $3$D remain lacking.

Recently, investigations into heterogeneous domains have generated considerable interest. Two types of heterogeneity are introduced in \cite{tony-defect-2d}. The first type is a perturbation of a spatially uniform feed rate, and the second involves the inclusion of a defect in the domain. The former type has been considered in some works \cite{VanGorder2021PatternFF}, \cite{Ishii2020StabilityOM}, while the latter has received less attention. The purpose of introducing the second type of heterogeneity is to create an asymmetric pattern, which provides a more general and practical research scenario relevant to real-world applications.

In dimensionless form, the Schnakenberg RD model (cf. \cite{schnakenberg}) is
\begin{equation}
    \begin{aligned}\label{or_mod}
    &\mathcal{V}_t = \varepsilon^2 \Delta \mathcal{V}  - \mathcal{V}+ b + \mathcal{UV}^2, \quad x \in \Omega; \quad \partial_n \mathcal{V} = 0, x \in \partial\Omega,\\
      & \tau\mathcal{U}_t = \mathcal{D} \Delta \mathcal{U} +A- \mathcal{UV}^2, \quad x \in \Omega; \quad \partial_n \mathcal{U} = 0, x \in \partial\Omega.
\end{aligned}
\end{equation}
Here, $\mathcal{V}$ and $\mathcal{U}$ denote the activator and inhibitor fields, respectively. The domain $\Omega \subset \mathbb{R}^3$ is bounded, while $b$ and $A$ are constant bulk activator and inhibitor feed rates. Additionally, $\mathcal{D} > 0$ and $0 < \varepsilon \ll 1$. To ensure that the amplitude of a local spot in model (\ref{or_mod}) remains $\mathcal{O}(1)$ as $\varepsilon \rightarrow 0$, we introduce the rescaling $\mathcal{V} = \varepsilon^{-3} b v$, $\mathcal{U} = \varepsilon^3 b^{-1} u$, $\tau = \varepsilon^3 b^{-2} \tau$, $\mathcal{D} = \varepsilon^{-4} b^{-2} D$, and $A = b^{-1} A$. We then obtain the rescaled, singularly perturbed Schnakenberg model.
\begin{equation}\label{mod}
   \begin{aligned}
    &v_t = \varepsilon^2 \Delta v- v + \varepsilon^3  + uv^2, \quad x \in \Omega; \quad \partial_n v = 0, x \in \partial\Omega,\\
      & \tau u_t = \frac{D}{\varepsilon} \Delta u +A- \varepsilon^{-3} uv^2, \quad x \in \Omega; \quad \partial_n u = 0, x \in \partial\Omega.
    \end{aligned} 
\end{equation}
In (\ref{mod}), the Hopf bifurcation parameter $\tau$ measures the rate at which the inhibitor responds to perturbations in the concentrations of the activator and inhibitor. As $\tau$ increases, the inhibitor’s response becomes progressively slower, leading to oscillatory instabilities via Hopf bifurcations.

In this paper, we consider the effect of increasing the parameter $\tau$ on a single spot in a bounded $3$D domain $\Omega$, which may be either a localized heterogeneous domain or a normal domain. Two types of oscillatory instabilities are presented. First, motivated by \cite{Tzou_2023}, we perform a stability analysis of the spot's motion (position). In particular, we analyze the effect of geometric asymmetry on the instability threshold and the preferred direction of the spot’s oscillation. The associated linearized eigenvalue problem has an eigenfunction of the form $\Psi(r)Y_n^m(\varphi,\theta)$, where $Y_n^m$ is the Spherical Harmonics. The translation mode is $n = \pm1$ for which the corresponding eigenvalue is small as $\varepsilon\rightarrow0$. We therefore refer this problem as $small$ eigenvalue problem. Second, we provide a detailed study of temporal oscillations in the spot amplitude, referred as $large$ eigenvalue problem, corresponding to the mode $n = 0$. One of our key findings is the existence of two regimes: the $\tau \sim \mathcal{O}(1)$ regime and $\tau \sim \mathcal{O}(\varepsilon^{-3})$ regime. Furthermore, there is an overlap between these two regimes for a certain range of $A$. We investigate the behavior of the Hopf bifurcation threshold $\tau_H$ within this interval and present how the stability status of the system changes as $\tau$ varies.

The outline of the paper is as follows. In \S\ref{equli}, we construct a one-spot equilibirum solution of the Schnakenberg PDE (\ref{mod}) basing on the work from \cite{jst-3d}. We use the correction terms in $\mathcal{O}(\varepsilon^3)$ in the stability analysis. 

In \S\ref{stability_samll}, we analyze the translational oscillatory instability of the one spot equilibrium. We perturb $\tau = \varepsilon^{-3}(\tau_0+\varepsilon\tau_1)$ and $\lambda = \varepsilon^2 (\lambda_0+\varepsilon\lambda_1)$. By applying the solvability conditions on the $\mathcal{O}(\varepsilon^2)$ of the small eigenvalue problem in the inner region, we obtain the leading order of the Hopf bifurcation threshold $\tau_0$. To compute $\tau_1$ and $\lambda_0$, the solvibility conditions are applied again on the $\mathcal{O}(\varepsilon^2)$ of the small eigenvalue problem in the inner region, deriving a $3 \times 3$ complex matrix-eigenvalue problem of the form $M\bold{a} = \lambda \mathbf{a}$. By setting $\lambda$ as purely imaginary, we obtain the targeted $\mathcal{O}(\varepsilon)$ correction term for the translational oscillatory instability threshold, the leading order of the corresponding oscillation frequency at onset, and the oscillation direction, which is indicated by the eigenvector $\bold{a}$. The matrix $M$ contains the information of domain geometry, which involves the quadratic terms of the local behavior of Helmholtz Green's function $G_\mu(\bold{x};\bold{x}_j)$ satisfying 
\begin{equation}\label{Helm}
    \begin{aligned} 
        &\Delta G_\mu - \mu^2 G_\mu = \delta(\bold{x}-\bold{x}_j),\quad \bold{x},\bold{x}_j \in \Omega,\quad \partial_n G_\mu = 0 ,\quad \bold{x} \in \partial\Omega,\\
        &G_\mu (\bold{x};\bold{x}_j) \sim -\frac{1}{4\pi|\bold{x}-\bold{x}_j|} + R_{\mu}(\bold{x}_j;\bold{x}_j)+ (\bold{x}-\bold{x}_j) \cdot \nabla_{\bold{x}} R_\mu(\bold{x};\bold{x}_j)|_{\bold{x}=\bold{x}_j}\\
     &\qquad \qquad\quad+\frac{1}{2}(\bold{x}-\bold{x}_j)^T \mathcal{H}_{\mu jj}(\bold{x}-\bold{x}_j)-
     \frac{\mu^2}{8\pi}|\bold{x}-\bold{x}_j|, \quad \text{as}\quad \bold{x}\rightarrow \bold{x}_j,\\
     &G_\mu (\bold{x};\bold{x}_i) \sim   G_{\mu}(\bold{x}_j;\bold{x}_i)+ (\bold{x}-\bold{x}_j) \cdot \nabla_{\bold{x}} G_\mu(\bold{x};\bold{x}_i)|_{\bold{x} = \bold{x}_j}+\frac{1}{2}(\bold{x}-\bold{x}_j)^T \mathcal{H}_{\mu ji}(\bold{x}-\bold{x}_j), \\
     &\qquad \qquad \qquad \qquad \qquad \qquad \qquad \qquad \qquad \qquad \qquad \qquad \text{as} \quad \bold{x} \rightarrow \bold{x}_i \neq \bold{x}_j,
    \end{aligned} 
\end{equation}
where $\bold{x}_j$ is the source center, and $\mathcal{H}_{\mu jj}$ and $\mathcal{H}_{\mu ji}$ are the $3\times3$ Hessian matrices. In \S\ref{unit}, we apply our result on a unit sphere and compute $\tau$ up to $\mathcal{O}(\varepsilon^2)$ correction and $\lambda$ to $\mathcal{O}(\varepsilon)$ correction, which provides a high accuracy in the numerical experiment. In \S\ref{perturbed}, we examine how the geometry select the dominant mode of oscillation. The result shows that the preferred direction of oscillation is along the major axis for which $\tau$ is the smallest among all three directions. In \S\ref{defected}, we further break the domain symmetry by removing a ball-shaped defect from which the chemicals leak out. Following the technique from \cite{tony-defect-2d}, the defect can be treated as a localized pinned spot and an extra term is introduced into the matrix $M$ in this geometry. Consequently, the dominant mode of oscillation differs from that in the spherical case. The equilibrium solution for the one-spot pattern on a defected domain is provided in Appendix \ref{equl_defect}.

In \S\ref{large_stability}, we analyze the stability of one spot equilibrium with respect to the amplitude oscillatory instability. Specifically, we study the behavior of $\tau_H$ when the feed rate $A$ changes. Two regimes of $\tau_H$ are discussed. In \S\ref{regime1}, $\tau_H$ is $\mathcal{O}(1)$ and a hybrid asymptotic-numerical method is used to determine the threshold. We find a unique threshold when $A<13.56$, where the system becomes unstable to amplitude oscillatory instability when $\tau$ exceeds the threshold and remains stable when it is below the threshold. For $13.56 \leq A < A_\varepsilon$, where $A_\varepsilon$ is a saddle point weakly depending on $\varepsilon$, two distinct HB thresholds exist. No bifurcation occurs for $A > A_\varepsilon$ in the $\tau \sim \mathcal{O}(1)$ regime. In \S\ref{regime2}, we study the $\mathcal{O}(\varepsilon^{-3})$ regime, where the values of $\tau_H$ depend on the spectrum of the leading order of the eigenvalue problem in the inner region. In this regime, there is a unique threshold for each value of $A>13.56$. For small $\tau_0$ where $\tau_0 = \varepsilon^3 \tau$, a reduced problem is introduced to investigate the values of $\tau_H$ near $A = 13.56$. There is an overlap between these two regimes for a small interval of $A$. In other words, three HB thresholds appear when $13.56 \leq A < A_\varepsilon$: two at $\mathcal{O}(1)$ and one at $\mathcal{O}(\varepsilon^{-3})$. 
\begin{figure}[h]
    \centering
    \begin{subfigure}[b]{0.45\textwidth}
         \includegraphics[width=\linewidth]{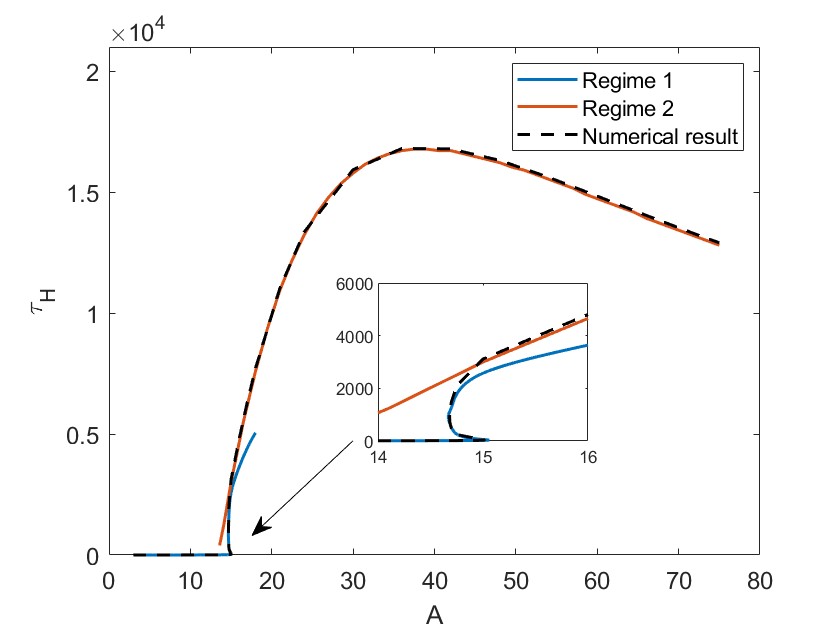}
        \caption{}
        \label{fig:full_large}
    \end{subfigure}
   \begin{subfigure}[b]{0.45\textwidth}
       \includegraphics[width =\linewidth]{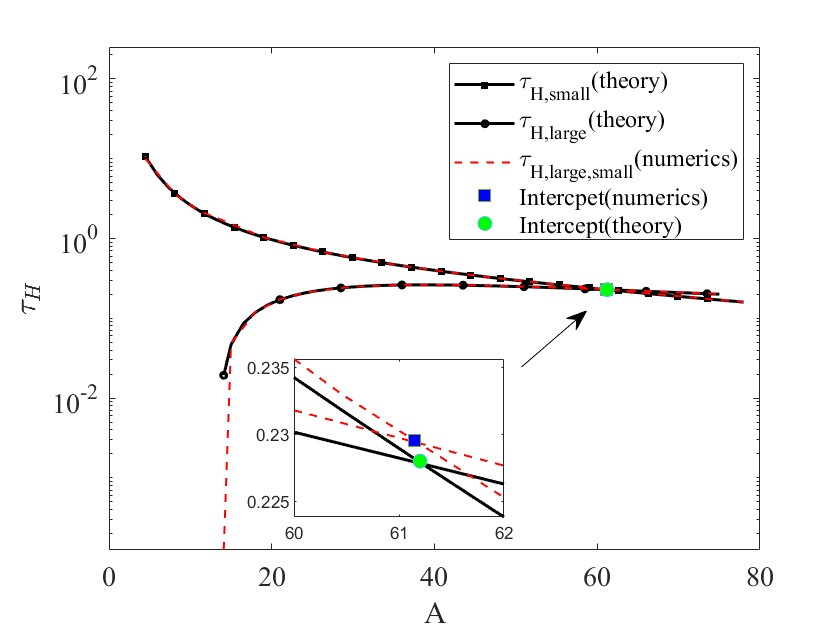}
       \caption{}
       \label{fig:smallvslarge_tau}
   \end{subfigure}
   \caption{(a) Theoretical value of $\tau_H$ in the large eigenvalue problem as a function of $A$. Two different regimes are found: $\tau_H \sim \mathcal{O}(1)$ regime and $\tau_H \sim \mathcal{O}(\varepsilon^{-3})$ regime.  The zoom-in box shows that there are 3 distinct HB thresholds when $13.56<A<13.98$. (b) We plot the HB values for amplitude and position oscillations for $\varepsilon = 0.03$. The numerical results coincide with our theoretical results in both figures. }
\end{figure}
In Figure \ref{fig:full_large}, a full theoretical result for the large eigenvalue problem is present. The line of regime $1$ is the $\mathcal{O}(1)$ HB threshold while the line of regime $2$ refers to $\mathcal{O}(\varepsilon^{-3})$ threshold. As shown in the zoom-in box, When $A$ is approaching to $13.56$ from the right, the value of $\tau_H$ tends to infinity, and there are 3 distinct thresholds when $13.56 \leq A < 13.98$ as we claimed. 
In Figure \ref{fig:smallvslarge_tau}, we plot the Hopf bifurcation threshold for both large eigenvalue problems and small eigenvalue problems. The figure shows that they intercept at $A = 61.2$. 

A full numerical solution of (\ref{mod}) is performed in \S\ref{numerical} to confirm our theoretical results. Finally, in \S\ref{conclusion}, we conclude with a briefly discussion.

\section{Equilibrium}\label{equli}
In this section, we briefly review the construction process of the equilibrium solution, as completed in \cite{jst-3d}. We start by obtaining the one-spot equilibrium solution of (\ref{mod}) using the method of matched asymptotic expansions.

\begin{figure}[h]
    \centering
    \includegraphics[width=0.4\linewidth]{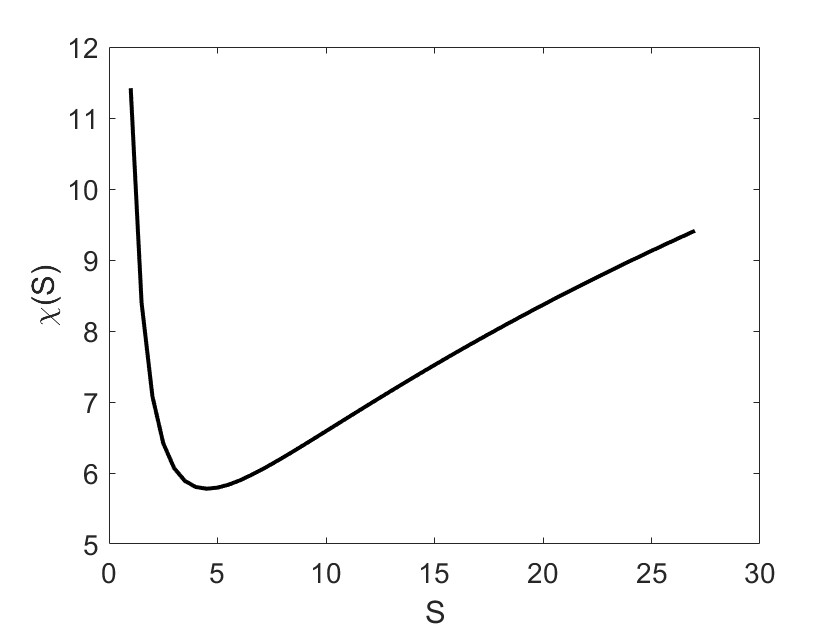}
    \caption{A plot of $\chi(S_\varepsilon)$ versus $S_\varepsilon$. The fold point at ($S_\varepsilon, \chi) \approx (4.52,5.78)$, where $\chi'(4.52) = 0$.}
    \label{fig:chi}
\end{figure}

In the inner region near the spot, we set $\bold{x}= \bold{x}(\sigma)$ where $\sigma = \varepsilon^3 t$ represents the slow time scale, and introduce the inner variables
\begin{equation}\label{inner_var}
    \begin{aligned} 
   & \bold{x} = \varepsilon \bold{y},\quad \bold{y} = \left(
    \begin{array} {cc}
    y_{1}\\
    y_{2}\\
    y_{3}
    \end{array}\right) = \rho\mathbf{e}, \quad \mathbf{e} = \left(
    \begin{array} {cc}
    \sin\theta \cos\varphi\\
   \sin\theta \sin\varphi\\
   \cos\theta
    \end{array}\right),\\
&v_e \sim \sqrt{D} (V_{\varepsilon}(\rho)+\varepsilon^3 V_{3}(\bold{y})+\dots), \quad u_e \sim \frac{1}{\sqrt{D}} (U_{\varepsilon}(\rho)+\varepsilon^3 U_{3}(\bold{y})+\dots),   
\end{aligned} 
\end{equation}
where $U_{\varepsilon}$ and $V_{\varepsilon}$ satisfy the radially symmetric core problem

\begin{equation}\label{core}
    \begin{aligned}
           \Delta_{\rho} V_{\varepsilon} - V_{\varepsilon} + U_{\varepsilon}V_{\varepsilon}^2 &= 0, \quad V'_{\varepsilon}(0) =0,\quad V_{\varepsilon} \rightarrow 0, \quad as \quad  \rho \rightarrow \infty,\\
    \Delta_{\rho} U_{\varepsilon} -U_{\varepsilon}V_{\varepsilon}^2 &=0, \quad U'_{\varepsilon}(0) = 0 
    \end{aligned} 
\end{equation}
with far-field behavior
\begin{equation}\label{core_farfield}
    U_{\varepsilon} \sim \chi(S_{\varepsilon})- S_{\varepsilon}/\rho+..., \quad as \quad \rho \rightarrow \infty,
\end{equation}
and noted that $\Delta_{\rho} \equiv \partial_{\rho\rho}+2\rho^{-1}\partial_{\rho}$. The $\mathcal{O}(\varepsilon^2)$ terms in the expansion of $u_e$ and $v_e$ are absent under the assumption that the spot is stationary over time.  Numerical solutions of (\ref{core}) are presented in \cite{jst-3d}. In Figure \ref{fig:chi}, the nonlinear function $\chi(S_\varepsilon)$ is generated for varies values of $S_\varepsilon$. 
To avoid the local (mode-2) self-replication instability of the spot, we impose $S_\varepsilon < \Sigma_2 \approx 20.16$. Applying the divergence theorem on the second equation of (\ref{core}), we obtain
\begin{equation}\label{Sj}
    \begin{aligned} 
    S_\varepsilon = \int_0^\infty U_{\varepsilon}V_{\varepsilon}^2\rho^2 d\rho.
\end{aligned} 
\end{equation}

In the outer region, we express the term $\varepsilon^{-3}uv^2$ in (\ref{mod}) in the sense of distributions as

\begin{equation}
    \begin{aligned} 
    \varepsilon^{-3}uv^2 \rightarrow 4\pi\sqrt{D}\left(\int_0^\infty U_{\varepsilon}V_{\varepsilon}^2\rho^2 d\rho\right)\delta(\bold{x}) = 4\pi\sqrt{D}S_\varepsilon\delta(\bold{x}-\bold{x}_1),
\end{aligned} 
\end{equation}
where $\bold{x}_1$ is the source location. Therefore, the quasi-equlibrium solution for $u$ in the outer region satisfies
\begin{equation}\label{outer_equlibrium}
    \begin{aligned} 
    \frac{D}{\varepsilon}\Delta u_e + A =4\pi\sqrt{D}S_\varepsilon\delta(\bold{x}-\bold{x}_1), \quad \bold{x}\in \Omega; \quad \partial_nu_e=0,\quad \bold{x} \in \partial \Omega.
\end{aligned} 
\end{equation}
We directly solve (\ref{outer_equlibrium}) to obtain the solution for $u$ in the outer region, namely
\begin{equation}\label{outer_sol}
    u_e = \bar{\mu}-\frac{4\pi\varepsilon}{\sqrt{D}}S_\varepsilon G(\bold{x};\bold{x}_1),
\end{equation}
where $G(\bold{x};\bold{x}_j)$ the Nuemann Green's function satisfying
\begin{equation}\label{Neumann_Green}
    \begin{aligned}
    \Delta G = \frac{1}{|\Omega|}-\delta(\bold{x}-\bold{x}_j)&, \quad \bold{x}\in\Omega;\quad \partial_n G = 0, \quad \bold{x}\in \partial\Omega,\\
    G(\bold{x};\bold{x}_j) \sim \frac{1}{4\pi|\bold{x}-\bold{x}_j|} &+R_{jj} + \nabla_\bold{x} R(\bold{x};\bold{x}_j)|_{\bold{x}=\bold{x}_j} \cdot (\bold{x}-\bold{x}_j) \\
    &+\frac{1}{2} (\bold{x}-\bold{x}_j)^T \mathcal{H}_{jj}(\bold{x}-\bold{x}_j),\quad as \quad \bold{x}\rightarrow \bold{x}_j; \\
    G(\bold{x};\bold{x}_i) \sim G_{ij} + \nabla_\bold{x} G&(\bold{x};\bold{x}_j)|_{\bold{x}=\bold{x}_i} \cdot (\bold{x}-\bold{x}_i) \\
    &+\frac{1}{2} (\bold{x}-\bold{x}_i)^T \mathcal{H}_{ij}(\bold{x}-\bold{x}_i),\quad as \quad \bold{x}\rightarrow \bold{x}_i \neq \bold{x}_j,
\end{aligned}
\end{equation}

Integrating (\ref{outer_equlibrium}) over $\Omega$ and applying the divergence theorem yields the solvability condition
\begin{equation}\label{s}
    \begin{aligned} 
    S_\varepsilon = \frac{A|\Omega|}{4\pi\sqrt{D}},
\end{aligned} 
\end{equation}
where $|\Omega|$ denotes the volume of the domain $\Omega$.

We then match the far-field behavior of the spot in the inner region (\ref{core_farfield}) to the leading order term of the outer solution (\ref{outer_sol}) using the local behavior of $G(\bold{x};\bold{x}_1)$ given in (\ref{Neumann_Green}), yielding an equation
\begin{equation}\label{N-nonlin-eqs}
    \bar{\mu} = 4\pi \varepsilon  R_{11} S_\varepsilon+\chi(S_\varepsilon),
\end{equation}
where $R_{11} = R(\bold{x}_1;\bold{x}_1)$ is the regular part of $G(\bold{x};\bold{x}_1)$ when $\bold{x}\rightarrow\bold{x}_1$.

We emphasize that the Green's function is of $\mathcal{O}(\varepsilon)$ in the outer solution, indicating the absence of $\mathcal{O}(\varepsilon^2)$ terms in the inner region due to the slow spot dynamics described in \cite{jst-3d}, 
\begin{equation}\label{steady-state}
   S_{\varepsilon} \nabla_\bold{x} R(\bold{x};\bold{x}_1)|_{\bold{x}=\bold{x}_1}  = \bold{0}.
\end{equation}

By solving (\ref{s}), (\ref{N-nonlin-eqs}) and (\ref{steady-state}) simultaneously, we obtain the source strength of the spot $S_{\varepsilon}$, the equilibrium position of the spot, and all other constants.

Expanding the outer solution (\ref{outer_sol}) and collecting the $\mathcal{O}(\varepsilon^3)$ terms the system of $V_3$ and $U_{3}$ is obtained, which is
\begin{equation}\label{o3}
    \begin{aligned} 
  & \Delta V_{3} - V_{3} +\frac{1}{\sqrt{D}} + U_{3}V_{\varepsilon}^2+2U_{\varepsilon}V_{\varepsilon}V_{3}  = 0, \quad V'_{3}(0) =0,\\
    &\Delta U_{3} +\frac{A}{\sqrt{D}} -U_{3}V_{\varepsilon}^2-2U_{\varepsilon}V_{\varepsilon}V_{3}=0, \quad U'_{3}(0) = 0,\\
   &V_{3} \rightarrow \frac{1}{\sqrt{D}}, \quad U_{3} \sim  -2\pi S_{\varepsilon}\bold{y}^T \mathcal{H}_{11} \bold{y}, \quad \text{as} \quad  \bold{y} \rightarrow \infty,
\end{aligned} 
\end{equation}
where $\mathcal{H}_{11}$ is the Hessian matrix of the regular part of $G(\bold{x};\bold{x}_1)$ as $\bold{x}$ is approaching to the source location.
With the inner coordinates $\rho$ and $\bold{e}$ defined in (\ref{inner_var}), the far-field condition can be written as 
\begin{equation}\label{o3-farfield}
    \begin{aligned} 
U_{3} = -2\pi S_\varepsilon \rho^2\bold{e}^T \mathcal{H}_{11}\bold{e} , \quad as \quad  \rho \rightarrow \infty.
\end{aligned} 
\end{equation}
This completes the construction of one spot equilibrium solution.

\section{The stability analysis: Small eigenvalues}\label{stability_samll}
We now analyze the instability of the small eigenvalue problem for a one-spot pattern. Let $|\psi|,|\phi|\ll\mathcal{O}(1)$, we perturbed the equilibrium solution
\begin{equation}\label{perturb}
    \begin{aligned} 
        v=v_e(r)+e^{\lambda t}\phi(x),\quad u = u_e(r) + e^{\lambda t}\psi(x),
\end{aligned} 
\end{equation}
which results a eigenvalue problem,
\begin{equation}\label{eigen_problem}
    \begin{aligned}
        &\varepsilon^2 \Delta \phi - \phi + 2u_e v_e \phi + v_e^2 \psi = \lambda \phi,\quad
        \frac{D}{\varepsilon}\Delta\psi - \varepsilon^{-3}(2u_ev_e\Phi + v_e^2\psi ) = \tau\lambda \psi, \quad x \in \Omega;\\
        &\partial_n \phi = \partial_n \psi = 0 \quad \text{on} \quad \partial\Omega.
    \end{aligned}
\end{equation}

In the inner region, we expand $\phi$ and $\psi$ in the power of $\varepsilon$, namely
\begin{equation}\label{eig-inner}
    \begin{aligned} 
         &\Phi \sim \sqrt{D}(\Phi_{\varepsilon}(\rho)+\varepsilon^2 \Phi_{2}(\bold{y}) + \varepsilon^3\Phi_{3}(\bold{y})+\cdots),\\
        &\Psi \sim \frac{1}{\sqrt{D}}(\Psi_{\varepsilon}(\rho)+\varepsilon^2\Psi_{2}(\bold{y})+\varepsilon^3\Psi_{3}(\bold{y})+\cdots),
\end{aligned} 
\end{equation}
where $\rho = |\bold{y}| = \varepsilon^{-1}|\bold{x}|=\varepsilon^{-1}r$, and $\bold{y}$ is defined in the equation (\ref{inner_var}).

Substituting (\ref{eig-inner}) into (\ref{eigen_problem}), we derive the leading order system
\begin{equation}\label{leading-order}
    \begin{aligned} 
        \Delta_{\bold{y}}\left(
        \begin{array}{c}
             \Phi_{\varepsilon}  \\
             \Psi_{\varepsilon}
        \end{array}
        \right)+ \mathcal{M}\left(
        \begin{array}{c}
             \Phi_{\varepsilon}  \\
             \Psi_{\varepsilon}
        \end{array}
        \right)=\bold0,
\end{aligned} 
\end{equation}
where 
\begin{equation}
    \mathcal{M} = \left( \begin{array}{cc}
         -1+2U_\varepsilon V_\varepsilon & V_\varepsilon^2  \\
         -2U_\varepsilon V_\varepsilon & -V_\varepsilon^2 
    \end{array}\right).
\end{equation}

It is readily observed that the solution to the first equation in (\ref{leading-order}) is a linear combination of $\partial_{y_{1}}V_{\varepsilon}$, $\partial_{y_{2}}V_{\varepsilon}$, and $\partial_{y_{3}}V_{\varepsilon}$. Therefore the solution to (\ref{leading-order}) is given by
\begin{equation}\label{lead-sol}
    \begin{aligned} 
        \Phi_{\varepsilon} = \bold{a}\cdot \nabla_{\bold{y}}V_{\varepsilon} = (\bold{a}^T \bold{e})\partial_{\rho} V_{\varepsilon},\quad \Psi_{\varepsilon} = \bold{a}\cdot \nabla_{\bold{y}}U_{\varepsilon} = (\bold{a}^T \bold{e})\partial_{\rho} U_{\varepsilon}; \quad \bold{a} = \left(
        \begin{array}{c}
             a_{1}  \\
             a_{2} \\
             a_{3}
        \end{array}
        \right),
\end{aligned} 
\end{equation}
and they satisfies the far-field of
\begin{equation}\label{leading_farfield_Psi}
        \Phi_{\varepsilon} \rightarrow 0, \quad \Psi_{\varepsilon} \sim (\bold{a}^T \bold{e})\frac{S_\varepsilon}{\rho^2}, \quad \text{as} \quad \rho \rightarrow \infty.
\end{equation}
The direction of oscillation of the spot at the onset of instability is determined by the value of $\bold{a}$. If $\bold{a}$ is real, the spot oscillates through the equilibrium position along a line. However, if $\bold{a}$ is complex, the motion exhibits rotation about the center. The value of $\bold{a}^T \bold{e}$ is obtained by matching with the outer solution.

In equation (\ref{leading_farfield_Psi}), the far-field behavior of $\Psi_{\varepsilon}$ must match with the singular behavior of the leading order term of $\psi$ in the outer region near the spot center, while the regular part of $\psi$ must be matched by a constant term in the far-field of $\Psi_{2}$.

Assume that $\tau \sim \mathcal{O}(\varepsilon^{-3})$, and $\lambda \sim \mathcal{O}(\varepsilon^2)$, thus $\tau \lambda \sim \mathcal{O}(\varepsilon^{-1})$ when $\tau$ is near the instability threshold. We scale $\tau$ and $\lambda$ as
\begin{equation}\label{scaling_tau}
       \lambda = \varepsilon^2 \lambda_0+\varepsilon^3\lambda_1, \quad \tau = \varepsilon^{-3} \tau_0 +\varepsilon^{-2}\tau_1.
\end{equation}

At $\mathcal{O}(\varepsilon^2)$ in the inner region, we have
\begin{equation}\label{oder 2 inner}
    \begin{aligned} 
        \Delta_{\bold{y}}\left(\begin{array}{c}
             \Phi_{2}  \\
             \Psi_{2}
        \end{array}\right) + \mathcal{M}\left(\begin{array}{c}
             \Phi_{2}  \\
             \Psi_{2}
        \end{array}\right) = \left(\begin{array}{c}
             \lambda_0 \Phi_{\varepsilon}  \\
             \frac{\tau_0 \lambda_0}{D}\Psi_{\varepsilon}
        \end{array}\right).
    \end{aligned} 
\end{equation} 
To analyze this system further, we begin by identifying the corresponding homogeneous solution to (\ref{oder 2 inner}), which is readily found as
\begin{equation}\label{sol_order_2} 
        \Phi_{2h} = \kappa \partial_{S_{\varepsilon}}V_{\varepsilon}, \quad \Psi_{2h} = \kappa \partial_{S_\varepsilon}U_{\varepsilon}.
\end{equation}
Here, $\kappa$ is a scaling constant determined by matching the regular part of $\psi$ at the spot location. Moving forward, we focus on the behavior of $\Psi_2$ as $\rho \rightarrow \infty$ by solving the following equivalent equation, which simplifies our analysis of the asymptotic behavior
\begin{equation}\label{perticular_Psi2}                \Delta_{\bold{y}}\Psi_{p2}=\frac{\bold{a}^T \bold{e}S_\varepsilon\tau_0 \lambda_0}{D\rho^2},
\end{equation}
The solution to (\ref{perticular_Psi2}) is then given by
\begin{equation}
    \Psi_{p2} = \bold{a}^T\bold{e}S_\varepsilon\frac{\lambda_0\tau_0}{D}\left( -\frac{1}{2} + \frac{c_1}{\rho^2} + c_2 \rho\right).
\end{equation}
Examining this solution, we note that the singular term of $\Psi_{p2}$ is of order $\mathcal{O}(\varepsilon^4)$ in the outer region and can therefore be ignored. The $c_2 \rho_j$ term must match the $\mathcal{O}(\varepsilon)$ term in the outer region. However, since the leading order in the outer region should be $\mathcal{O}(\varepsilon^2)$, we impose $c_2 = 0$ to maintain consistency.

As a result, the far-field condition of $\Psi_2$ is given by
\begin{equation}\label{inner_farfield_Psi2} 
        \Psi_{2} \sim \kappa(\chi'(S_\varepsilon)-\frac{1}{\rho})-\bold{a}^T\bold{e}S_\varepsilon\frac{\lambda_0\tau_0}{2D}. 
\end{equation}

In the outer region, we expand $\psi \sim \varepsilon^2 \psi_1+\varepsilon^3 \psi_2$. Thus $\psi_1$ satisfies
\begin{equation}\label{psi1}
    \begin{aligned} 
       & \Delta\psi_1 - \mu^2 \psi_1 = 0,\quad \bold{x} \in \Omega; \quad \partial_n \psi_1 =  0, \quad \bold{x} \in \partial \Omega.\\
        &\psi_1 \sim \frac{\bold{a}^T (\bold{x}-\bold{x}_1)S_{\varepsilon}}{|\bold{x}-\bold{x}_1|^3}+\kappa\chi'(S_{\varepsilon})-\frac{\bold{a}^T(\bold{x}-\bold{x}_1)S_{\varepsilon}}{|\bold{x}-\bold{x}_1|}\frac{\mu^2}{2}, \quad \text{as} \quad \bold{x}\rightarrow\bold{x}_1,
    \end{aligned} 
\end{equation}
where $\mu^2 = \frac{\tau_0\lambda_0}{D}$. Therefore, we determine $\psi_1$ as
\begin{equation}\label{out_psi1}
    \begin{aligned} 
        \psi_1  = -\frac{4\pi}{D} \Big[S_{\varepsilon}\bold{a}^T \nabla_{\bold{x}_1} G_\mu (\bold{x};\bold{x}_1)\Big],
    \end{aligned} 
\end{equation}
where $G_\mu(\bold{x};\bold{x}_1)$ is the Helmholtz Green's function defined in (\ref{Helm}), and $\nabla_{\bold{x}_1} G_\mu (\bold{x};\bold{x}_1)$ is the gradient of $G_\mu (\bold{x};\bold{x}_1)$ with respect to the second variable. The local behavior of $\nabla_{\bold{x}_1} G_\mu(\bold{x};\bold{x}_1)$ near the spot is
\begin{equation}\label{grad_G_localbhv}
    \begin{aligned} 
     &\nabla_{\bold{x}_1} G_\mu (\bold{x};\bold{x}_1) \sim -\frac{(\bold{x}-\bold{x}_1)}{4\pi|\bold{x}-\bold{x}_1|^3} + \bold{F}_{\mu}+ (\bold{x}-\bold{x}_1) \cdot \mathcal{Q}_{\mu1} +\frac{\mu^2}{8\pi}\frac{(\bold{x}-\bold{x}_1)}{|\bold{x}-\bold{x}_1|}, \quad \text{as} \quad \bold{x}\rightarrow\bold{x}_1,
    \end{aligned} 
\end{equation}
where we define the quantities
\begin{equation}
    \begin{aligned} 
       &\bold{F}_{\mu} = \left( \begin{array}{c}
             F^{(1)}_{\mu}  \\
             F^{(2)}_{\mu}  \\ 
             F^{(3)}_{\mu}  
        \end{array}\right ) = \nabla_{\bold{x}} R_\mu(\bold{x};\bold{x}_1)|_{\bold{x}=\bold{x}_1},
    \end{aligned} 
\end{equation}
and $\mathcal{Q}_{\mu1}$ is a $3 \times 3$ matrix computing by differentiating on $R_\mu(\bold{x};\bold{x}_1)$ respect to the first variable, then differentiating each element of the resulting vector respect to the second variable, and lastly substitute $\bold{x} = \bold{x}_1$, yielding
\begin{equation}\label{Qmu}
    \mathcal{Q}_{\mu1}  = \Big(\nabla_{\bold{x}_1} \partial_{x_1}R_\mu(\bold{x};\bold{x}_1)\quad \nabla_{\bold{x}_1} \partial_{x_2}R_\mu(\bold{x};\bold{x}_1)\quad \nabla_{\bold{x}_1} \partial_{x_3}R_\mu(\bold{x};\bold{x}_1)\Big)\Big|_{\bold{x}=\bold{x}_1}.
\end{equation}
At $\mathcal{O}(\varepsilon^3)$, $\psi_2$ satisfies 
\begin{equation}\label{psi2}
    \begin{aligned}
        &\Delta\psi_2 - \mu^2\psi_2 = \mu_1^2\psi_1,\quad \bold{x} \in \Omega; \quad \partial_n \psi_2 = 0, \quad \bold{x} \in \partial \Omega.\\
        &\psi_2 \sim -\frac{\kappa}{|\bold{x}-\bold{x}_1|},\quad\text{as}\quad \bold{x} \rightarrow \bold{x}_1,
    \end{aligned} 
\end{equation}
where $\mu_1^2 = \frac{\tau_0\lambda_1+\tau_1\lambda_0}{D}$. We observed that the particular solution to (\ref{psi2}) is the derivative of $\psi_1$ respect to $\mu^2$. Hence the expression of $\psi_2$ is
\begin{equation}\label{out_psi2}
    \psi_2 = \frac{4\pi}{D}\kappa G_\mu(\bold{x};\bold{x}_1) + \mu_1^2\partial_{\mu^2}\psi_1.
\end{equation}
Now we generate the outer solution $\psi_\varepsilon = \psi_1+\varepsilon \psi_2$ by adding (\ref{out_psi1}) and (\ref{out_psi2}), yielding
\begin{equation}\label{outer_psi}
    \begin{aligned}
        \psi_\varepsilon = -\frac{4\pi}{D} \Big[S_\varepsilon\bold{a}^T \nabla_{\bold{x}_1}& G_\mu (\bold{x};\bold{x}_1) + \varepsilon\mu_1^2S_\varepsilon\bold{a}^T\partial_{\mu^2} \nabla_{\bold{x}_1} G_\mu (\bold{x};\bold{x}_1)-\varepsilon\kappa G_\mu(\bold{x};\bold{x}_1)\Big].
    \end{aligned}
\end{equation}

We then match (\ref{outer_psi}) with (\ref{leading_farfield_Psi}) and (\ref{inner_farfield_Psi2}), using the local behavior of $G_\mu(\bold{x},\bold{x}_1)$ and $\nabla_{\bold{x}_1}G_\mu(\bold{x};\bold{x}_1)$ given in (\ref{Helm}) and (\ref{grad_G_localbhv}), to obtain the constants $\kappa$, which is
\begin{equation}\label{kappa}
    \begin{aligned}
        \kappa = -\frac{4\pi S_{\varepsilon}}{D\chi'(S_{\varepsilon})}\bold{a}^T\nabla_\bold{x}R_\mu(\bold{x};\bold{x}_1)|_{\bold{x}=\bold{x}_1}.
    \end{aligned}
\end{equation}
In highly symmetric geometries, such as a cube, sphere, or ellipsoid, the value of $\nabla_\bold{x}R_\mu(\bold{x};\bold{x}1)|{\bold{x} = \bold{x}1}$ coincides with $\nabla\bold{x}R(\bold{x};\bold{x}1)|{\bold{x} = \bold{x}_1}$, which is zero at equilibrium. Consequently, $\kappa$ is zero according to (\ref{kappa}).

To determine the Hopf stability threshold $\tau$, the frequency of oscillations at onset $\lambda$, and the direction of oscillation of the spot $\bold{a}$, solvability conditions will be applied at $\mathcal{O}(\varepsilon^2)$ for $\Phi_{2}$ and $\Psi_{2}$, as well as at $\mathcal{O}(\varepsilon^3)$ for $\Phi_{3}$ and $\Psi_{3}$. In the following analysis, we combine the results of the $\mathcal{O}(\varepsilon^2)$ and $\mathcal{O}(\varepsilon^3)$ analyses in the inner region to obtain
\begin{equation}\label{W1}
\begin{aligned}
        \Delta_\bold{y} \bold{W}_1 + \mathcal{M} \bold{W}_1 = \left(\begin{array}{c}
             (\lambda_0+\varepsilon\lambda_1)\Phi_{\varepsilon}  \\
             \mu_\varepsilon^2\Psi_{\varepsilon}
        \end{array}\right )+\varepsilon\mathcal{N}\left(\begin{array}{c}
             \Phi_{\varepsilon}  \\
             \Psi_{\varepsilon}
        \end{array} \right),
\end{aligned}
\end{equation}
where
\begin{equation}
    \bold{W}_1 = \left(\begin{array}{c}
         \Phi_{2}+\varepsilon \Phi_{3}  \\
         \Psi_{2}+\varepsilon \Psi_{3}
    \end{array} \right),\quad 
    \mathcal{N} = \left(\begin{array}{cc}
             -2V_{\varepsilon}U_{3} - 2V_{3}U_{\varepsilon} & -2V_{\varepsilon}V_{3}  \\
             2V_{\varepsilon}U_{3} + 2V_{3}U_{\varepsilon} & 2V_{\varepsilon}V_{3} 
        \end{array}\right),\quad \mu_\varepsilon = \mu + \varepsilon \mu_1.
\end{equation}
According to (\ref{Helm}) and (\ref{grad_G_localbhv}), expanding (\ref{outer_psi}) yields the far-field behavior of $\bold{W}_1$ as $\rho \rightarrow \infty$, namely
\begin{equation}
    \bold{W}_1 \sim \left(\begin{array}{c}
         0  \\
         -\frac{4\pi}{D}\Big[S_{\varepsilon}\bold{a}^T\big(\varepsilon\mathcal{Q}_{\mu1}\rho+\frac{\mu_\varepsilon^2}{8\pi}\big) \bold{e}-\kappa\big(\chi'(S_{\varepsilon})-\frac{1}{\rho}\big) - \varepsilon \kappa R_\mu(\bold{x}_1;\bold{x}_1)\Big] 
    \end{array}\right).
\end{equation}

Next, we apply solvability conditions on (\ref{W1}). The vector $\bold{W}_1$ can be decomposed into four components as 
\begin{equation}
    \begin{aligned} 
        \bold{W}_1 = \bold{W}_{sc}\sin{\theta}\cos{\varphi}+\bold{W}_{ss}\sin{\theta}\sin{\varphi}+\bold{W}_{cc}\cos{\theta}+\bold{\bar{W}},
    \end{aligned} 
\end{equation}
where $\bold{W}_{sc}$,$\bold{W}_{ss}$ , $\bold{W}_{cc}$ and $\bold{\bar{W}}$ are $2\times1$ vectors satisfying

\begin{equation}
    \begin{aligned} 
        &\Delta_{11} \bold{W}_{sc} - \mathcal{M} \bold{W}_{sc} = a_1\left[ \begin{array}{c}
             (\lambda_0+\varepsilon\lambda_1)\partial_{\rho} V_{\varepsilon} \\
             \mu_\varepsilon^2\partial_{\rho} U_{\varepsilon} 
        \end{array}\right],\quad \Delta_{11} \bold{W}_{ss} - \mathcal{M} \bold{W}_{ss} = a_2\left[ \begin{array}{c}
             (\lambda_0+\varepsilon\lambda_1)\partial_{\rho} V_{\varepsilon} \\
             \mu_\varepsilon^2\partial_{\rho} U_{\varepsilon} 
        \end{array}\right],\quad \\
        &\Delta_{11} \bold{W}_{cc} - \mathcal{M} \bold{W}_{cc} =  a_3\left[ \begin{array}{c}
             (\lambda_0+\varepsilon\lambda_1)\partial_{\rho} V_{\varepsilon} \\
             \mu_\varepsilon^2\partial_{\rho} U_{\varepsilon} 
        \end{array}\right]\quad ,\quad \Delta_{\rho} \bold{\bar{W}} - \mathcal{M} \bold{\bar{W}} = \bold{0},\qquad \qquad\quad 
    \end{aligned} 
\end{equation}
with far-field conditions
\begin{equation}\label{radially-farfield}
    \begin{aligned}
         &\bold{W}_{sc} \sim \left( \begin{array}{c}
            0 \\
            -\frac{4\pi}{D}\varepsilon S_\varepsilon a_{Q1}\rho-\frac{a_{1}S_{\varepsilon}\mu_\varepsilon^2}{2}
        \end{array}\right ),\quad \bold{W}_{ss} \sim \left( \begin{array}{c}
            0 \\
          -\frac{4\pi}{D}\varepsilon S_\varepsilon a_{Q2}\rho-\frac{a_{2}S_{\varepsilon}\mu_\varepsilon^2}{2}
        \end{array}\right ),\\
        &\bold{W}_{cc} \sim \left( \begin{array}{c}
            0 \\
            \frac{4\pi}{D}\varepsilon S_\varepsilon a_{Q3}\rho-\frac{a_{3}S_{\varepsilon}\mu_\varepsilon^2}{2}
        \end{array}\right ),\quad \bold{\bar{W}} \sim \left( \begin{array}{c}
            0 \\
-\frac{4\pi}{D}\Big[\kappa\big(\chi'(S_\varepsilon)-\frac{1}{\rho}\big) + \varepsilon \kappa R_\mu(\bold{x}_1;\bold{x}_1)\Big]
        \end{array}\right ).
    \end{aligned}   
\end{equation}
Here $\Delta_{11} = \partial_{\rho\rho}+2\rho^{-1}\partial_\rho-2\rho^{-2}$ is the translation mode, and $\mathcal{Q}^T_\mu \bold{a}= (a_{Q1},a_{Q2},a_{Q3})^T$.

The non-homogeneous terms of (\ref{W1}) must be orthogonal to the null space of the homogeneous adjoint operation, given by 
\begin{equation}\label{radially-adjoint}
    \begin{aligned} 
        \Delta_{\bold{y}} \bold{P} + \mathcal{M}^T \bold{P}= \bold{0}, \quad \bold{y} \in \mathbb{R}^3.
    \end{aligned} 
\end{equation}
Similarly, we seek three linearly independent solutions of the form $\bold{P}_{sc} = \bar{\bold{P}}\sin{\theta}\cos{\varphi},  \bold{P}_{ss} = \bar{\bold{P}}\sin{\theta}\sin{\varphi}$, $\bold{P}_{cc} = \bar{\bold{P}}\cos{\theta}$, and the radially symmetric $\bar{\bold{P}}$ satisfies
\begin{equation}
    \begin{aligned} 
        \Delta_1 \bar{\bold{P}} + \mathcal{M}^T \bar{\bold{P}} = \bold{0}, \quad \bar{\bold{P}} = \left( \begin{array}{c}
             \bar{P}_{1} \\
             \bar{P}_{2}
        \end{array}\right)
    \end{aligned} 
\end{equation}
with boundary and far-field conditions
\begin{equation}\label{radially-adjoint farfield}
    \begin{aligned} 
        \bar{\bold{P}}(\bold{0}) = \bold{0}, \quad \bar{\bold{P}} \sim \left(\begin{array}{c}
             0  \\
             1/\rho^2 
        \end{array}\right),\quad \text{as} \quad \rho \rightarrow \infty.
    \end{aligned} 
\end{equation}

To apply the solvability condition, we multiply (\ref{W1}) by $\bold{P}_{sc,ss,cc}$ on the left, respectively, and integrate over a ball with radius $R \gg 1$ centered at the origin, so we obtain
\begin{equation}\label{sovability}
    \begin{aligned}
                \int\int\int_{B_R} \bold{P}^T_{sc,ss,cc}\left[\Delta_{\bold{y}}\bold{W}_1+\mathcal{M}\bold{W}_1\right] d\bold{y}\\
        =\int\int\int_{B_R} \bold{P}^T_{sc,ss,cc}\left[\left(\begin{array}{c}
             (\lambda_0+\varepsilon\lambda_1)\Phi_{\varepsilon}  \\
             \mu_\varepsilon^2\Psi_{\varepsilon}
        \end{array}\right )+\mathcal{N}\left(\begin{array}{c}
             \Phi_{\varepsilon}  \\
             \Psi_{\varepsilon}
        \end{array} \right)\right] d\bold{y}.
    \end{aligned} 
\end{equation}

We now compute each term in (\ref{sovability}). On the left, we use Green's identity to obtain
\begin{equation}\label{left1}
    \begin{aligned} 
        &\int\int\int_{B_R} \bold{P}^T_{sc,ss,cc}\left[\Delta_{\bold{y}_1}\bold{W}_1+\mathcal{M}\bold{W}_1\right] d\bold{y}_1\\
        = &\int_0^\pi\int_0^{2\pi} \left(\bold{P}^T_{sc,ss,cc}\partial_{\rho}\bold{W}_1+ \bold{W}_1^T \partial_{\rho}\bold{P}_{sc,ss,cc} \right)\rho^2|_{\rho = R} \sin{\theta_1}d\varphi_1 d\theta_1.
    \end{aligned} 
\end{equation}
Therefore, with the far-filed in (\ref{radially-farfield}), (\ref{radially-adjoint}) and (\ref{radially-adjoint farfield}), we get

\begin{equation}\label{left}
    \begin{aligned} 
        &\int_0^\pi\int_0^{2\pi} \bold{P}^T_{sc}\partial_{\rho}\bold{W}_1+ \bold{W}_1^T \partial_{\rho}\bold{P}_{sc} \rho^2 |_{\rho = R}\sin{\theta_1}d\varphi_1 d\theta_1\sim -\frac{\varepsilon16\pi^2S_{\varepsilon}a_{Q1}}{D}-\frac{4\pi}{3}\frac{a_1S_{\varepsilon}\mu_\varepsilon^2}{R},\\
         &\int_0^\pi\int_0^{2\pi} \bold{P}^T_{ss}\partial_{\rho}\bold{W}_1+ \bold{W}_1^T \partial_{\rho}\bold{P}_{ss} \rho^2 |_{\rho = R}\sin{\theta_1}d\varphi_1 d\theta_1\sim  -\frac{\varepsilon16\pi^2S_{\varepsilon} a_{Q2}}{D}-\frac{4\pi}{3}\frac{a_2S_{\varepsilon}\mu_\varepsilon^2}{R},\\
        &\int_0^\pi\int_0^{2\pi} \bold{P}^T_{cc}\partial_{\rho}\bold{W}_1+ \bold{W}_1^T \partial_{\rho}\bold{P}_{cc} \rho^2 |_{\rho = R} \sin{\theta_1}d\varphi_1 d\theta_1\sim -\frac{\varepsilon16\pi^2S_{\varepsilon}a_{Q3}}{D}-\frac{4\pi}{3}\frac{a_3S_{\varepsilon}\mu_\varepsilon^2}{R}.
    \end{aligned} 
\end{equation}

\begin{figure}[h]
    \centering
    \includegraphics[width=0.4\linewidth]{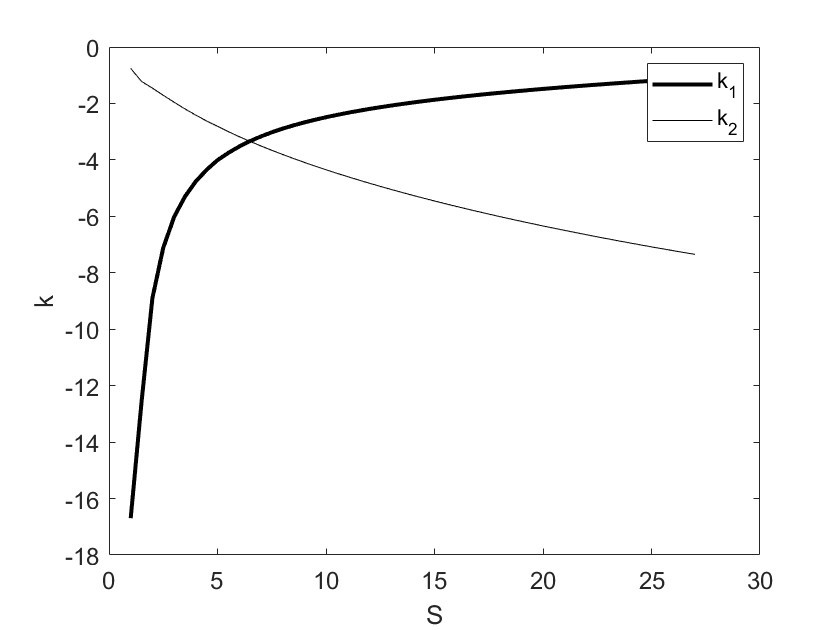}
    \caption{A plot of $k_1$ and $k_2$ versus $S_\varepsilon$ as defined in (\ref{k1k2}).}
    \label{fig:k1k2}
\end{figure}

On the right-hand side of (\ref{sovability_defect}), we apply integration by parts on the first term, and use (\ref{lead-sol}) to obtain
\begin{equation}\label{right_1}
    \begin{aligned} 
       \int\int\int_{B_R} \bold{P}^T_{sc}\left(\begin{array}{c}
             (\lambda_0+\varepsilon\lambda_1)\Phi_{\varepsilon}  \\
             \mu_\varepsilon^2\Psi_{\varepsilon}
        \end{array}\right )d\bold{y}_1 & \sim \frac{4\pi a_1}{3}\left((\lambda_0+\varepsilon\lambda_1)k_{1} - 
        \mu_\varepsilon^2k_{2}-\frac{\mu_\varepsilon^2 S_{\varepsilon}}{R}\right),\\
        \int\int\int_{B_R} \bold{P}^T_{ss}\left(\begin{array}{c}
             (\lambda_0+\varepsilon\lambda_1)\Phi_{\varepsilon}  \\
             \mu_\varepsilon^2\Psi_{\varepsilon}
        \end{array}\right )d\bold{y}_1 & \sim \frac{4\pi a_2}{3}\left((\lambda_0+\varepsilon\lambda_1)k_{1} - 
        \mu_\varepsilon^2k_{2}-\frac{\mu_\varepsilon^2 S_{\varepsilon}}{R}\right),\\
        \int\int\int_{B_R} \bold{P}^T_{cc}\left(\begin{array}{c}
             (\lambda_0+\varepsilon\lambda_1)\Phi_{\varepsilon}  \\
             \mu_\varepsilon^2\Psi_{\varepsilon}
        \end{array}\right )d\bold{y}_1 & \sim \frac{4\pi a_3}{3}\left((\lambda_0+\varepsilon\lambda_1)k_{1} -
        \mu_\varepsilon^2k_{2}-\frac{\mu_\varepsilon^2 S_{\varepsilon}}{R}\right),
    \end{aligned} 
\end{equation}
where $k_{1}$ and $k_{2}$ are defined by the integrals
\begin{equation}\label{k1k2}
        k_{1} = \int_0^\infty V'_{\varepsilon1} \bar{P}_{1} \rho^2 d \rho, \quad k_{2} = \int_0^\infty[U_{\varepsilon}-\chi(S_{\varepsilon})](\bar{P}_{2}\rho^2)'d\rho.
\end{equation}
We plot the value of $k_1$ and $k_2$ versus $S_\varepsilon$ in Figure \ref{fig:k1k2}.

For the last term of the right side of (\ref{sovability_defect}), we use (\ref{lead-sol}) to compute
\begin{equation}\label{N_UV}
    \mathcal{N}\left(\begin{array}{c}
             \Phi_{\varepsilon}  \\
             \Psi_{\varepsilon}
        \end{array} \right) = \mathcal{N} \bold{a}_1\cdot \nabla_{\bold{y}_1}\left(\begin{array}{c}
             V_{\varepsilon}\\
             U_{\varepsilon} 
        \end{array}\right).
\end{equation}
Expanding (\ref{N_UV}), we have
 \begin{equation}
    \begin{aligned} 
        \mathcal{N}\left(\begin{array}{c}
             \Phi_{\varepsilon}  \\
             \Psi_{\varepsilon}
        \end{array} \right) =\left(\begin{array}{c}
             -U_{3}\bold{a}_1\cdot \nabla_{\bold{y}_1}V_{\varepsilon}^2-V_{3}\bold{a}_1\cdot \nabla_{\bold{y}_1}(2V_{\varepsilon}U_{\varepsilon}) \\
            U_{3}\bold{a}_1\cdot \nabla_{\bold{y}_1}V_{\varepsilon}^2+V_{3}\bold{a}_1\cdot \nabla_{\bold{y}_1}(2V_{\varepsilon}U_{\varepsilon})
        \end{array}\right). 
    \end{aligned} 
\end{equation}

By constructing full derivatives, and passing the operator  $\bold{a}_1\cdot \nabla_{\bold{y}_1}$, we find
\begin{equation}
    \begin{aligned} 
        \mathcal{N}\left(\begin{array}{c}
             \Phi_{\varepsilon}  \\
             \Psi_{\varepsilon}
        \end{array} \right) =\left(\begin{array}{c}
             -\bold{a}_1\cdot \nabla_{\bold{y}_1}(V_{\varepsilon}^2U_{3}+2U_{\varepsilon}V_{\varepsilon}V_{3})+V_{\varepsilon}^2\bold{a}_1\cdot \nabla_{\bold{y}_1}U_{3}+2U_{\varepsilon}V_{\varepsilon}\bold{a}_1\cdot \nabla_{\bold{y}_1}V_{3}\\
             \bold{a}_1\cdot \nabla_{\bold{y}_1}(V_{\varepsilon}^2U_{3}+2U_{\varepsilon}V_{\varepsilon}V_{3})-V_{\varepsilon}^2\bold{a}_1\cdot \nabla_{\bold{y}_1}U_{3}-2U_{\varepsilon}V_{\varepsilon}\bold{a}_1\cdot \nabla_{\bold{y}_1}V_{3}
        \end{array}\right).
    \end{aligned} 
\end{equation}
From (\ref{o3}), we observe that
\begin{equation}
    \begin{aligned} 
        -\bold{a}_1\cdot \nabla_{\bold{y}_1}(V_{\varepsilon}^2U_{3}+2U_{\varepsilon}V_{\varepsilon}V_{3}) &= \Delta_{\bold{y}_1}\bold{a}_1\cdot \nabla_{\bold{y}_1}V_{3}-\bold{a}_1\cdot \nabla_{\bold{y}_1}V_{3},\\
         \bold{a}_1\cdot \nabla_{\bold{y}_1}(V_{\varepsilon}^2U_{3}+2U_{\varepsilon}V_{\varepsilon}V_{3}) &= \Delta_{\bold{y}_1}\bold{a}_1\cdot \nabla_{\bold{y}_1}U_{3}.
    \end{aligned} 
\end{equation}
Hence we have
\begin{equation}\label{fst-right_o3}
    \begin{aligned} 
       &\mathcal{N}\left(\begin{array}{c}
             \Phi_{\varepsilon}  \\
             \Psi_{\varepsilon}
        \end{array} \right)=\left[\Delta_{\bold{y}_1}+\mathcal{M}\right]\left(\bold{a}_1\cdot \nabla_{\bold{y}_1}\right)\bold{n};\quad \bold{n} = \left(\begin{array}{c}
            V_{3} \\
             U_{3}
        \end{array}\right),\\
        &\bold{a}_1\cdot \nabla_{\bold{y}_1}\bold{n} \sim \left(\begin{array}{c}
             0 \\
             -4\pi\bold{a}^TS_{\varepsilon} \mathcal{H}_{11}\bold{e}\rho
        \end{array}\right),\quad \text{as} \quad \rho \rightarrow \infty.
    \end{aligned} 
\end{equation}
Multiplying (\ref{fst-right_o3}) by $\bold{P}^T_{sc,ss,cc}$ and applying the divergence theorem, we get
\begin{equation}\label{right_2}
    \begin{aligned} \int\int\int_{B_R}\bold{P}^T_{sc,ss,cc}\mathcal{N}\left(\begin{array}{c}
             \Phi_{\varepsilon}  \\
             \Psi_{\varepsilon}
        \end{array} \right)d\bold{y}_1 \sim -16 \pi^2S_{\varepsilon}\Big(a_1\mathcal{H}^{(1n)}_{11}+a_2\mathcal{H}^{(2n)}_{11}+a_3\mathcal{H}^{(3n)}_{11}\Big) ,\quad n = 1,2,3,
    \end{aligned} 
\end{equation}
where $\mathcal{H}^{(mn)}_{ji}$ is the $(m,n)$th entry of $\mathcal{H}_{ji}$.

Now we substitute (\ref{left}), (\ref{right_1}) and (\ref{right_2}) to (\ref{sovability}) with $R\gg 1$ and cancel the term of $1/R$, yielding 
\begin{equation}\label{left=right}
    \begin{aligned} 
       -\frac{\varepsilon16\pi^2 S_{\varepsilon} a_{Qn}}{D} = \frac{4\pi a_{n}}{3}\left((\lambda_0+\varepsilon\lambda_1)k_{1} 
       - \mu_\varepsilon^2k_{2}\right)-16\varepsilon \pi^2S_{\varepsilon}\Big(a_1\mathcal{H}^{(1n)}_{11}+a_2\mathcal{H}^{(2n)}_{11}+a_3\mathcal{H}^{(3n)}_{11}\Big),\quad n = 1,2,3,
    \end{aligned} 
\end{equation}
To obtain the leading order of $\tau$, we collect the leading order terms from (\ref{left=right}) and use $\mu_\varepsilon^2 = \frac{1}{D}\Big(\tau_0\lambda_0 +\varepsilon (\tau_0\lambda_1+\tau_1\lambda_0)\Big)$ to have
\begin{equation}
    (k_2\tau_0-Dk_1)\lambda_0  = 0,
\end{equation}
where $k_1$ and $k_2$ are defined in (\ref{k1k2}). Assuming $\lambda_0$ is nonzero in (\ref{scaling_tau}), we obtain the leading order threshold is given by
    \begin{equation}\label{tau0}
    \tau_0 = \frac{Dk_1}{k_2},
\end{equation}
Similarly, by collecting $\mathcal{O}(\varepsilon)$ terms from (\ref{left=right}), we get
\begin{equation}\label{order_eps}
\begin{aligned}
        -\frac{16\pi^2 S_{\varepsilon} a_{Qn}}{D} = \frac{4\pi a_{n}}{3}\left(\lambda_1k_{1} 
       - (\tau_0\lambda_1+\tau_1\lambda_0)k_{2}\right)-16 \pi^2S_{\varepsilon}\Big(a_1\mathcal{H}^{(1n)}_{11}+a_2\mathcal{H}^{(2n)}_{11}+a_3\mathcal{H}^{(3n)}_{11}\Big),\quad n = 1,2,3,
\end{aligned}
\end{equation}
By (\ref{tau0}), we further simplify (\ref{order_eps}) to 
\begin{equation}\label{order_eps2}
    \begin{aligned}
            \frac{12\pi S_{\varepsilon} a_{Qn} }{D}= a_{n}\tau_1\lambda_0k_{2}+12 \pi S_{\varepsilon}\Big(a_1\mathcal{H}^{(1n)}_{11}+a_2\mathcal{H}^{(2n)}_{11}+a_3\mathcal{H}^{(3n)}_{11}\Big),\quad n = 1,2,3,
    \end{aligned}
\end{equation}
Substituting $a_{Qi}$, we arrive at a $3 \times 3$ matrix-eigenvalue problem for the oscillation modes $\bold{a}$, Hopf bifurcation frequency $\lambda_0$ and the first correction of the threshold $\tau_1$,
\begin{equation}\label{eig_matrix}
    \frac{12\pi}{\tau_1k_2}\left[S_{\varepsilon}(\mathcal{Q}_{\mu1}-D\mathcal{H}_{11})^T\right]\bold{a} = \lambda_0 \bold{a},
\end{equation}
where $\mathcal{Q}_{\mu1}$ is defined in (\ref{Qmu}), and $\mathcal{H}_{11}$ is given in (\ref{o3}). As shown in the matrix eigenvalue problem, the value of $\kappa$ does not contribute to the computation of solvability or to calculating the Hopf bifurcation (HB) threshold. Therefore, in the following analysis, we will not calculate the value of $\kappa$.

To determine the threshold of $\tau$ beyond which translatory oscillatory instability is triggered, we set $\lambda_0 = \lambda_{I}i$ and $\mu_I^2 = \tau_0 \lambda_{I}i$. By solving two transcendental equations, namely
\begin{equation}\label{eig}
    \begin{aligned}
        \bold{Re}\left\{\text{det}\left(\frac{12\pi}{\tau_1k_2}\left[S_{\varepsilon}(\mathcal{Q}_{\mu1}-\mathcal{H}_{11})^T-S_{\varepsilon0}\mathcal{H}_{10}^T\right]- \lambda_{I} i \mathcal{I}_{3}\right) \right\} = 0,\\
        \bold{Im}\left\{\text{det}\left(\frac{12\pi}{\tau_1k_2}\left[S_{\varepsilon}(\mathcal{Q}_{\mu1}-\mathcal{H}_{11})^T-S_{\varepsilon0}\mathcal{H}_{10}^T\right]- \lambda_{I} i \mathcal{I}_{3}\right) \right\} = 0,
    \end{aligned}
\end{equation}
we obtain the critical value $\tau^*_1$ and the corresponding eigenvalue $\lambda^*_I$. Together with (\ref{tau0}), we thus conclude that $\tau^* = \tau^*_0 + \varepsilon \tau^*_1$.

As $\tau$ increases beyond $\tau^*$, the real part of the dominant complex eigenvalue crosses the imaginary axis from left to right, entering the positive side and triggering translational oscillatory instability. In an asymmetric geometry, multiple solutions may be found in the system (\ref{eig}), each corresponding to a distinct oscillation mode with frequency $2\pi/\lambda^*_I$. However, the leading order Hopf bifurcation value $\tau^*_0$, governed by the source strength of the spot $S{\varepsilon}$, remains the same across different oscillation modes. Thus, the preferred oscillatory mode is determined by the first-order correction $\tau^*_1$. The minimum $\tau^*$ among all possible values represents the Hopf stability threshold, with the corresponding null space $\bold{a}^*$ of the matrix in (\ref{eig}) indicating the direction of spot oscillation at onset

In this section, we examine the problem on various types of domains, including a unit sphere, a perturbed sphere, and a defected sphere. The purpose of this section is to investigate how geometry affects the preferred direction of oscillation at onset.

\subsection{The unit sphere}\label{unit}
 In an unit sphere, the matrices $Q_{\mu1}$ and $\mathcal{H}_{11}$ are diagonal and can be expressed as a constant times an identity matrix. Therefore, the eigenvalue system (\ref{eig_matrix}) can be simplified to 
\begin{equation}\label{eigenmatrix_ball}
        12\pi S_\varepsilon\left(\mathcal{Q}_{\mu1}^{(1,1)} - D\mathcal{H}_{11}^{(1,1)}\right)-k_{2}\tau_{1}\lambda_{I} = 0.
\end{equation}
In this case, the vector $\bold{a}$, which indicates the direction of spot oscillation at onset, becomes arbitrary, and there is no preferred direction of oscillation.

We now solve the equation (\ref{eigenmatrix_ball}). We first need $\mathcal{Q}_{\mu1}$, the gradient with respect to the source location of the gradient of the regular part of the Helmholtz Green's function, and the Hessian matrix of the Neumann Green's function $\mathcal{H}_{11}$. 

We begin by computing $\mathcal{Q}_{\mu1}$. Rather than solving the Helmholtz Green's function on a unit sphere and then differentiating with respect to the source location, we directly solve PDE of the gradient of the Helmholtz Green's function. Let $\bold{g}_0 = \nabla_{\bold{x}_1}G_\mu(\bold{x};\bold{x}_1)$, which satisfies
\begin{equation}
    \begin{aligned}
            &\Delta\bold{g}_0 - \mu^2 \bold{g}_0 = \nabla_{\bold{x}_1}\delta(\bold{x}-\bold{x}_1), \quad \bold{x}\in \Omega; \quad \partial_n \bold{g}_0 = \bold{0},\quad \bold{x}\in \partial\Omega;\\
           & \bold{g}_0 \sim -\frac{(\bold{x}-\bold{x}_1)}{4\pi|\bold{x}-\bold{x}_1|^3}+\nabla_{\bold{x}_1}R_\mu(\bold{x};\bold{x}_1)+\frac{\mu^2}{8\pi}\frac{(\bold{x}-\bold{x}_1)}{|\bold{x}-\bold{x}_1|},\quad \text{as} \quad \bold{x}\rightarrow \bold{x}_1.
    \end{aligned}
\end{equation}
We highlight that $\bold{g}_0$ is a $3\times1$ vector, so its far-filed behavior of $\bold{g}_0$ contains three components.
The explicit solution is 
\begin{equation}\label{g0}
    \bold{g}_0 = -\left(\frac{\mu}{2\pi}\right)^{\frac{3}{2}}F(r)\bold{e}, \quad F(r) = r^{-\frac{1}{2}}\Big(Q(\mu)I_{\frac{3}{2}}(\mu r)+K_{\frac{3}{2}}(\mu r)\Big),
\end{equation}
where $r = |\bold{x}-\bold{x}_1|$, $\bold{e}$ is defined in (\ref{inner_var}), $I_n(z)$ and $K_n(z)$ are the modified Bessel functions
\begin{equation}\label{Q(mu)}
    Q(\mu) = -\frac{2\mu K'_{3/2}(\mu)-K_{3/2}(\mu)}{2\mu I_{3/2}'(\mu)-I_{3/2}(\mu)}.
\end{equation}
Using $\nabla_{\bold{x}} = \bold{e}\partial_r +\frac{1}{r \sin^2\theta} \bold{e}_\varphi \partial_\varphi+\frac{1}{r}\bold{e}_\theta \partial_\theta$ with 
\begin{equation}\label{de}
    \bold{e}_\varphi = \left(\begin{array}{c}
         -\sin\theta\sin\varphi  \\
         \sin\theta\cos\varphi \\
         0
    \end{array}\right),\quad  \bold{e}_\theta = \left(\begin{array}{c} \cos\theta\cos\varphi\\
    \cos\theta\sin\varphi\\
    -\sin\theta
    \end{array}\right),
\end{equation}
and setting $\bold{x}=\bold{x}_1$, we obtain
\begin{equation}\label{grad_g}
    \begin{aligned}
            \nabla_{\bold{x}}g_0^{(1)}|_{\bold{x} = \bold{x}_1} =  -\left(\frac{\mu}{2\pi}\right)^{\frac{3}{2}}\Big[F'(r_0) \sin{\theta_0}\cos{\varphi_0}\bold{e}_0-\frac{1}{r_0\sin^2{\theta_0}}F(r_0)\sin{\theta_0}\sin{\varphi_0}\bold{e}_{\varphi_0}\\
            +\frac{1}{r_0}F(r_0)\cos{\theta_0}\cos{\varphi_0}\bold{e}_{\theta_0}\Big].
    \end{aligned}
\end{equation}
From the fact that for a small argument, $|z|\ll1$, 
\begin{equation}
    \begin{aligned}
            &I_{\frac{3}{2}}(z)z^{-\frac{1}{2}} \sim \frac{1}{3}\sqrt{\frac{2}{\pi}}z+\mathcal{O}(z^3),\\
           & K_{\frac{3}{2}}(z)z^{-\frac{1}{2}} \sim \sqrt{\frac{\pi}{2}}z^{-2}-\frac{1}{2}\sqrt{\frac{\pi}{2}}+\frac{1}{3}\sqrt{\frac{\pi}{2}}z+\mathcal{O}(z^2),
   \end{aligned}
\end{equation}
we find 
\begin{equation}\label{F}
    \begin{aligned}
            &\lim_{r\rightarrow0^+}F'(r) = -\frac{\sqrt{2\pi}}{\mu^{3/2}r^3}+\frac{\big(\pi+2Q(\mu)\big)\mu^{3/2}}{3\sqrt{2\pi}},\\
           & \lim_{r\rightarrow0^+}F(r)/r = \sqrt{\frac{\pi}{2}}\frac{1}{\mu^{3/2}r^3}-\sqrt{\frac{\pi\mu}{2}}\frac{1}{2r}+\frac{\big(\pi+2Q(\mu)\big)\mu^{3/2}}{3\sqrt{2\pi}}.
    \end{aligned}
\end{equation}
Recalling that $\mathcal{Q}_{\mu1}$ is computed from the regular part of $G_\mu(\bold{x};\bold{x}_1)$, we extract the regular part from (\ref{F}) and substitute into (\ref{grad_g}), yielding the first column of $\mathcal{Q}_{\mu1}$,
\begin{equation}\label{1stcolumnQ}
    \left(\begin{array}{c}
         \mathcal{Q}_{\mu1}^{(1,1)}  \\
         \mathcal{Q}_{\mu1}^{(2,1)} \\
         \mathcal{Q}_{\mu1}^{(3,1)}
    \end{array} \right) = -\frac{\big(\pi+2Q(\mu)\big)\mu^{3}}{12\pi^2}\left(\begin{array}{c}
         1  \\
         0 \\
         0
    \end{array} \right).
\end{equation}
As expected, the second and third elements of the first column are zero, indicating the diagonal structure of $\mathcal{Q}_{\mu1}$ due to the symmetric properties of an unit sphere. The similar results hold when computing the second and third columns of $\mathcal{Q}_{\mu1}$. 

The Neumann Green's function $G(r)$ satisfying (\ref{Neumann_Green}) with source at the origin is given by
\begin{equation}\label{H}
    \begin{aligned} 
        G(r) = \frac{1}{4\pi r} +\frac{r^2}{8\pi}-\frac{9}{20\pi}.
    \end{aligned} 
\end{equation}
Thus the Hessian term is $\mathcal{H}_{11}^{(1,1)} = \frac{1}{4\pi}$.

From (\ref{1stcolumnQ}) and using $\mathcal{H}_{11}^{(1,1)} = (4\pi)^{-1}$, we derive, from (\ref{eigenmatrix_ball}),
\begin{equation}\label{tau1lamb0}
        12\pi S_\varepsilon\left(- \frac{\big(\pi+2Q(\mu)\big)\mu^{3}}{12\pi^2} - \frac{D}{4\pi}\right)-k_{2}\tau_{1}\lambda_{I} = 0.
\end{equation}

\begin{figure}[h]
    \centering
    \begin{subfigure}[b]{0.45\textwidth}
        \centering
        \includegraphics[width=\textwidth]{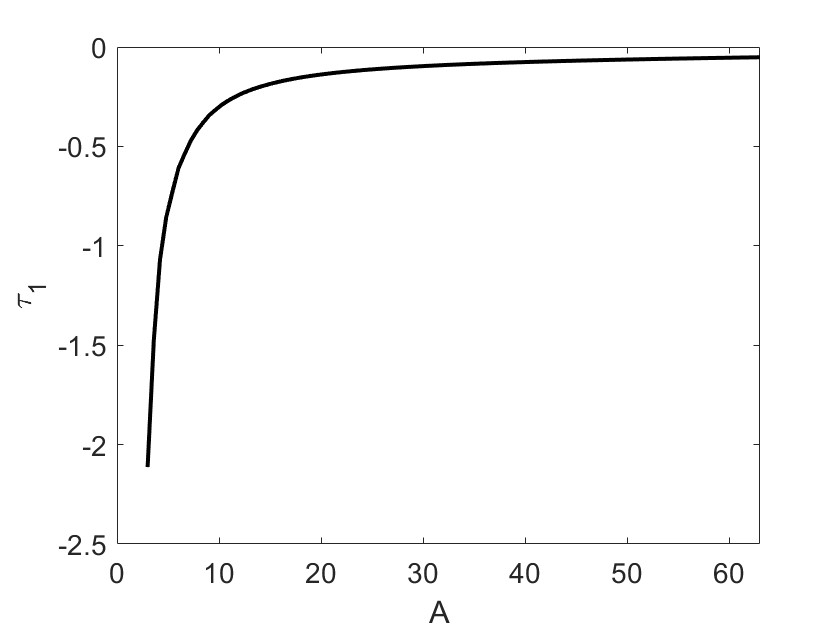} 
        \caption{$A$ versus $\tau_1$}
        \label{fig:AvsTau1_samll}
    \end{subfigure}
    \begin{subfigure}[b]{0.45\textwidth}
        \centering
        \includegraphics[width=\textwidth]{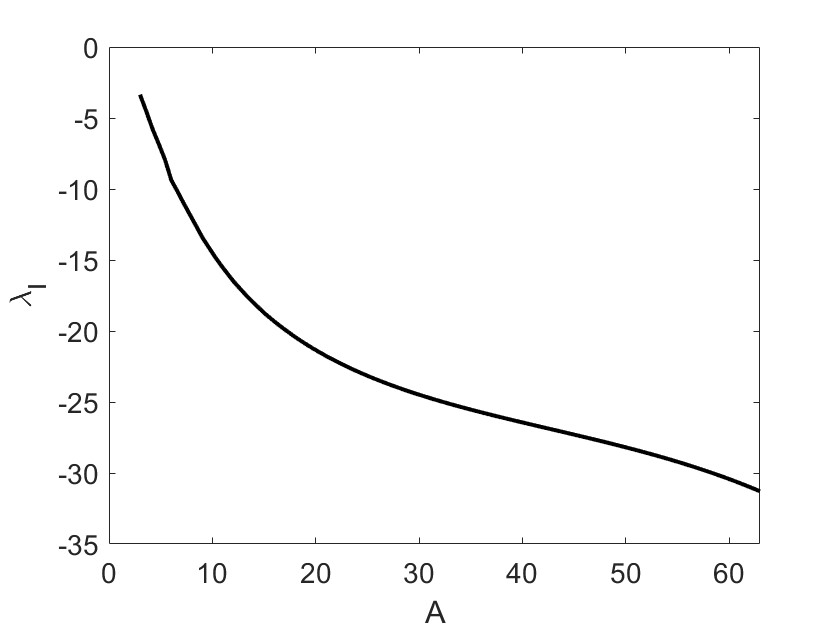} 
        \caption{$A$ versus $\lambda_I$}
        \label{fig:AvsLamb0_small}
    \end{subfigure}
    \caption{(a) First order correction of Hopf bifurcation threshold and (b) the leading order of the corresponding eigenvalue obtained by solving (\ref{tau1lamb0}). $\tau_1$ is approaching zero as $A$ increase. $\lambda_I$ is the pure imaginary eigenvalue. }
    \label{fig:eigenvalue_small1}
\end{figure}

By solving (\ref{tau1lamb0}), we obtain $\tau_1$ and the corresponding leading-order eigenvalue $\lambda_I$. We plot $\tau_1$ and $\lambda_I$ as functions of the feed rate $A$ in Figure \ref{fig:eigenvalue_small1}. As shown in (\ref{fig:AvsTau1_samll}), the value of $\tau_1$ monotonically approaches zero as $A$ increases. Hence, as $A$ becomes large, the value of $\tau_1$ is too small to contribute as a correction term. Therefore, we compute the next-order correction, $\mathcal{O}(\varepsilon^2)$, for $\tau$.

Let $\tau = \varepsilon^{-3}(\tau_0 + \varepsilon \tau_1 + \varepsilon^2 \tau_2)$ and $\lambda = \varepsilon^2 ( \lambda_0 + \varepsilon \lambda_1+\varepsilon^2\lambda_2)$. We now aim to derive an eigenvalue system similar to (\ref{eig_matrix}) in Section \ref{stability_samll}. The process of computing $\tau_2$ and $\lambda_1$ follows that in Section \ref{stability_samll}, therefore, we focus only on the differences in the following analysis.

In the outer region, we expand $\psi \sim \varepsilon^2 \psi_1 + \varepsilon^3 \psi_2 + \varepsilon^4 \psi_3$, where $\psi_3$ satisfies
\begin{equation}\label{psi3}
    \Delta\psi_3-\mu^2 \psi_3 = \mu_1^2\psi_2 + (\tau_0\lambda_2+\tau_1\lambda_1+\tau_2\lambda_0)\psi_1,\quad \bold{x} \in \Omega;\quad \partial_n \psi_3 = 0, \bold{x}\in\partial\Omega.
\end{equation}
In a unit ball, we let
\begin{equation}
    \psi_i = \bold{a}^T\bold{e}\tilde{\psi_i}(r), \quad r = |\bold{x}-\bold{x}_1|,\quad i = 1,2,3,
\end{equation}
so that (\ref{psi3}) simplifies to
\begin{equation}\label{radially_psi3}
    \Delta_{11}\tilde{\psi_3} - \mu^2 \tilde{\psi_3} = \mu_1^2\tilde{\psi_2} + \frac{\tau_0\lambda_2+\tau_1\lambda_1+\tau_2\lambda_0}{D}\tilde{\psi_1},
\end{equation}
where $\Delta_{11} = \partial_{\rho\rho} + 2 \rho^{-1} \partial_{\rho} - 2 \rho^{-2}$. We solve (\ref{radially_psi3}) to find
\begin{equation}\label{h3h4}
    \tilde{\psi_3} \sim \frac{1}{r^2}\Big[h_1e^{\mu r}(\mu r-1)+h_2e^{-\mu r}(\mu r+1)\Big]+\frac{\tau_0\lambda_2+\tau_1\lambda_1+\tau_2\lambda_0}{2D}\Big[h_3 e^{\mu r}-h_4 e^{-\mu r}\Big] + \mathcal{O}(r),
\end{equation}
where $h_3 = 1 / (Q(\mu) - 1)$, $h_4 = Q(\mu) / (Q(\mu) - 1)$, and $Q(\mu)$ is defined in (\ref{Q(mu)}). Thus, the local behavior of $\tilde{\psi_3}$ is
\begin{equation}\label{local_psi3}
    \tilde{\psi_3} \sim \frac{h_1-h_2}{r^2} + \frac{\tau_0\lambda_2+\tau_1\lambda_1+\tau_2\lambda_0}{2D}(h_3 - h_4) - \frac{\mu^2}{2}(h_2-h_1).
\end{equation}
The first term in (\ref{local_psi3}) matches the singular term of $\Psi_2$ in the inner region, as described in Section \ref{stability_samll}, while the second term matches $\Psi_4$ in the inner region. The last term in (\ref{local_psi3}) corresponds to the $\mathcal{O}(\varepsilon)$ term in the inner region, which combines with $\mathcal{O}(1)$ terms to form $\Psi_\varepsilon$. However, in a single-spot pattern, the $\mathcal{O}(\varepsilon)$ term is actually zero because $U_1$ is zero according to \cite{jst-3d}, and $\Psi_1$ is derived from the partial derivative of $U_1$. This results in $h_1 = h_2$, causing the last term to vanish, as well as the singular term in (\ref{local_psi3}).

Using the local behavior of $\psi_\varepsilon$ in (\ref{outer_psi}) and $\tilde{\psi_3}$ in (\ref{local_psi3}), we derive the far-field behavior of $\Psi_4$ in the inner region, yielding
\begin{equation}\label{far-field-psi4}
    \Psi_4 \sim \bold{a}^T\bold{e}S_\varepsilon\Big\{-\frac{1}{8}\mu^4\rho^2++\frac{2d_1}{3}\rho+\frac{\mu\mu_1^2}{2}(h_3+h_4)\rho-\frac{\tau_0\lambda_2+\tau_1\lambda_1+\tau_2\lambda_0}{2D}\Big\},\quad \text{as}\quad \rho\rightarrow\infty,
\end{equation}
where $d_1 = -\frac{e^{2\mu}\mu_1^2(\mu^6+2\mu^4)}{\big(2+2\mu+\mu^2-e^{2\mu}(2-2\mu+\mu^2)\big)^2}$. In the inner region of $\mathcal{O}(\varepsilon^4)$, $\Phi_4$ and $\Psi_4$ satisfy
\begin{equation}\label{inner_psi4}
    \Delta_{\bold{y}}\left(\begin{array}{c}
         \Phi_4 \\
         \Psi_4
    \end{array}\right) + \mathcal{M}\left(\begin{array}{c}
         \Phi_4 \\
         \Psi_4
    \end{array}\right) = \mathcal{N}_1\left(\begin{array}{c}
         \Phi_\varepsilon \\
         \Psi_\varepsilon
    \end{array}\right)+\left(\begin{array}{c}
         \lambda_2\Phi_\varepsilon+\lambda_0\Phi_2 \\
         \frac{\tau_0\lambda_2+\tau_1\lambda_1+\tau_2\lambda_0}{D}\Psi_\varepsilon+\mu^2\Psi_2
    \end{array}\right).
\end{equation}
Here $\mathcal{N}_1$ is
\begin{equation}
    \mathcal{N}_1 = \left(\begin{array}{cc}
      2U_4V_\varepsilon+2U_\varepsilon V_4  & 2V_\varepsilon V_4 \\
      -2U_4V_\varepsilon-2U_\varepsilon V_4   & -2V_\varepsilon V_4
   \end{array}\right),
\end{equation}
where $V_4$ and $U_4$ are the $\mathcal{O}(\varepsilon^4)$ correction to the equilibrium solution in the inner region, satisfying
\begin{equation}
    \begin{aligned}
            \Delta_{\bold{y}}\left( \begin{array}{c}
         V_4 \\
         U_4
    \end{array}\right) + \mathcal{M} \left( \begin{array}{c}
         V_4 \\
         U_4
    \end{array}\right) =\bold{0},\\
    V_4 \rightarrow 0, \quad U_4 \rightarrow c_3 -\frac{c_4}{\rho},\quad \text{as} \quad \rho \rightarrow 0.
    \end{aligned}
\end{equation}
Here, $c_3$ and $c_4$ are constants obtained by matching to the outer solution. However,When applying the solvability conditions to (\ref{inner_psi4}), we compute $\bold{a} \cdot \nabla_{\bold{y}} U_4$, following the process from (\ref{N_UV}) to (\ref{right_2}), which causes the constant $c_3$ to disappear. Additionally, the singular term in the far-field of $U_4$ must be balanced by other terms in the solvability conditions, so we do not need to compute $c_4$ either.

Now we apply the solvability conditions on (\ref{inner_psi4}). We multiply (\ref{inner_psi4}) by $\bold{P}_{sc,ss,cc}$ on both side and integrate over a large ball, yielding
\begin{equation}
    \begin{aligned}
            \int\int\int_{R_B}\bold{P}_{sc,ss,cc}^T\left( \Delta_{\bold{y}}\left(\begin{array}{c}
         \Phi_4 \\
         \Psi_4
    \end{array}\right) + \mathcal{M}\left(\begin{array}{c}
         \Phi_4 \\
         \Psi_4
    \end{array}\right)\right)d\bold{y} =  \\
    \int\int\int_{R_B}\bold{P}_{sc,ss,cc}^T\left(\mathcal{N}_1\left(\begin{array}{c}
         \Phi_\varepsilon \\
         \Psi_\varepsilon
    \end{array}\right)+\left(\begin{array}{c}
         \lambda_2\Phi_\varepsilon+\lambda_0\Phi_2 \\
         \frac{\tau_0\lambda_2+\tau_1\lambda_1+\tau_2\lambda_0}{D}\Psi_\varepsilon+\mu^2\Psi_2
    \end{array}\right)\right) d\bold{y}.
    \end{aligned}
\end{equation}
Following the steps from (\ref{left1}) to (\ref{left=right}), we arrive at an equation
\begin{equation}\label{tau2lamb1}
    S_\varepsilon\Big(2d_1+\frac{3\mu}{2}\mu_1^2(h_3+h_4)\Big) = (\tau_2\lambda_0+\tau_1\lambda_1)k_2+\lambda_0 H_1 - \mu^2 H_2,
\end{equation}
where $\mu^2 = \tau_0 \lambda_0$, $\mu_1^2 = \tau_0 \lambda_1 + \tau_1 \lambda_0$, $k_2$ is defined in (\ref{k1k2}), $h_3$ and $h_4$ are defined in (\ref{h3h4}), and $d_1$ is given in (\ref{far-field-psi4}). Additionally, $H_1 = \int_0^\infty P_1 \tilde{\Phi}_2(\rho) \rho^2 d\rho$ and $H_2 = \int_0^\infty \tilde{\Psi}_2(\rho) \rho^2 d\rho$, where $\tilde{\Phi}_2$ and $\tilde{\Psi}_2$ are the radial parts of $\Phi_2$ and $\Psi_2$, respectively.

We now summarize the solution for the problem in a unit sphere. From (\ref{tau0}), we calculate the leading order of $\tau$. By solving (\ref{tau1lamb0}), we obtain the first order correction to $\tau$, namely $\tau_1$, and the leading order of $\lambda_I$. By setting $\lambda$ as purely imaginary in (\ref{tau2lamb1}) and solving it, we obtain $\tau_2$ and $\lambda_1$. Consequently, we compute the HB threshold $\tau$ to the $\mathcal{O}(\varepsilon^2)$ correction and the oscillation frequency $\lambda_I$ to the $\mathcal{O}(\varepsilon)$ correction.

\begin{figure}[h]
    \centering
    
\end{figure}
\begin{figure}[H]
    \centering
    \begin{subfigure}[b]{0.45\textwidth}
        \centering
        \includegraphics[width=\textwidth]{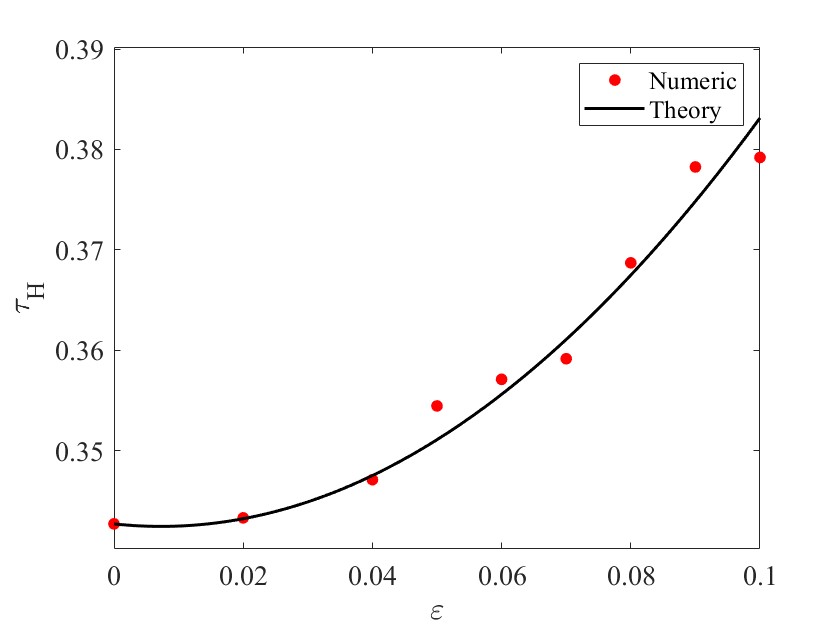}
    \caption{$\tau_H$ versus $\varepsilon$}
    \label{fig:tauvseps}
    \end{subfigure}
    \begin{subfigure}[b]{0.45\textwidth}
        \centering
        \includegraphics[width=\textwidth]{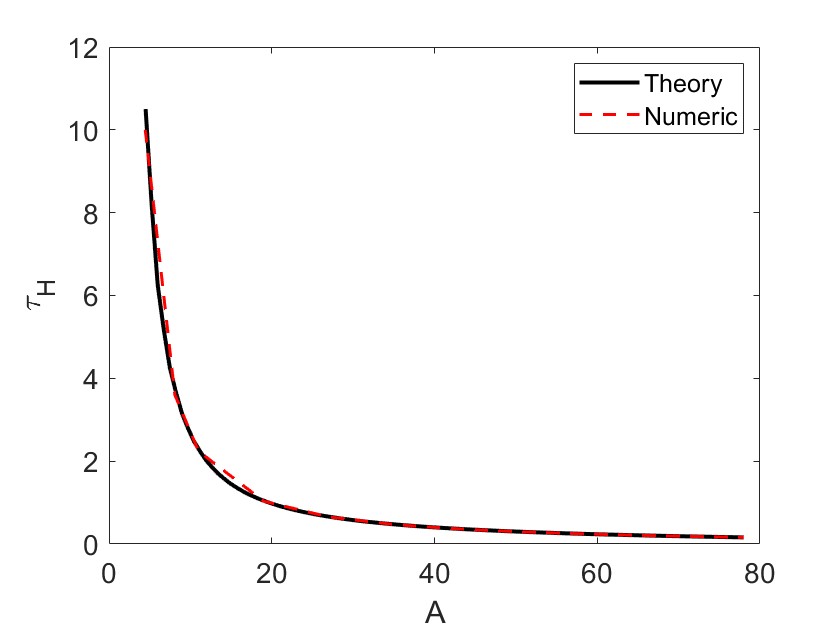} 
        \caption{$\tau_H$ versus $A$}
        \label{fig:AvsTau_small}
    \end{subfigure}
    \caption{(a) A plot of $\tau_H$, including $\mathcal{O}(\varepsilon^2)$ corrections, versus $\varepsilon$ for $S_\varepsilon = 15$. Numerical data are generated using FlexPDE.  (b) A plot of $\tau_H$ as a function of $A$ with $\varepsilon = 0.03$. Numerical data are generated by solving the full eigenvalue problem in (\ref{eigen_problem}), and the analytical results show an excellent fit with the numerical data.}
    \label{fig:eigenvalue_small}
\end{figure}
In Figure \ref{fig:tauvseps}, we use FlexPDE to simulate the problem and generate numerical results for the threshold across various values of $\varepsilon$, ranging from 0 to 0.1, with $S_\varepsilon = 15$. The analytical results show an excellent fit with the numerical data. In Figure \ref{fig:AvsTau_small}, we plot $A$ versus $\tau_H$ for $\varepsilon = 0.03$. These comparisons validate the accuracy of the theoretical predictions against computational simulations.

\subsection{The perturbed sphere}\label{perturbed}
In this section, we consider the problem on a perturbed sphere with boundary parameterized by $(x_1, x_2, x_3) = (1 + \sigma f(\theta, \varphi))\bold{e}$, where $\varepsilon \ll \sigma \ll 1$, $\bold{e} = (\sin{\theta} \cos{\varphi}, \sin{\theta} \sin{\varphi}, \cos{\theta})^T$, and $f(\theta, \varphi)$ is a periodic function over the interval $\varphi \in [0, 2\pi)$ and $\theta \in [0, \pi)$. The function $f(\theta, \varphi)$ can be expressed in terms of spherical harmonics as
\begin{equation}
    f(\theta,\varphi) = P_n^m(\cos{\theta}) \cos{m \varphi},\quad 0\leq m \leq n \quad \text{and} \quad n \in \mathbb{N} ,
\end{equation}
where $P_n^m(\cos{\theta})$ is an associated Legendre polynomial. The negative harmonics result from a rotation of the positive harmonics about the $x_3$-axis by $\pi/(2m)$. Thus, we analyze the non-negative harmonics as representatives. In this problem, we aim to gain insight into how the geometry influences the preferred direction of oscillation at onset.

In the following analysis, we show that the $n = 2$ modes affect the bifurcation threshold, oscillatory frequency, and oscillation mode at $\mathcal{O}(\sigma)$. The effect of the perturbation on the domain is to stretch it in one direction, transforming the geometry into an ellipsoid. When $m=0$, the domain stretches along the $x_3$-axis. We focus on the case $n = 2$ and $m=0$, in which the $x_3$-axis is semi-major axis.

We begin by computing the perturbed boundary conditions. The effect of the boundary perturbation on $v(\bold{x})$ is minimal due to its localized properties; therefore, we focus on its effect on $u(\bold{x})$.

We expand 
\begin{equation}
    u \sim u_0(\rho) + \sigma u_1(\rho,\theta,\varphi),
\end{equation}
and compute the boundary conditions for $u_1$. The outward normal on $\partial\Omega$ is
\begin{equation}
    \vec{\bold{n}} = (1+\sigma f) \bold{e} -\sigma \frac{f_\varphi \bold{e}_\varphi}{\sin^2\theta}-\sigma f_\theta \bold{e}_\theta,
\end{equation}
where $f_\varphi$ is $\partial_\varphi f(\theta,\varphi)$, $f_\theta$ is $\partial_\theta f(\theta,\varphi)$, $\bold{e}_\varphi$ and $\bold{e}_\theta $ are defined in (\ref{de}).

The boundary condition $\nabla u \cdot \vec{\bold{n}} = 0$ becomes
\begin{equation}\label{perturbedBds}
    u_{0\rho}+\sigma u_{1\rho} - \sigma^2 \frac{f_\varphi u_{1\varphi}}{(1+\sigma f)^2\sin^2{\theta}}-\sigma^2\frac{f_\theta u_{1\theta}}{(1+\sigma f)^2}=0.
\end{equation}

By collecting the first order of $\sigma$ in (\ref{perturbedBds}) and letting $f(\theta,\varphi) =  \frac{1}{2}(3\cos^2{\theta}-1) $, we obtain the boundary conditions
\begin{equation}\label{boundary_condition}
    u_{0\rho}(1) = 0, \quad u_{1\rho}(1,\theta,\varphi) = -\frac{1}{2}(3\cos^2{\theta}-1)u_{0\rho \rho}(1).
\end{equation}

We now compute the Hessian term of the Neumann Green's function. We expand $G \sim G_0 + \sigma G_1$, where $G_0$ is the solution given in (\ref{H}), defined in an unperturbed unit ball with the source at the origin, and $G_1$ satisfies
\begin{equation}\label{G1}
    \Delta G_1 = 0,\quad \bold{x} \in \Omega; \quad G_{1\rho}(1,\theta,\varphi) = -\frac{3}{8\pi}(3\cos^2{\theta}-1).
\end{equation}
Accordingly, the solution is 
\begin{equation}
    G_1 = -\frac{3}{16\pi}\rho^2(3\cos^2{\theta}-1).
\end{equation}
Recalling that $x_1 = \rho \sin{\theta}\cos{\varphi}$, $x_2 = \rho \sin{\theta}\sin{\varphi}$, $x_3 = \rho \cos{\theta}$, we write $G_1 = \frac{3}{16\pi}(x_1^2+x_2^2-2x_3^2)$ in Cartesian coordinates. Therefore, the Hessian matrix $\mathcal{H}_{11}$ is given by
\begin{equation}\label{H11}
    \begin{aligned}
        \mathcal{H}_{11}\sim \frac{1}{4\pi}\mathcal{I}_3+\sigma\frac{3}{8\pi}\left(\begin{array}{ccc}
           1  & 0 & 0 \\
           0  & 1 & 0 \\
           0  & 0 & -2
        \end{array}\right).
    \end{aligned}
\end{equation}

For the Helmholtz Green's function $G_\mu$, we expand $G_\mu \sim G_{\mu0} + \sigma G_{\mu1}$. Then we compute $\nabla_{\bold{x}_1} G\mu \sim \bold{g}_0 + \sigma \bold{g}_1$, where $\bold{g}_0$ is given in (\ref{g0}), and $\bold{g}_1$ satisfies
\begin{equation}\label{g_1}
    \begin{aligned}
            &\Delta \bold{g}_1 - \mu^2 \bold{g} = 0,\quad \bold{x},\bold{x}_0\in\Omega;\\
            &\bold{g}_{1\rho}(1,\theta,\varphi) = -f(\theta,\varphi) \bold{g}_{0\rho\rho}(1,\theta,\varphi).
    \end{aligned}
\end{equation}

Therefore, the boundary condition of $\bold{g}_1$ is
\begin{equation}\label{boundary_g1}
    \bold{g}_{1\rho} 
     = \left(\begin{array}{c}
         \bold{g}_{1\rho}^{(1)} \\
         \bold{g}_{1\rho}^{(2)} \\
         \bold{g}_{1\rho}^{(3)}
     \end{array}\right) =  \frac{1}{2}(3\cos^2{\theta}-1)F''(1)\left(\begin{array}{c}
         \sin{\theta}\cos{\varphi}  \\
         \sin{\theta}\sin{\varphi}\\
         \cos{\theta}
    \end{array}\right).
\end{equation}
Further, we use normalized real Spherical Harmonics $Y_n^m$ to express the boundary conditions, which is
\begin{equation}\label{bondary_Y}
    \begin{aligned}
       &\bold{g}_{1\rho}^{(k)} =\frac{3}{4} (c_{1k} Y_3^m + c_{2k} Y_1^m)\left(\frac{\mu}{2\pi} \right)^{\frac{3}{2}}F''(1),\quad k = 1 , 2 , 3,\quad m = -1, 0 ,1,
    \end{aligned}
\end{equation}
where 
\begin{equation}\label{boundar_sphericalHarmonics}
    \begin{aligned}
        &Y_n^1 = -\sqrt{2}\sqrt{\frac{2n+1}{4\pi n(n+1)}}P_n^1(\cos{\theta})\cos{\varphi},\quad Y_n^0 = \sqrt{\frac{2n+1}{4\pi}}P_n^0(\cos{\theta}),\\
       &Y_n^{-1} = -\sqrt{2}\sqrt{\frac{2n+1}{4\pi n(n+1)}}P_n^1(\cos{\theta})\sin{\varphi},\quad n = 1,3,\\
    &c_{11} = c_{12} = \frac{12}{5}\sqrt{\frac{2\pi}{21}},\quad c_{21} = c_{22}= -6\sqrt{\frac{\pi}{3}},\quad c_{13} = \frac{12}{5}\sqrt{\frac{\pi}{7}},\quad c_{33} = -\frac{12}{5}\sqrt{\frac{\pi}{3}}.
    \end{aligned}
\end{equation}

We then solve $\bold{g}_1$ in (\ref{g_1}) with boundary conditions given in (\ref{boundar_sphericalHarmonics}) and get 
\begin{equation}
    \begin{aligned}
        &\bold{g}_1^{(1)} = A_1(\rho)Y_3^1 + B_1(\rho) Y_1^1,\\
        &\bold{g}_1^{(2)} = A_1(\rho)Y_3^{-1} + B_1(\rho) Y_1^{-1},\\
        &\bold{g}_1^{(3)} = A_2(\rho)Y_3^{0} + B_2(\rho) Y_3^0,
    \end{aligned}
\end{equation}
where 
\begin{equation}
    \begin{aligned}
        &A_1(\rho) = \frac{12\sqrt{\frac{2\pi}{21}}\left(\frac{\mu}{2\pi}\right)^\frac{3}{2}F''(1)I_{7/2}(\mu \rho)\rho^{-\frac{1}{2}}}{5(\mu I'_{7/2}(\mu)-\frac{1}{2}I_{7/2}(\mu))},\quad B_1(\rho) =  -\frac{6\sqrt{\frac{\pi}{3}}\left(\frac{\mu}{2\pi}\right)^\frac{3}{2}F''I_{3/2}(\mu \rho)\rho^{-\frac{1}{2}}}{(\mu I'_{3/2}(\mu)-\frac{1}{2}I_{3/2}(\mu))},\\
        &A_2(\rho) = \frac{12\sqrt{\frac{\pi}{7}}\left(\frac{\mu}{2\pi}\right)^\frac{3}{2}F''(1)I_{7/2}(\mu \rho)\rho^{-\frac{1}{2}}}{5(\mu I'_{7/2}(\mu)-\frac{1}{2}I_{7/2}(\mu))},\quad B_2(\rho) =  -\frac{12\sqrt{\frac{\pi}{3}}\left(\frac{\mu}{2\pi}\right)^\frac{3}{2}F''(1)I_{3/2}(\mu \rho)\rho^{-\frac{1}{2}}}{5(\mu I'_{3/2}(\mu)-\frac{1}{2}I_{3/2}(\mu))}.
    \end{aligned}
\end{equation}

Now we take gradient on $\bold{g}_1$ respect to $\bold{x}$ using $\nabla_{\bold{x}} = \partial_\rho \bold{e} + \rho^{-1}\sin^{-2}{\theta}\partial_\varphi \bold{e}_\varphi+ \rho^{-1}\partial_\theta \bold{e}_\theta$ to obtain
\begin{equation}
    \begin{aligned}
        \nabla_{\bold{x}}\bold{g}_1^{(1)} = \left(A_1'(\rho)Y_3^1-B_1'(\rho)Y_1^1\right)\bold{e}
        +\frac{1}{\rho\sin^2{\theta}}\left(A_1(\rho)\partial_\varphi Y_3^1-B_1(\rho)\partial_\varphi Y_1^1\right)\bold{e}_\varphi\\
        +\frac{1}{\rho}\left(A_1(\rho)\partial_\theta Y_3^1-B_1(\rho)\partial_\theta Y_1^1\right)\bold{e}_\theta,
    \end{aligned}
\end{equation}
where
\begin{equation}
    \begin{aligned}
        &A_1'(\rho) = \frac{12\sqrt{\frac{2\pi}{21}}\left(\frac{\mu}{2\pi}\right)^\frac{3}{2}F''(1)}{5(\mu I'_{7/2}(\mu)-\frac{1}{2}I_{7/2}(\mu))}\left(\mu I'_{7/2}(\mu \rho)\rho^{-\frac{1}{2}}-\frac{1}{2}I_{7/2}(\mu \rho)\rho^{-\frac{3}{2}}\right),\\
        &B_1'(\rho) =  -\frac{6\sqrt{\frac{\pi}{3}}\left(\frac{\mu}{2\pi}\right)^\frac{3}{2}F''}{(\mu I'_{3/2}(\mu)-\frac{1}{2}I_{3/2}(\mu))}\left(\mu I'_{3/2}(\mu \rho)\rho^{-\frac{1}{2}}-\frac{1}{2}I_{3/2}(\mu \rho)\rho^{-\frac{3}{2}}\right).
    \end{aligned}
\end{equation}
By setting $\rho = \rho_0 \rightarrow 0^+$ and using the properties of the Bessel function of the second kind for small arguments, we obtain
\begin{equation}
    \nabla_{\bold{x}} \bold{g}_1^{(1)} = -\frac{F''(1)\mu^3}{2\pi^2(\mu I'_{3/2}(\mu)-\frac{1}{2}I_{3/2}(\mu))}\left(\begin{array}{c}
         1 \\
         0 \\
         0
    \end{array}\right)
\end{equation}
Following the same process, we can compute $\nabla_{\bold{x}}\bold{g}_1^{(2)}$ and $\nabla_{\bold{x}}\bold{g}_1^{(3)}$, and then we end up with
\begin{equation}
    \nabla_{\bold{x}}\bold{g}_1 = -\frac{F''(1)\mu^3}{\pi^2(\mu I'_{3/2}(\mu)-\frac{1}{2}I_{3/2}(\mu))}\left(\begin{array}{ccc}
        \frac{1}{2} & 0 & 0  \\
        0 & \frac{1}{2} & 0 \\
        0 & 0 & \frac{1}{5}
    \end{array}\right).
\end{equation}
Note that the leading order of $\mathcal{Q}{\mu1}$ is the same as in an unperturbed unit ball. Therefore, $\mathcal{Q}_{\mu1}$ in the perturbed ball is given by
\begin{equation}\label{Q_mu_perturbed}
    \mathcal{Q}_{\mu1} =  -\frac{\big(\pi+2Q(\mu)\big)\mu^{3}}{12\pi^2}\mathcal{I}_3-\sigma \frac{F''(1)\mu^3}{\pi^2(\mu I'_{3/2}(\mu)-\frac{1}{2}I_{3/2}(\mu))}\left(\begin{array}{ccc}
        \frac{1}{2} & 0 & 0  \\
        0 & \frac{1}{2} & 0 \\
        0 & 0 & \frac{1}{5}
    \end{array}\right).
\end{equation}

We now substitute (\ref{H11}) and (\ref{Q_mu_perturbed}) into (\ref{eig_matrix}) yielding a system containing $\tau$ and $\lambda$. We perturb $\lambda_0 = (\lambda_{00}+\sigma\lambda_{01})i$, $\lambda_1 = (\lambda_{10}+\sigma \lambda_{11})i$, $
\tau_0 = \tau_{00}+\sigma \tau_{01}$ and $\tau_1 = \tau_{10} + \sigma \tau_{11}$. From (\ref{tau0}) and given that $\lambda_{00}$ is not zero, the value of $\tau_{00}$ and $\tau_{01}$ are
\begin{equation}\label{leading_tau_perturbed}
    \tau_{00} = \frac{k_1}{k_2}, \quad \tau_{01} = 0,
\end{equation}
where $k_1$ and $k_2$ are defined in (\ref{k1k2}). Since $\lambda_0$ is purely imaginary, we write $\mu^2 = \tau_0\lambda_0 = \mu_0^2i+\sigma \mu_1^2i$, where $\mu_0^2 = \tau_{00}\lambda_{00}$ and $\mu_1^2 = \tau_{00}\lambda_{01}$. Next we substitute (\ref{H11}) and (\ref{Q_mu_perturbed}) into (\ref{eig_matrix}) and collect the leading order as well as $\mathcal{O}(\sigma)$ terms, respectively, to obtain
\begin{subequations}\label{eigen-system}
    \begin{align}
        &12\pi S_\varepsilon \left(- \frac{\big(\pi+2Q(\mu)\big)\mu^{3}}{12\pi^2} - \frac{D}{4\pi}\right)\left(\begin{array}{c}
         a_1  \\
         a_2 \\
         a_3
    \end{array}\right) = k_2\tau_{10}\lambda_{00}i\left(\begin{array}{c}
         a_1  \\
         a_2 \\
         a_3
    \end{array}\right),\label{tau_10}\\
        &12\pi S_\varepsilon \left(Q_{\mu\sigma}-D\mathcal{H}_{\sigma}\right)\left(\begin{array}{c}
         a_1  \\
         a_2 \\
         a_3
    \end{array}\right) = k_2(\tau_{10}\lambda_{01}+\tau_{11}\lambda_{00})i\left(\begin{array}{c}
         a_1  \\
         a_2 \\
         a_3
    \end{array}\right)\label{tau_11},
    \end{align}
\end{subequations}
where 
\begin{equation*}
    Q_{\mu\sigma} =-\frac{F''(1)\mu_1^3}{\pi^2(\mu_1 I'_{3/2}(\mu_1)-\frac{1}{2}I_{3/2}(\mu_1))}\left(\begin{array}{ccc}
        \frac{1}{2} & 0 & 0  \\
        0 & \frac{1}{2} & 0 \\
        0 & 0 & \frac{1}{5}
    \end{array}\right),\quad \mathcal{H}_{\sigma} = \frac{3}{8\pi}\left(\begin{array}{ccc}
           1  & 0 & 0 \\
           0  & 1 & 0 \\
           0  & 0 & -2
        \end{array}\right).
\end{equation*}
From (\ref{leading_tau_perturbed}) and (\ref{tau_10}), we find that the leading order and the $\mathcal{O}(\varepsilon)$ terms of the threshold are the same for all three directions, while the  $\mathcal{O}(\varepsilon\sigma)$ terms of $\tau$ differ. Thus, we conclude that the preferred mode of oscillation is determined by the value of $\tau_{11}$ solved in (\ref{tau_11}). In this geometry, we numerically observe that the value of $\tau_{11}$ corresponding to the $(0,0,1)$-direction is smaller than in the other two directions. Therefore, the threshold for the $(0,0,1)$-direction is reached first as $\tau$ increases, indicating that the dominant mode of translational oscillation is along the major axis. Furthermore, the threshold in the perturbed domain is smaller than that in the unperturbed sphere due to the negative value of $\tau_{11}$ in the perturbed sphere.

We now briefly discuss the effect of perturbations of the form $f(\theta,\varphi)$ when $n \neq 2$. For $n = 0$, the domain is a sphere with radius $(1+\sigma)$ since $f(\theta,\varphi) = 1$. For $n = 1$, $m = 0$ or $m = 1$, the perturbation constitutes a translation of the unit ball by $\sigma$ along one axis. For example, the equation for the perturbed sphere, when $m = 0$ and $f(\theta,\varphi) = \cos{\theta}$, is
\begin{equation}
    x_1 ^2+x_2^2 + (x_3- \sigma)^2 = [1+\frac{\sigma^2}{8}(3-\cos{2\varphi}+2\cos^2{\varphi}\cos{2\theta})]^2.
\end{equation} 
This perturbation is equivalent to translating the sphere in the $x_3$-direction, accompanied by a $\mathcal{O}(\sigma^2)$ dilation of the sphere's shape. For $n\geq 3$, the leading order correction to the Neumann Green's function is an $\mathcal{O}(\rho^n)$ function, which has no impact on the Hessian. For the gradient of the Helmholtz Green's function with respect to the source location, $\nabla_{\bold{x}_1}G_\mu(\bold{x};\bold{x}_1)$, the first order correction is a function of  $(I_{n-\frac{1}{2}}(z) z^{-\frac{1}{2}}+I_{n+\frac{3}{2}}(z)z^{-\frac{1}{2}}) \sim \mathcal{O}(z^{n-1})$ and they are not affecting the value of $\mathcal{Q}_{\mu1}$ in the eigenvalue problem.

\subsection{The defected sphere}\label{defected}
We now discuss the translational oscillatory instability of one spot pattern on a defected domain. In \cite{tony-defect-2d}, a technique for converting the defect to a localized pinned spot is introduced for a $2$D problem. We apply the same technique and extend it to a $3$D domain. We firstly define the defected domain. In this geometry, we remove a ball-shaped defect with a radius of $\mathcal{O}(\varepsilon)$ from the domain, from which the chemicals are leaking. The defect is modeled by
\begin{equation}
    \Omega_\varepsilon = \{ \bold{x}\in \Omega: |\bold{x}-\bold{x}_0|\leq \varepsilon C\},
\end{equation}
 where $C >0$ is an $\mathcal{O}(1)$ constant that controls the size of the defect. The equilibrium construction of this problem is provided in Appendix \ref{equl_defect}.

In this analysis, we treat the defect as a pinned spot, hence we can extend the result in Section \ref{stability_samll}. In the inner region near the defect, it is readily to see that $\Psi_{\varepsilon 0}$ is zero, and the leading order to $\Psi_0$ is $\mathcal{O}(\varepsilon^2)$. Therefore, we have
\begin{equation}
    \Delta_{\bold{y}} \Psi_{20} - (2U_{\varepsilon 0}V_{\varepsilon 0} \Phi_{2 0}+V_{\varepsilon 0}^2\Psi_{2 0}) = 0;\quad
    \Psi_{2 0} = 0 \quad \text{on} \quad |\bold{y}| = C.
\end{equation}
The solution of $\Psi_{20}$ is given by
\begin{equation}
    \Psi_{2 0} = \kappa_0\partial_{s_{\varepsilon0}} U_{\varepsilon 0},
\end{equation}
thus the far-filed behavior is 
\begin{equation}
    \Psi_{20}\sim \kappa_0\left(\frac{1}{C}-\frac{1}{\rho_0}\right),
\end{equation}
where $U_{\varepsilon0}$ is computed in(\ref{leadingU0}). Here $\kappa_0$ is a constant determined by matching with the outer solution.

The behavior of the spot in the inner region is similar to that of the one spot problem on a general domain. We change the notation of (\ref{lead-sol}) and (\ref{inner_farfield_Psi2}), yielding
\begin{equation}
    \begin{aligned}
        &\Psi_{\varepsilon1} = \bold{a}_1^T\bold{e}_1\partial_{\rho_1}U_{\varepsilon1}, \quad \Psi_{\varepsilon 1} \sim \bold{a}_1^T\bold{e}_1\frac{S_{\varepsilon 1}}{\rho_1^2} \quad \text{as} \quad \rho_1 \rightarrow \infty;\\
       & \Psi_{21} \sim \kappa_1(\chi'(S_{\varepsilon1})-\frac{1}{\rho_1}) - \bold{a}_1^T\bold{e}_1S_{\varepsilon1}\frac{\tau_0\lambda_0}{2} \quad \text{as} \quad \rho_1 \rightarrow \infty.
    \end{aligned}
\end{equation}

In the outer region, we use the same notation from Section \ref{stability_samll} and have
\begin{equation}\label{psi_2}
    \begin{aligned}
        &\Delta \psi_{\varepsilon}-\mu_\varepsilon^2\psi_{\varepsilon} = 0,\quad x\in \Omega;\\
        &\psi_{\varepsilon}\sim \kappa_0\left(\frac{1}{C}-\frac{\varepsilon}{|\bold{x}-\bold{x}_0|}\right),\quad \text{as} \quad\bold{x}\rightarrow \bold{x}_0;\\
        &\psi_{\varepsilon}\sim \frac{S_{\varepsilon 1}\bold{a}^T(\bold{x}-\bold{x}_1)}{|\bold{x}-\bold{x}_1|^3}+\kappa_1\left(\chi'(S_{\varepsilon1})-\frac{\varepsilon}{|\bold{x}-\bold{x}_1|}\right)-\frac{\bold{a}^T\bold{e}_1S_{\varepsilon1}\mu^2}{2} ,\quad \text{as} \quad\bold{x}\rightarrow \bold{x}_1,
    \end{aligned}
\end{equation}
Hence the new $\psi_{\varepsilon}$ is
\begin{equation}\label{outer}
    \begin{aligned}
        \psi_\varepsilon = -4\pi \Big[S_\varepsilon\bold{a}^T \nabla_{\bold{x}_1}& G_\mu (\bold{x};\bold{x}_1) -\varepsilon \kappa_1 G_\mu(\bold{x};\bold{x}_1) \\
     &+ \varepsilon(\tau_0\lambda_1+\tau_1\lambda_0)S_\varepsilon\bold{a}^T\partial_{\mu^2} \nabla_{\bold{x}_1} G_\mu (\bold{x};\bold{x}_1)-\kappa_0\varepsilon G_\mu(\bold{x};\bold{x}_0)\Big],
    \end{aligned}
\end{equation}
which includes one additional term compared to (\ref{outer_psi}). The local behavior of $\nabla_{\bold{x}_1}G_\mu(\bold{x};\bold{x}_1)$ near the spot location is given in (\ref{grad_G_localbhv}) and (\ref{Neumann_Green}). While near the defect, the local behavior of $\nabla_{\bold{x}_1}G_\mu(\bold{x};\bold{x}_1)$ is
\begin{equation}\label{gradGmu_defect}
    \begin{aligned}
         \nabla_{\bold{x}_1} G_\mu(\bold{x};\bold{x}_1)\sim \nabla_{\bold{x}_1}G_\mu(\bold{x}_0;\bold{x}_1) ,\quad \text{as} \quad \bold{x}\rightarrow\bold{x}_0.
    \end{aligned}  
\end{equation}
Matching (\ref{outer}) with (\ref{psi_2}) and using (\ref{grad_G_localbhv}) and (\ref{gradGmu_defect}), we obtain the constants $\kappa_0$ and $\kappa_1$,
\begin{equation}
    \begin{aligned}
        &\kappa_0 = -4\pi C S_{\varepsilon1}\bold{a}^T\nabla_{\bold{x}_1}G_\mu(\bold{x}_0;\bold{x}_1);\\
        &\kappa_1 = -\frac{4\pi S_{\varepsilon1}}{\chi'(S_{\varepsilon1})}\bold{a}^T\nabla_\bold{x}R_\mu(\bold{x};\bold{x}_1)|_{\bold{x}=\bold{x}_1}.
    \end{aligned}
\end{equation}
In particular, the presence of the defect breaks the symmetric of the geometry. Therefore, the value of $\nabla_\bold{x}R_\mu(\bold{x};\bold{x}_1)|_{\bold{x}=\bold{x}_1}$, which is zero in a unit sphere, is non-zero here.

We next apply solvability conditions on the system of $\mathcal{O}(\varepsilon^2)$ and $\mathcal{O}(\varepsilon^3)$ in the inner region. 
By expanding (\ref{outer}), we obtain the far-field behavior of $\bold{W}_1$, as $\rho_1 \rightarrow \infty$,
\begin{equation}\label{newW1_farfield}
    \bold{W}_1 \sim \left(\begin{array}{c}
         0  \\
         -4\pi\Big[S_{\varepsilon1}\bold{a}^T\big(\varepsilon\mathcal{Q}_{\mu1}\rho_1+\frac{\mu_\varepsilon^2}{8\pi}\big) \bold{e}_1-\kappa_1\big(\chi'(S_{\varepsilon1})-\frac{1}{\rho_1}\big) - \varepsilon \kappa_1 R_\mu(\bold{x}_1;\bold{x}_1)-\varepsilon\kappa_0G_\mu(\bold{x}_1;\bold{x}_0)\Big] 
    \end{array}\right).
\end{equation}
We then multiply (\ref{W1}) by $\bold{P}_{sc,ss,cc}$, which is defined in (\ref{radially-adjoint}), on both side of (\ref{W1}) and then integrate over a ball with radius $R \gg 1$ centered at the origin. Hence we have
\begin{equation}\label{sovability_defect}
    \begin{aligned}
                \int\int\int_{B_R} \bold{P}^T_{sc,ss,cc}\left[\Delta_{\bold{y}}\bold{W}_1+\mathcal{M}_1\bold{W}_1\right] d\bold{y}\\
        =\int\int\int_{B_R} \bold{P}^T_{sc,ss,cc}\left[\left(\begin{array}{c}
             (\lambda_0+\varepsilon\lambda_1)\Phi_{\varepsilon1}  \\
             \frac{\mu_\varepsilon^2}{D}\Psi_{\varepsilon1}
        \end{array}\right )+\mathcal{N}_1\left(\begin{array}{c}
             \Phi_{\varepsilon1}  \\
             \Psi_{\varepsilon1}
        \end{array} \right)\right] d\bold{y}.
    \end{aligned} 
\end{equation}
Following the process from (\ref{left1}) to (\ref{left=right}) in Section \ref{stability_samll}, but using the far-field condition of $W_1$ given in (\ref{newW1_farfield}), we derive an eigenvalue system,
\begin{equation}\label{eigen_system_defect}
    \frac{12\pi}{\tau_1k_2}\left[S_{\varepsilon1}(\mathcal{Q}_{\mu1}-\mathcal{H}_{11})^T-S_{\varepsilon0}\mathcal{H}_{10}^T\right]\bold{a} = \lambda_0 \bold{a},
\end{equation}
where $\mathcal{Q}_{\mu1}$ is defined in (\ref{Qmu}), and $\mathcal{H}_{11}$, $\mathcal{H}_{10}$ are given in (\ref{Helm}). Note that $\mathcal{Q}_{\mu1}$ here differs from that in a sphere or a perturbed sphere where the spot is located at the domain center. A general form of the regular part of the Helmholtz Green's function is provided in Appendix \ref{Rmu}.

\begin{figure}[h]
    \centering
    \includegraphics[width=0.5\linewidth]{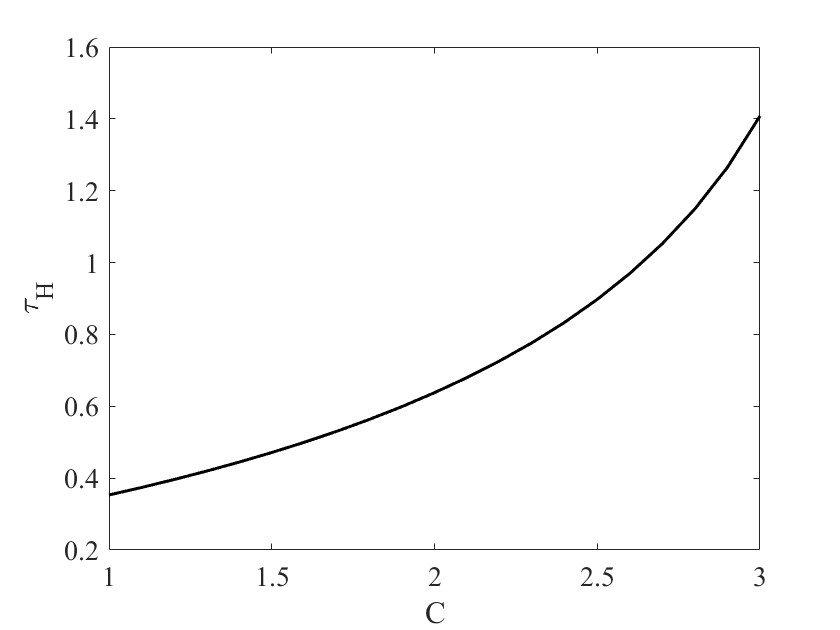}
    \caption{Plot of the translational instability threshold $\tau_H$ versus $C$ in the defected domain, where the defect is a small disk radius $\varepsilon C$. }
    \label{fig:CvsTau}
\end{figure}

From (\ref{tau0}) and (\ref{eigen_system_defect}), we observe that the HB threshold is closely related to the source strength $S_{\varepsilon1}$. However, for a fixed feed rate $A$, the defect competes with the spot for resources according to (\ref{S_defect}). Furthermore, as the defect size, measured by $C$, increases, more chemicals leak out, resulting in a smaller $S_{\varepsilon1}$ at equilibrium. The value of the HB threshold increases with increasing $C$ (see Figure \ref{fig:CvsTau}), indicating that for a large $C$, the HB threshold is more difficult to reach. Therefore, for a fixed feed rate $A$, the translational oscillatory instability is more difficult to trigger in the defected domain than in a normal domain (without a defect). Moreover, when $S_{\varepsilon1}$ falls below a certain critical value, an amplitude oscillatory instability occurs. Further details are provided in Section \ref{large_stability}.

\section{The stability analysis: Large eigenvalues}\label{large_stability}
In this section, we discuss the amplitude instability of the spot on a general domain. We assume $\lambda \sim \mathcal{O}(1)$ and investigate how the Bifurcation threshold $\tau$ changes with the feed rate $A$ in the system. When $A$ is less than $13.56$, the scaling of $\tau$ is $\mathcal{O}(1)$. As $A$ increases, the scaling of $\tau$ becomes $\mathcal{O}(\varepsilon^{-3})$. Lastly, we compare the result with the translational oscillatory instability threshold and obtaining the intercept of the HB threshold, at which the dominant instability changes. 

\subsection{The $\tau \sim \mathcal{O}(1)$ regime}\label{regime1}
In this subsection, we compute the $\mathcal{O}(1)$ HB threshold. The eigenvalue problem is given in (\ref{eigen_problem}). In the inner region, we expand
\begin{equation}\label{inner_large_eigen}
    \phi = P_n^m(\cos{\theta})e^{im\varphi}(\Phi_\varepsilon(\rho)+\dots),\quad \psi = P_n^m(\cos{\theta})e^{im\varphi}(\Psi_\varepsilon(\rho)+\dots).
\end{equation}
Substituting (\ref{inner_large_eigen}) into (\ref{eigen_problem}), we derive the radially symmetric eigenvalue problem
\begin{equation}\label{leading_eigen}
    \begin{aligned}
        \lambda \Phi_\varepsilon &= \Delta_n \Phi_\varepsilon - \Phi_\varepsilon + 2U_\varepsilon V_\varepsilon \Phi_\varepsilon + V_\varepsilon^2 \Psi_\varepsilon,\\
        0&= \Delta_n \Psi_\varepsilon - (2U_\varepsilon V_\varepsilon \Phi_\varepsilon + V_\varepsilon^2 \Psi_\varepsilon),
    \end{aligned}
\end{equation}
where $\Delta_n = \partial_{\rho\rho}+2\rho^{-1}\partial_\rho-n(n+1)\rho^{-2}$.

We now consider $n = 0$ which corresponding to the instability of the spot's amplitude. Since $V_\varepsilon$ decays exponentially as $\rho$ tends to infinity, by rescalling the eigenfunction, we get the far-field of $\Psi_\varepsilon$ as 
\begin{equation}\label{farfield_Psi}
    \Psi_\varepsilon \sim B(\lambda,S_\varepsilon) - \frac{1}{\rho}, \quad \rho \rightarrow \infty.
\end{equation}
Using the divergence theorem on the second equation of (\ref{leading_eigen}), we have
\begin{equation}\label{impose}
    \int_\Omega 2U_\varepsilon V_\varepsilon\Phi_\varepsilon + V_\varepsilon^2 \Psi_\varepsilon d\textbf{y} = 4\pi.
\end{equation}
Using (\ref{impose}), we have
\begin{equation}
    \int_\Omega \frac{1}{\varepsilon^3}(2u_ev_e\phi + v_e^2 \psi) d\textbf{x} \sim 4\pi,
\end{equation}
and the outer problem is
\begin{equation}\label{outer_eigen}
    \Delta \psi -\frac{\tau\lambda\varepsilon}{D}\psi = \frac{4\pi \varepsilon}{D}\delta(\textbf{x}).
\end{equation}
The solution to (\ref{outer_eigen}) is 
\begin{equation}
    \psi = -\frac{\varepsilon}{\sqrt{D}r(1+\bar{Q}(\mu))}\left(e^{\mu r}+\bar{Q}(\mu)e^{-\mu r}\right), \quad r = |\textbf{x}|,\quad \bar{Q}(\mu) = \frac{e^{2\mu}(\mu-1)}{\mu+1}, \quad \mu = \sqrt{\frac{\tau\lambda\varepsilon}{D}}.
\end{equation}
Expanding $\psi$ for small $r$ and matching with (\ref{farfield_Psi}), we obtain
\begin{equation}\label{BTauLamb}
    B(\lambda,S_\varepsilon) =  - \frac{\varepsilon \mu(1-\bar{Q}(\mu))}{\sqrt{D}(1+\bar{Q}(\mu))}.
\end{equation}
Assume that $\mu\ll 1$, the equation (\ref{BTauLamb}) can be simplified as
\begin{equation}\label{simpliedB}
     B(\lambda,S_\varepsilon) =  - \frac{3\varepsilon}{\mu^2\sqrt{D}} + \frac{9\varepsilon}{5\sqrt{D}} + \mathcal{O}(\mu^2).
\end{equation}
By substituting $\mu = \sqrt{\varepsilon\tau\lambda}$ into (\ref{simpliedB}) and collecting the leading order, we obtain
\begin{equation}\label{B_leading}
    \begin{aligned}
        B(\lambda,S_\varepsilon) = -\frac{3\sqrt{D}}{\tau\lambda}.
    \end{aligned}
\end{equation}

To seek the possibility of the instability arises by Hopf bifurcation, we let $\lambda = \lambda_I i$, and separate (\ref{B_leading}) into real and imaginary parts, yielding
\begin{subequations}\label{B_complex}
    \begin{align}
        &\mathbf{Re}\Big(B(\lambda_I,S_\varepsilon) \Big)= 0,\label{reB}\\
        &\mathbf{Im}\Big(B(\lambda_I,S_\varepsilon) \Big)= \frac{3\sqrt{D}}{\tau\lambda_I}.\label{imB}
    \end{align}
\end{subequations}

We now numerically compute the value of $ B(\lambda_I, S_\varepsilon)$ by solving (\ref{leading_eigen}) with the far-field behavior in (\ref{farfield_Psi}).

\begin{figure}[h]
    \centering
    \begin{subfigure}[b]{0.45\textwidth}
        \centering
        \includegraphics[width=\textwidth]{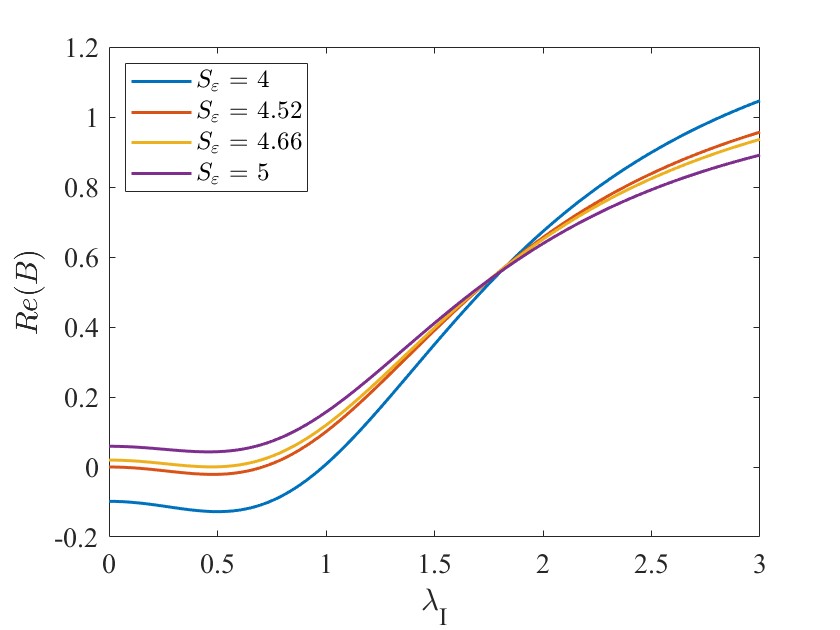} 
        \caption{$\mathfrak{R}(B(\lambda_I,S_\varepsilon))$ versus $\lambda_I$}
        \label{fig:Rb}
    \end{subfigure}
    \begin{subfigure}[b]{0.45\textwidth}
        \centering
        \includegraphics[width=\textwidth]{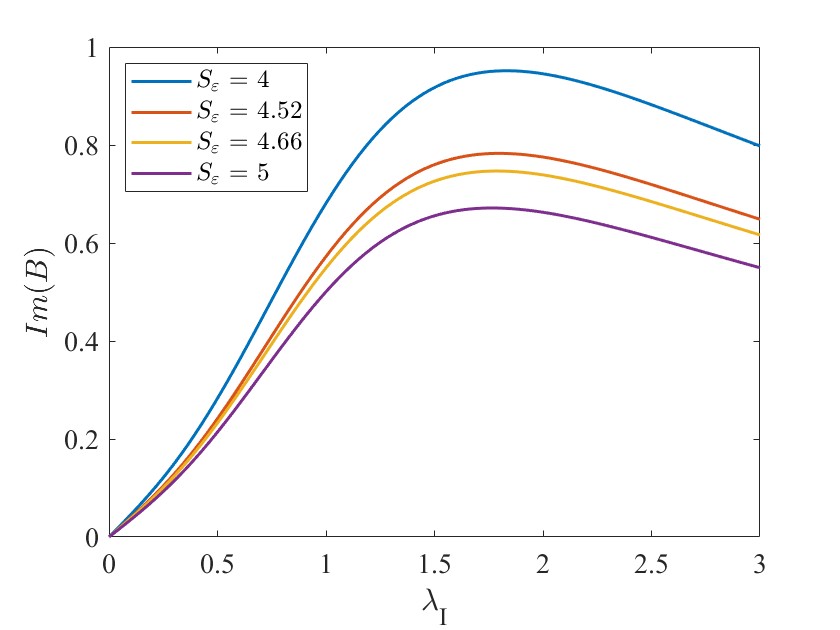} 
        \caption{$\mathfrak{I}(B(\lambda_I,S_\varepsilon))$ versus $\lambda_I$}
        \label{fig:Ib}
    \end{subfigure}
    \caption{(a) Real part of $B(\lambda_I,S_\varepsilon)$ and (b) imaginary part of $B(\lambda_I,S_\varepsilon)$ for $S_\varepsilon = 4, 4.52, 4.66 $ and $5$. Here $\lambda_I$ is the pure imaginary eigenvalue.}
    \label{fig:B}
\end{figure}
In Figure \ref{fig:Rb}, the value of $ \mathfrak{R}(B) $ increases as $S_\varepsilon$  increases when $\lambda_I < 1.5$. When $S_\varepsilon < 4.52,$ the curve crosses the $\mathfrak{R}(B) = 0$ once, indicating a single solution for (\ref{reB}). When $S_\varepsilon = 4.52$ , we observe that  $\lambda_I = 0 $ is one of the solutions according to (\ref{reB}). To verify this, we set  $\lambda_I = 0 $ in (\ref{leading_eigen}) and find that the corresponding solution is obtained by differentiating the solution to the core problem with respect to  $S_\varepsilon$. Thus, we have $\mathfrak{R}(B(0, S_\varepsilon)) = \chi'(S_\varepsilon)$. Given the required condition $ \mathfrak{R}(B) = 0 $, we conclude that the critical value of $ S_\varepsilon$ is 4.52 where $\chi'(S_\varepsilon) = 0$ from Figure \ref{fig:chi}. For $ 4.52 \leq S_\varepsilon < 4.66 $, two solutions can be found for (\ref{reB}), indicating that two distinct thresholds exist of $ \tau $ in this range of $ S_\varepsilon $. As $ S_\varepsilon $ increases beyond 4.66, the curve no longer touches the line $\mathfrak{R}(B) = 0$, so there is no solution when $ S_\varepsilon > 4.66 $.

By using the Taylor expansion on $B(\lambda_I,S_\varepsilon)$ around the point $(\lambda_I,S_\varepsilon) = (0,4.52)$, we have
\begin{equation}\label{asymptotic_tau}
        \lambda_{IH} \sim \frac{|\mathfrak{R}(B_\lambda)|-\alpha}{\mathfrak{R}(B_{\lambda\lambda})},\quad \tau_H \sim \frac{3\mathfrak{R}(B_{\lambda\lambda})^2}{\beta_2(S_\varepsilon-4.52)-\alpha\beta_1 },\quad \text{as}\quad S_\varepsilon \rightarrow 4.52^+,
\end{equation}
where $B_\lambda$ is $\partial_{\lambda_I}B(\lambda_I,S_\varepsilon)$, $B_{\lambda\lambda} = \partial_{\lambda_I\lambda_I}B(\lambda_I,S_\varepsilon)$ and $\alpha = \sqrt{\mathfrak{R}(B_\lambda)^2-2\mathfrak{R}(B_{\lambda\lambda})\mathfrak{R}(B_s)(S_\varepsilon-4.52)}$ with $B_s=\partial_{S_\varepsilon}B(\lambda_I,S_\varepsilon)$, $\beta_1 = 2\mathfrak{R}(B_\lambda)\mathfrak{I}(B_\lambda)-\mathfrak{I}(B)$ and $\beta_2 = 2\mathfrak{R}(B_s)\mathfrak{I}(B_\lambda)\mathfrak{R}(B_{\lambda\lambda})^2$. Note that $B(\lambda_I,S_\varepsilon) = B(-\lambda_I,S_\varepsilon)$, hence $\mathfrak{R}(B_\lambda)$ tends to zero when $\lambda_I\rightarrow 0^+$, which implies $\lambda_{IH}$ to approach zero and $\tau_{H}$ to diverge.

\begin{figure}[ht]
    \centering
    \begin{subfigure}[b]{0.45\textwidth}
        \centering
        \includegraphics[width=\textwidth]{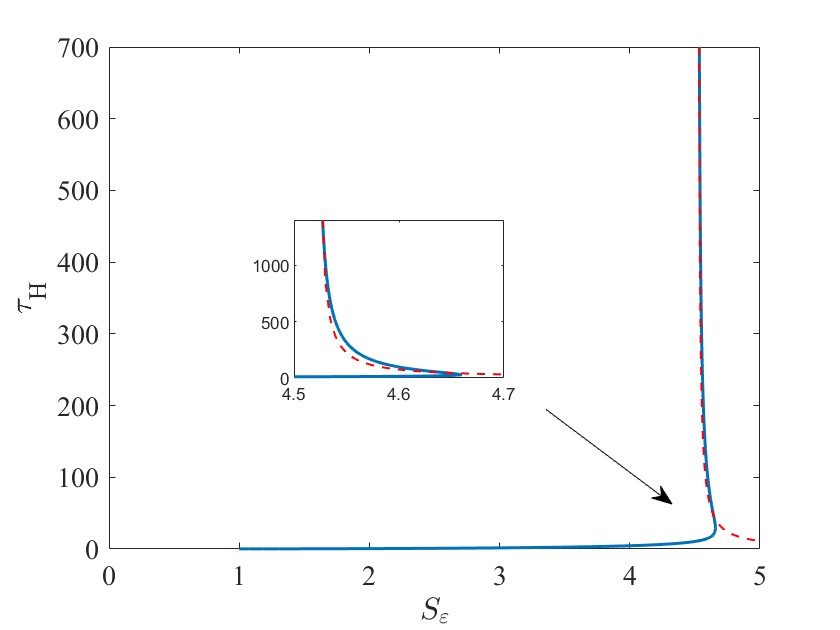} 
        \caption{$\tau_H$ versus $S_\varepsilon$}
        \label{fig:tau_H}
    \end{subfigure}
    \begin{subfigure}[b]{0.45\textwidth}
        \centering
        \includegraphics[width=\textwidth]{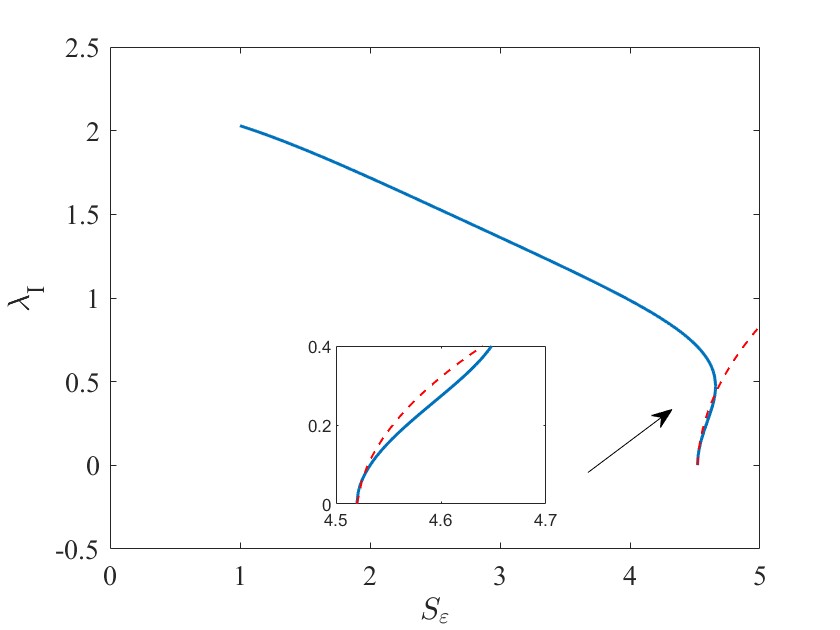} 
        \caption{$\lambda_{IH}$ versus $S_\varepsilon$}
        \label{fig:lambda_IH}
    \end{subfigure}
    \caption{(a) The HB threshold $\tau_H$ and (b) eigenvalues $\lambda_{IH}$ are plotted versus $S_\varepsilon$ in the $\tau \sim \mathcal{O}(1)$ regime. In (a), $\tau_H$ diverges as $S_\varepsilon \rightarrow 4.52^+$. Two thresholds exist within the range $4.52 < S_\varepsilon < 4.66$, while no thresholds are found for $S_\varepsilon > 4.66$. In (b), the corresponding eigenvalue approaches zero as $S_\varepsilon \rightarrow 4.52^+$. The red dashed line in both plots represents the asymptotic results for $\tau_H$ and $\lambda_{IH}$, respectively, as given in (\ref{asymptotic_tau}).}
    \label{fig:taulamb_regime1}
\end{figure}

We now solve the transcendental equations in (\ref{B_complex}) to obtain the values of the amplitude oscillatory frequency and the oscillatory threshold. The resulting bifurcation threshold $\tau_H$ and the corresponding eigenvalues $\lambda_{IH}$ are shown in Figure \ref{fig:taulamb_regime1}.
It shows that there is a unique HB threshold for $ S_\varepsilon < 4.52 $ and two thresholds when $ S_\varepsilon $ is between 4.52 and 4.66. The limiting asymptotic behavior given in (\ref{asymptotic_tau}) agrees well with the numerical results from (\ref{B_complex}) as $S_\varepsilon \rightarrow 4.52^+$. There is no HB threshold for $S_\varepsilon>4.66$ in this regime.

\begin{figure}[ht]
    \centering
    \includegraphics[width=0.5\linewidth]{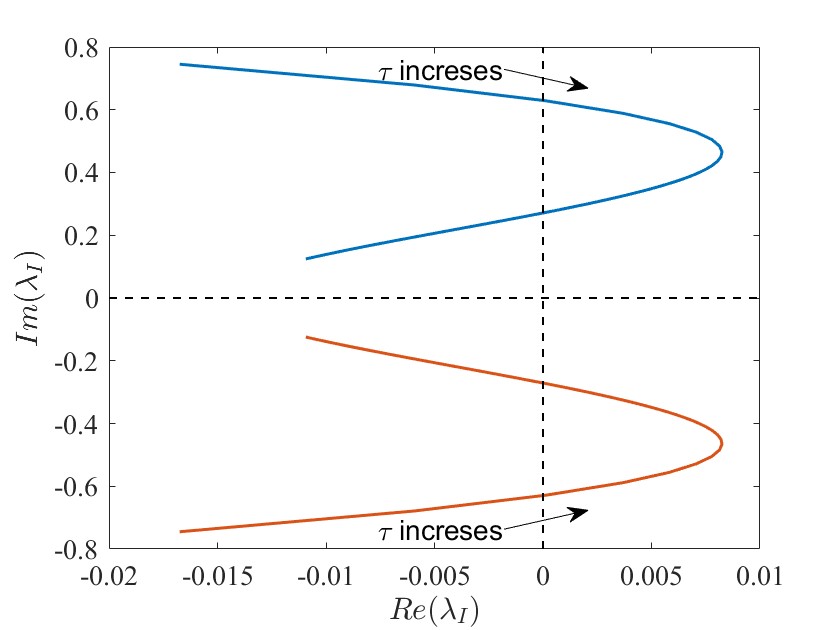}
    \caption{Trajectory of a pair of complex eigenvalues for $S_\varepsilon = 4.6$ as $\tau$ increases. As $\tau$ increases, the eigenvalues cross from the left half-plane (LHP) to the right half-plane (RHP) when $\tau$ exceeds the smaller $\tau_H$. They later return to the LHP as $\tau$ continues to increase beyond the larger $\tau_H$.}
    \label{fig:eig}
\end{figure}

In Figure \ref{fig:taulamb_regime1}, two distinct critical values of $\tau$ can be found when $4.52\leq S_\varepsilon < 4.66$. To investigate how a pair of conjugate eigenvalues cross the imaginary axis, we track the behavior of the eigenvalues by setting $S_\varepsilon = 4.6$. Starting with a small $ \tau $ (below the smaller HB threshold), we gradually increase $\tau$ until it surpasses the larger HB threshold. In Figure \ref{fig:eig}, the eigenvalues cross from the left half-plane to the right half-plane when $ \tau $ exceeds the smaller $ \tau_H $, making the system unstable to amplitude oscillations. Later they return to the left half-plane as $ \tau $ continues to increase beyond the larger HB threshold, and the system becomes stable again. As $\tau$ increases further, two complex eigenvalues collide and combine at zero-imaginary axis, becoming real and negative. As a result, the system transitions through stable-unstable-stable phases with respect to amplitude oscillatory instability as $\tau$ increases.

From Figure \ref{fig:taulamb_regime1}, no Hopf bifurcation occurs when $S_\varepsilon > 4.66$. We denote this point as a saddle node. Next, we examine how the saddle-node varies with changes in $\varepsilon$. From (\ref{B_complex}), the saddle-node values are identified as $(\lambda_{I}^0, S_{\varepsilon}^0, \tau_{H}^0) = (0.4848, 4.66, 27.66)$. We then perturb 
\begin{equation}\label{perturb_saddle}
    \lambda_{I}^s \sim \lambda_{I}^0 + \varepsilon^{\delta_1} \lambda_{I}^1,\quad S_{\varepsilon }^s \sim S_{\varepsilon }^0+\varepsilon^{\delta_2}S_{\varepsilon }^1,\quad \tau_H^s \sim \tau_H^0+\varepsilon^{\delta_3}\tau_H^1,
\end{equation}
where $\lambda_I^s, S_\varepsilon^s, \tau_H^s$ are the values at the saddle-node for $\varepsilon>0$. To compute the values of $\delta_{1,2,3}$, we apply Taylor expansion on $B(\lambda_I, S_\varepsilon)$ around the point $(\lambda_{I}^0, S_{\varepsilon}^0, \tau_{H}^0)$ and substitute to (\ref{simpliedB}), yielding
\begin{equation}\label{taylor_saddle}
    \begin{aligned}
            B(\lambda_{I}^0, S_{\varepsilon}^0) + \varepsilon^{\delta_2} S_\varepsilon^1 B_s(\lambda_{I}^0, S_{\varepsilon}^0) + \varepsilon^{\delta_1} \lambda_I^1 B_\lambda(\lambda_{I}^0, S_{\varepsilon}^0) + \varepsilon ^{\delta_1+\delta_2} \lambda_I^1 S_\varepsilon^1 B_{\lambda s}(\lambda_{I}^0, S_{\varepsilon}^0)\\
            = \frac{3 i}{(\tau_H^0+\varepsilon^{\delta_3}\tau_H^1)(\lambda_I^0+\varepsilon^{\delta_1}\lambda_I^1)}+\frac{9\varepsilon}{5}.
    \end{aligned}
\end{equation}
Thus we obtain $\delta_1 = \delta_2 = \frac{1}{2}$ and $\delta_3 = 1$. To compute $(\lambda_I^1, S_\varepsilon^1, \tau_H^1)$, in addition to solving the real and imaginary parts of (\ref{taylor_saddle}), one additional condition is required. Since the saddle-node is defined as the fold point of $B$ at which the real part of $B(\lambda_I^s, S_\varepsilon^s)$ is tangent to the horizontal line $y = \frac{9\varepsilon}{5}$, we determine the additional condition as follows:
\begin{equation}\label{condition_saddle}
    \mathfrak{R}\big(B_\lambda(\lambda_I^s, S_\varepsilon^s) \big)= 0.
\end{equation}
By solving (\ref{taylor_saddle}) and (\ref{condition_saddle}) and incorporating (\ref{perturb_saddle}), we derive $\lambda_I^s$, $S_\varepsilon^s$ and $\tau_H^s$ as a function of $\varepsilon$, respectively.
\renewcommand{\arraystretch}{1.2}
\begin{table}[h]
    \centering
    \begin{tabular}{c|cccc|cccc|c} 
        &\multicolumn{4}{|c|}{Theoretical results}&\multicolumn{3}{|c}{Numerical results}\\
        $\varepsilon$ & 0 & 0.025 & 0.03 & 0.04  & 0.025 & 0.03 & 0.04\\ \hline\hline
        
        $\lambda_I^s$ & 0.485 &  0.458 & 0.453 & 0.443 & 0.456 & 0.453 & 0.444\\ 
        $S_\varepsilon^s$ & 4.660 &  5.027 & 5.124 & 5.376 & 5.006 & 5.091 & 5.277\\ 
        $\tau_H^s$ & 27.660 &  34.467 & 35.887 & 39.540 & 34.468 & 35.888 & 39.550
     \\ \end{tabular}
    \caption{Theoretical values and numerical values of the saddle nodes for various $\varepsilon$. The theoretical values are computed by solving (\ref{taylor_saddle}) and (\ref{condition_saddle}). The numerical values are derived from the full eigenvalue problem in regime 1, where there is only one value of $\lambda_I$ corresponds to $S_\varepsilon$.}
    \label{tab:saddel_node}
\end{table}

Table \ref{tab:saddel_node} compares the theoretical and numerical results. The numerical results are obtained by solving the full eigenvalue problem in Equation (\ref{eigen_problem}). Near $S_\varepsilon =4.66$, two imaginary eigenvalues correspond to a single value of $S_\varepsilon$. Gradually increasing $S_\varepsilon$, these two eigenvalues eventually collide into one. The corresponding $S_\varepsilon$ is the saddle node, and the eigenvalue at this point is the saddle eigenvalue. During this process, $\tau$ is adjusted to ensure that the eigenvalue(s) remain purely imaginary. The numerical results coincide with the predicted theoretical values.

\subsection{The $\tau \sim \mathcal{O}(\varepsilon^{-3})$ regime}\label{regime2}
In this section, we investigate the instability of the eigenvalue problem for large $\tau$. We assume that $\tau = \varepsilon^{-\alpha} \tau_0$, where $1 \leq \alpha < 3$. With this assumption, we have $\mu = \varepsilon^{\frac{1 - \alpha}{2}} \sqrt{\tau_0 \lambda} = \varepsilon^{\frac{1 - \alpha}{2}} \mu_0$, where $\mu_0 \sim \mathcal{O}(1)$. Consequently, (\ref{BTauLamb}) simplifies to
\begin{equation}
    B(\lambda,S_\varepsilon) = -\frac{\varepsilon^{\frac{3-\alpha}{2}}\mu_0(1/\bar{Q}(\mu)-1)}{\sqrt{D}(1/\bar{Q}(\mu)+1)} = \varepsilon^{\frac{3-\alpha}{2}}\mu_0,
\end{equation}
as $1/\bar{Q}(\mu) $ is exponentially small for $\mu\gg 1$. In this case, the leading order of $B(\lambda_I,S_\varepsilon)$ is zero. Therefore, according to the Figure \ref{fig:B}, there is no solution to this problem.

Therefore, we consider the scaling of $ \tau \sim \mathcal{O}(\varepsilon^{-3}) $. Assume that $ \tau = \varepsilon^{-3} \tau_0 $, so $ \mu = \varepsilon^{-1} \sqrt{\tau_0 \lambda} = \varepsilon^{-1} \mu_0 $. When $ \mu_0 \ll 1 $, the leading order of the eigenvalue problem in the inner region is the same as in (\ref{leading_eigen}). This is the reduced $\tau \sim \mathcal{O}(\varepsilon^{-3})$ regime. Thus, the far-field behavior of $ \Psi_\varepsilon $ in (\ref{farfield_Psi}) and the matching conditions in (\ref{BTauLamb}) hold here. We further simplify the matching condition for $\tau_0 \ll 1$, yielding
\begin{equation}\label{small_tau0}
    B(\lambda,S_\varepsilon) = \mu_0 = \sqrt{\tau_0\lambda}.
\end{equation}
Letting $\lambda = \lambda_I i$, equation (\ref{small_tau0}) becomes
\begin{equation}\label{small_tau_imaginary}
    \mathfrak{R}\big(B(\lambda_I,S_\varepsilon)\big) = \mathfrak{I}\big(B(\lambda_I,S_\varepsilon)\big)=\frac{\sqrt{2\tau_0\lambda_I}}{2}.
\end{equation}
To Solve (\ref{small_tau_imaginary}) for a given $S_\varepsilon$, we first solve for $\lambda_I$ from the first equation where the real part of $B$ equals its imaginary part. In Figure \ref{fig:B}, we observe that when $S_\varepsilon > 4.52$, there is one intercept between the real and imaginary curves, located on the branch where $\lambda_I < 1$. We then obtain the value of $\tau$ by solving the second equation in (\ref{small_tau_imaginary}). When $S_\varepsilon = 4.52$, from Figure \ref{fig:B}, $\lambda_I = 0$ is the solution to the first equation of (\ref{small_tau_imaginary}). Letting $\lambda_I = \sigma \lambda_{I1}$ where $\sigma \ll 1$, we then expand $B$ around $(0,4.52)$, yielding
\begin{equation}
    B(0,4.52) +\sigma\lambda_{I1}B_\lambda(0,4.52) = \frac{\sqrt{2\tau_0\sigma\lambda_{I1}}}{2}.
\end{equation}
By collecting the $\mathcal{O}(\sigma)$ terms, we readily see that $\tau_0 = 0$. There is another intercept at $\lambda_I > 1$ for each value of $S_\varepsilon$, as shown in Figure \ref{fig:B}. The corresponding $\tau$, however, is large so that the assumption $\mu_0 \ll 1$ is not satisfied. Therefore, these values of $\tau$ do not represent an HB threshold in this context. As a result, no HB for $S_\varepsilon<4.52$ in this regime.

\begin{figure}[h]
    \centering
    \begin{subfigure}[b]{0.45\textwidth}
        \centering
        \includegraphics[width=\textwidth]{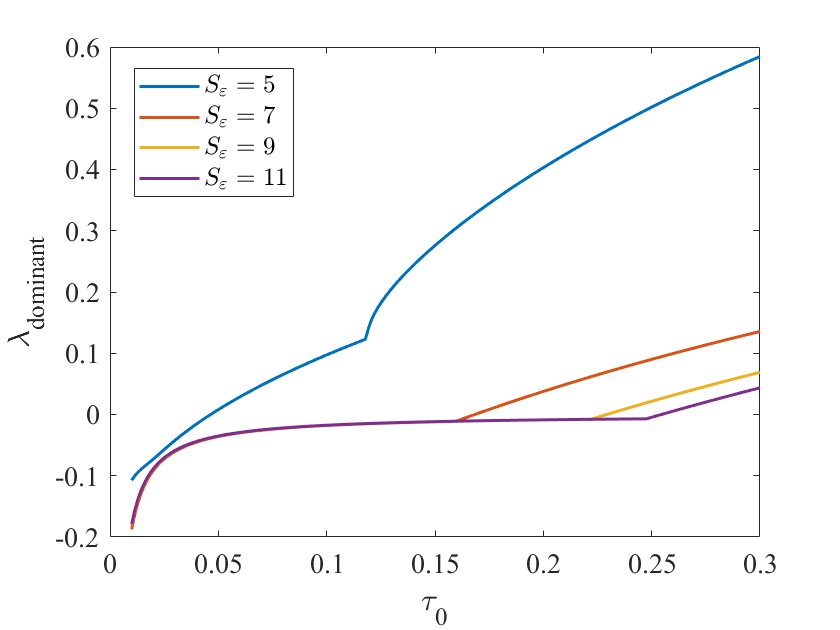} 
        \caption{$\tau_0$ versus $\lambda$(dominant)}
        \label{fig:tauvslamb}
    \end{subfigure}
    \begin{subfigure}[b]{0.45\textwidth}
        \centering
        \includegraphics[width=\textwidth]{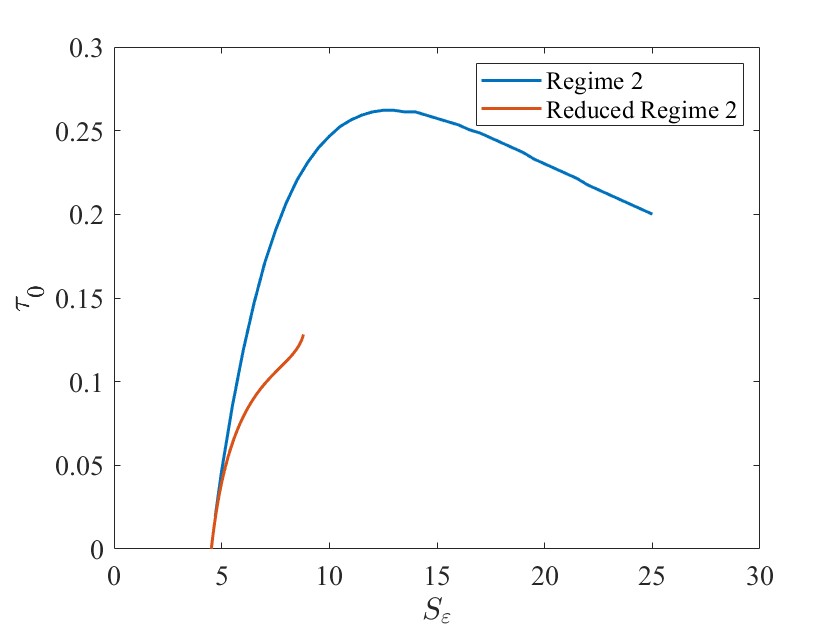} 
        \caption{$\tau_0$  versus $S_\varepsilon$}
        \label{fig:SvsTau_large_regime2}
    \end{subfigure}
    \caption{(a) Dominant (largest real) eigenvalues of $\mathcal{L}_0$ computed numerically, showing a zero-crossing for each value of $S_\varepsilon$. (b) Leading order of the HB threshold versus $S_\varepsilon$ for both the reduced $\mathcal{O}(\varepsilon^{-3})$ regime and the full $\mathcal{O}(\varepsilon^{-3})$ regime. The HB threshold in the reduced regime converges to that of the full regime as $S_\varepsilon$ decreases.}
    \label{fig:taulamb_regime2}
\end{figure}

We now consider the case where $\mu_0 \sim \mathcal{O}(1)$. Under this assumption, $\tau_0\lambda$ is raised to the leading order in the inner region, and the eigenvalue problem becomes
\begin{equation}\label{assumpt_leading_eig}
    \begin{aligned}
        &\Delta_0 \Phi_\varepsilon - \Phi_\varepsilon + 2U_\varepsilon V_\varepsilon \Phi_\varepsilon + V_\varepsilon^2 \Psi_\varepsilon = \lambda \Phi_\varepsilon,\\
        &\Delta_0 \Psi_\varepsilon - (2U_\varepsilon V_\varepsilon \Phi_\varepsilon + V_\varepsilon^2 \Psi_\varepsilon) = \tau_0\lambda\Psi_\varepsilon,\\
        &\Phi'_\varepsilon(0) = \Psi'_\varepsilon(0) = 0; \quad \Phi_\varepsilon \rightarrow 0, \quad \Psi_\varepsilon \rightarrow 0 \quad \text{as}\quad \rho \rightarrow \infty.
    \end{aligned}
\end{equation}
We define the operator
\begin{equation}
    \mathcal{L}_0 =  \left(\begin{array}{c}
         \Delta_0  \\
         \Delta_0/\tau_0
    \end{array} \right)+ \left(\begin{array}{cc}
         -1+2U_\varepsilon V_\varepsilon & V_\varepsilon^2  \\
         -2U_\varepsilon V_\varepsilon/\tau_0& V_\varepsilon^2/\tau_0
    \end{array} \right),
\end{equation}
thus (\ref{assumpt_leading_eig}) can be written as
\begin{equation}
    \mathcal{L}_0 \left(\begin{array}{c}
         \Phi_\varepsilon  \\
         \Psi_\varepsilon
    \end{array}\right) = \lambda \left(\begin{array}{c}
         \Phi_\varepsilon  \\
         \Psi_\varepsilon
    \end{array}\right).
\end{equation}
A key feature here is that $\Psi_\varepsilon$ approaches zero in the far-field, indicating the spot's instability is determined solely by the spectrum of the operator $\mathcal{L}_0$. In Figure \ref{fig:tauvslamb}, we plot the numerically computed dominant (largest real) eigenvalue of $\mathcal{L}_0$ for various values of $S_\varepsilon$. It is clearly that a zero-crossing as $\tau_0$ increases for each $S_\varepsilon$. The values of $\tau_0$ at which the eigenvalue becomes purely imaginary are then collected, giving the critical threshold $\tau_H$ for each $S_\varepsilon$, as shown in Figure \ref{fig:SvsTau_large_regime2}. Thus, for a given $S_\varepsilon$, the system is stable to amplitude oscillations when $\tau_0$ is below the critical threshold $\tau_H$ but becomes unstable when $\tau_0$ exceeds $\tau_H$. In Figure \ref{fig:SvsTau_large_regime2}, we also plot the critical threshold computed from the reduced $\tau \sim \mathcal{O}(\varepsilon^{-3})$ regime. The results from the reduced regime converge increasingly close to those from the full $\mathcal{O}(\varepsilon^{-3})$ regime as $S_\varepsilon$ decreases, where $\mu_0 = \tau_0 \lambda \ll 1$.

\begin{figure}[h]
    \centering
    \includegraphics[width=0.5\linewidth]{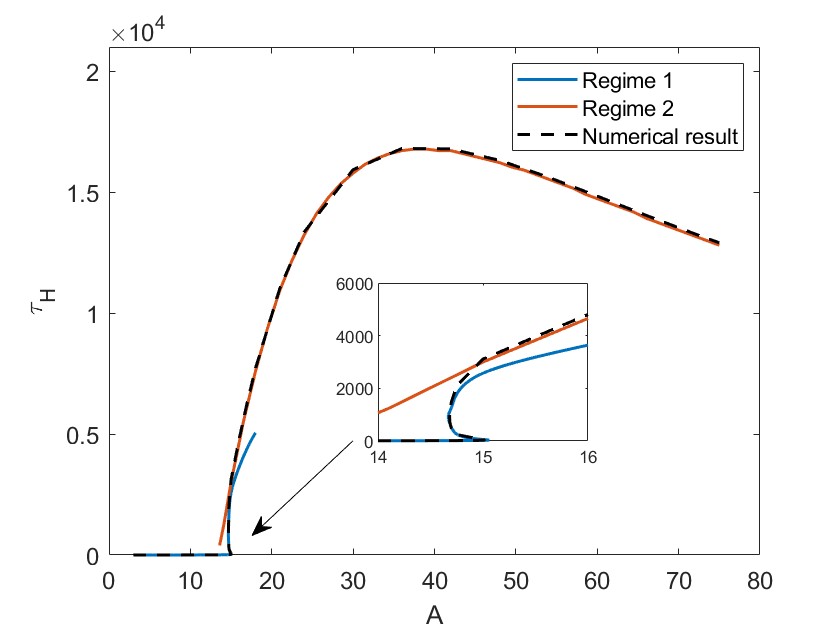}
    \caption{Full theoretical Hopf bifurcation threshold and the full numerical solution by solving (\ref{eigen_problem}) for $\varepsilon = 0.025$.}
    \label{fig:full_large&numeric}
\end{figure}

We now conclude the Hopf bifurcation analysis in the large eigenvalue problem. For the feed rate $A$ ranging from $1$ to $60$, two regimes of the Hopf bifurcation threshold exist. The first is $\tau \sim \mathcal{O}(1)$, valid for $1 < A < A_\varepsilon$, where $A_\varepsilon > 13.56$ is a saddle point. The second regime is $\tau \sim \mathcal{O}(\varepsilon^{-3})$, valid for $13.56 < A < 60$. Notably, three thresholds are observed when $13.56 < A < A_\varepsilon$. A numerical experiment validates this behavior, showing clearly how the instability changes as $\tau$ increases within this interval of $A$. For $A > 60$, the splitting instability occurs and is therefore not considered here. A numerical solution of the full eigenvalue problem in (\ref{eigen_problem}) is provided in Figure \ref{fig:full_large&numeric} to verify our theoretical result.

\section{Numerical validation}\label{numerical}
In this section, we numerically validate our theoretical result in (\ref{eig_matrix}) for the small eigenvalue problem and the HB threshold for the large eigenvalue problem by solving the full time-dependent PDE system in (\ref{mod}) using the finite element solver FlexPDE 7.

The procedure for verifying the threshold is as follows. We begin by setting $\tau$ below any Hopf thresholds, then run the PDE until a sufficiently large time is reached, such that no further changes in the solution are observed. Using this equilibrium solution as an initial condition, we trial various values of $\tau$ to test stability. The numerical lower bound $\tau_{lower}$ of the Hopf bifurcation threshold is identified when the oscillation amplitude decays and stabilizes at the end, while the numerical upper bound $\tau_{upper}$ is identified when the oscillation amplitude grows gradually. In the asymmetric domain, only the smallest critical $\tau$ is tested.

\subsection{Hopf bifurcation for small eigenvalue problem}
In this section, we examine the theoretical result in (\ref{eig_matrix}) for the small eigenvalue problem on three domains: a unit sphere, a perturbed sphere, and a sphere with a single ball-shaped defect.

\begin{figure}[h]
    \centering
        \begin{subfigure}[b]{0.45\textwidth}
            \centering
            \includegraphics[width=\textwidth]{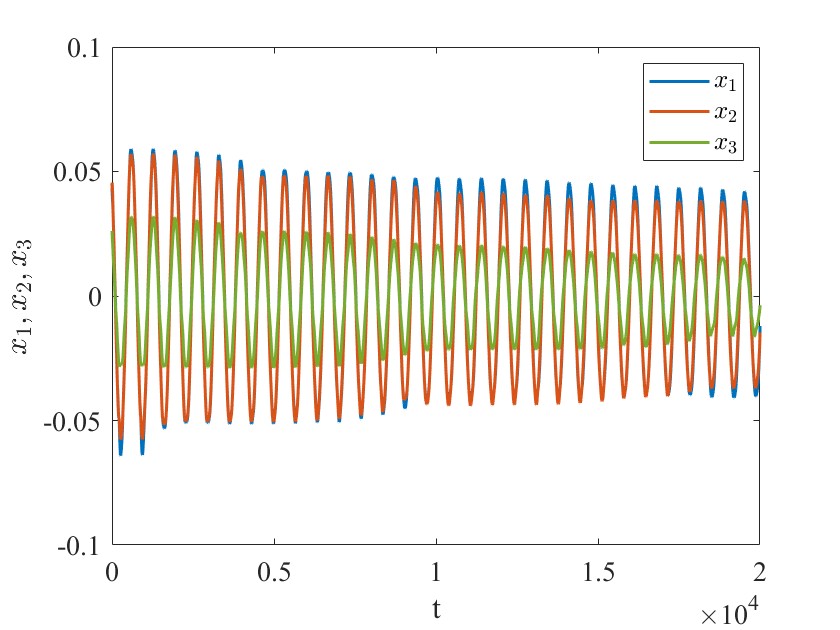}
            \caption{}
        \end{subfigure}
        \begin{subfigure}[b]{0.45\textwidth}
            \centering
            \includegraphics[width=\textwidth]{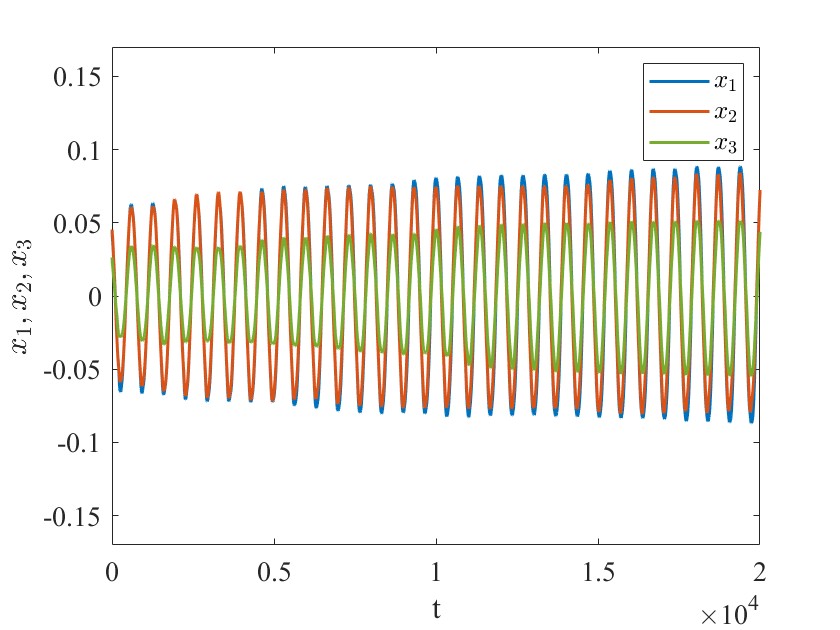}
            \caption{}
        \end{subfigure}
        \caption{\label{fig:unit}Numerical results from FlexPDE 7 with $\varepsilon = 0.03$ and $S_\varepsilon = 15$. (a) for $\tau = 0.3448$, a decaying oscillation is observed, indicating that $\tau$ is below the HB threshold. (b) for $\tau = 0.3460$, a growing oscillation is observed, indicating that $\tau$ has exceeded the HB threshold.}
\end{figure}

\textbf{Experiment 1}. In this example, we place the problem on a simple, highly symmetric geometry - a unit sphere. In this context,the thresholds are identical for all three directions, resulting in no preferred oscillatory mode. Other parameters are $\varepsilon = 0.03$ and $S_\varepsilon = 15$. The theoretical values computed from Section \ref{unit} are as follows:
\begin{equation}
    \tau_0 = 0.3427,\quad \tau_1 = -0.0688,\quad \tau_2 = 4.7448,\quad \tau = 0.3449.
\end{equation}
We test the stability when $\tau = 0.3448$ and $\tau = 0.3460$ in FlexPDE. The results in Figure \ref{fig:unit} show a decaying oscillation for $\tau = 0.3448$ and a growing oscillation for $\tau = 0.3460$.

\textbf{Experiment 2.} In this example, we chose $\sigma = 0.05$, resulting the perturbation of the domain given by
\begin{equation}
    r = (1+0.05)f(\theta,\varphi),
\end{equation}
where $ f(\theta,\varphi) =  \frac{1}{2}(3\cos^2{\theta}-1) $. This configuration sets the major axis of the geometry along the $x_3$-direction

We set $S_\varepsilon = 19$, $\varepsilon = 0.03$ and the initial position of the spot is at the origin. We then solve (\ref{eigen-system}) in MATLAB and compare the results with those obtained using FlexPDE under the same parameters and initial position. Due to the geometry's shape, the thresholds $\tau$ for the $x_1$- and $x_2$-directions are identical, while the threshold in the $x_3$-direction differs. As expected, two distinct values of $\tau_{11}$ are found, corresponding to different oscillatory modes. The positive value indicates oscillation starting in the $x_{1,2}$-plane, while the negative value corresponds to oscillation along the $x_3$-axis.

The theoretical result is
\begin{equation}
    \tau_{00} = 0.2512,\quad \tau_{10} = -0.0565, \quad \tau_{11} = -5.1169(3.1779),\quad \tau = 0.2419(0.2543).
\end{equation}
Therefore, as $\tau$ increases to reach the $(0,0,1)^T$ mode threshold, the spot begins oscillating along the $x_3$-axis, which is the preferred mode of oscillatory translational instability.
\begin{figure}[h]
    \centering
        \begin{subfigure}[b]{\textwidth}
                \centering
                \includegraphics[width=0.32\textwidth]{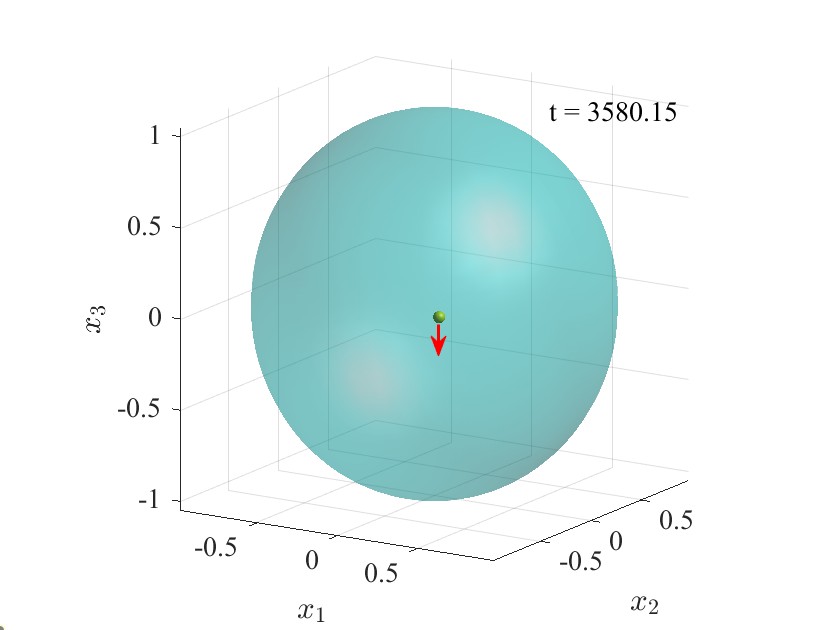} 
                \includegraphics[width=0.32\textwidth]{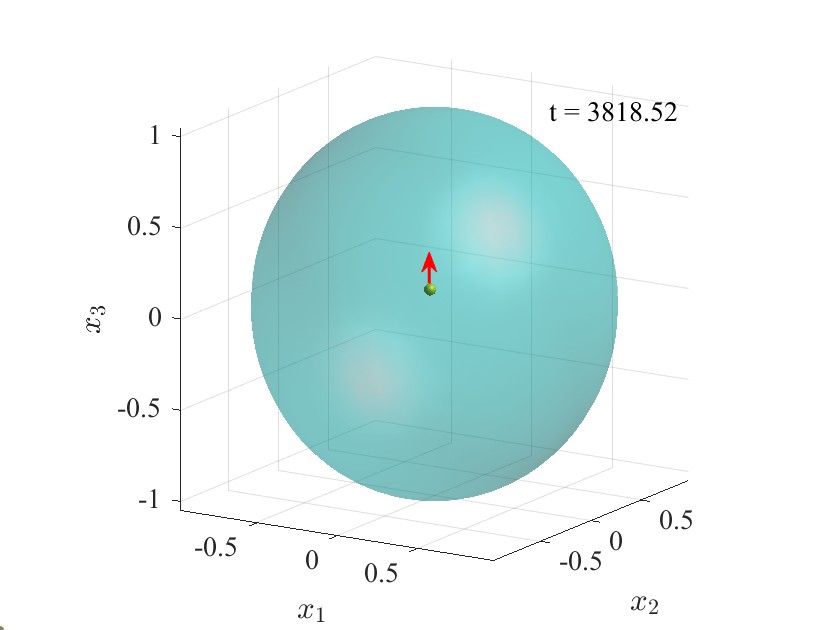}
                \includegraphics[width=0.32\textwidth]{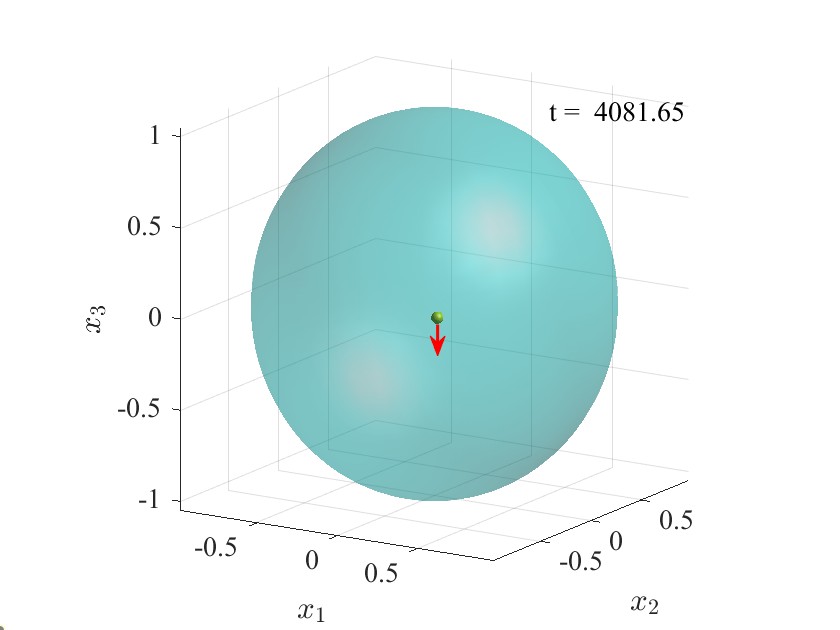}
                \caption{}
                \label{fig:peturbed_domain1}
            \end{subfigure}
            \begin{subfigure}[b]{\textwidth}
                \centering
                \includegraphics[width=0.45\textwidth]{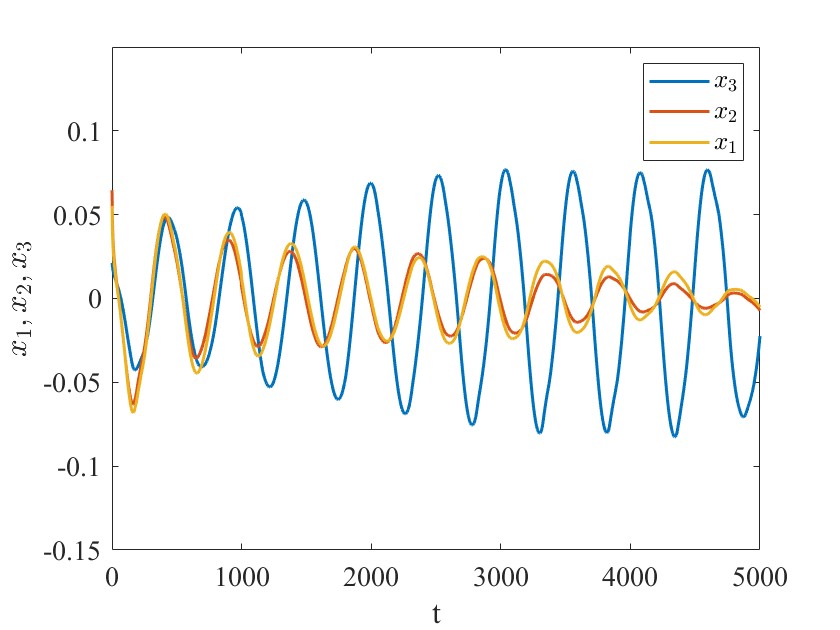} 
                \includegraphics[width=0.45\textwidth]{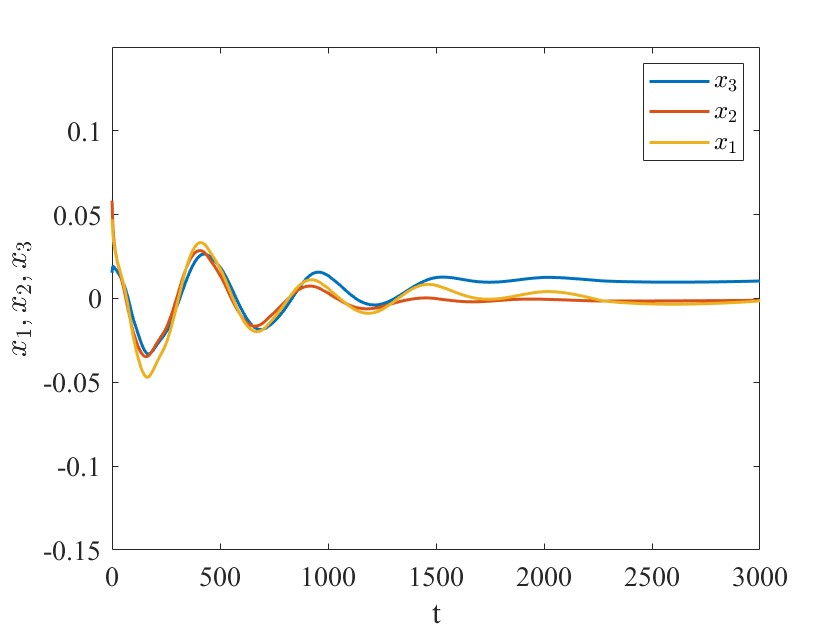} 
                \caption{}
                \label{fig:peturbed1}
            \end{subfigure}
    \caption{\label{fig:ellipsoid}In the top half, we figure out the direction of spot motion with arrows at the perticular time. In the bottom half, we plot all three coordinates as functions of time, respectively. Parameters: $S_\varepsilon = 19$, $\varepsilon = 0.03$, $\sigma = 0.05$. Left: $\tau = 0.2540$, $\tau_{11} = 3$. Right: $\tau = 0.2420$, $\tau_{11} = -5$. For increasing time, the $x_3$-direction mode (major axis of the domain) emerges as dominant.}
\end{figure}

\begin{figure}[h]
    \centering
    \begin{subfigure}[b]{0.45\textwidth}
        \centering
        \includegraphics[width=\textwidth]{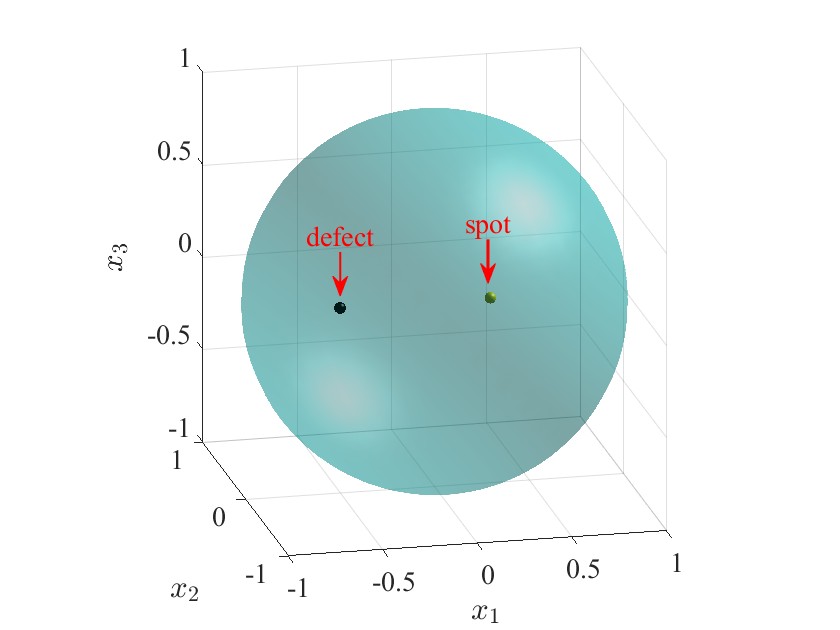} 
        \caption{}
        \label{fig:domain_initial}
    \end{subfigure}
    \begin{subfigure}[b]{0.45\textwidth}
    \centering
        \includegraphics[width=\textwidth]{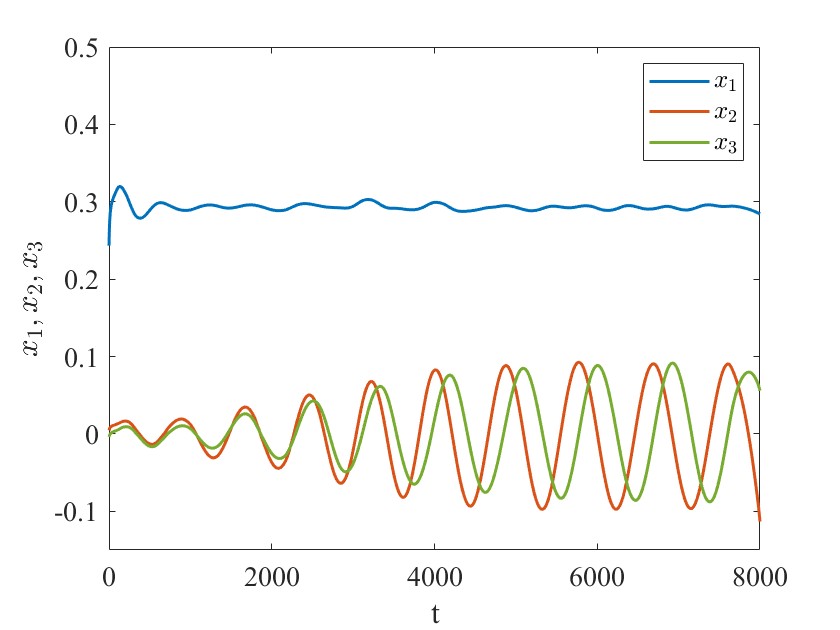} 
        \caption{}
       \label{fig:defect1}
    \end{subfigure}
    \caption{Numerical results for a single-spot pattern with one defect are presented. Parameters are set as $\varepsilon = 0.03$, $A = 69$, and $C = 1$. Figure (a) shows the initial pattern for this experiment: the defect is placed at $(-0.5, 0, 0)$, and the spot is initialized at $(0.2956, 0, 0)$. In (b), with $\tau = 0.3530$ and $\tau_1 = 1$, the oscillation in the $x_3$ direction is decaying, while oscillations in the $x_{2}$-, $x_3$-directions are growing, indicating that the oscillation mode along $x_{2,3}$-plane is selected as the dominant mode.}
    \label{fig:defect}
\end{figure}

\textbf{Experiment 3}. In this experiment, we present numerical results for the translational oscillatory instability on a sphere with a single defect. We compare our analytical results obtained using MATLAB with the numerical solution provided by FlexPDE for the same parameters. Here, we set $\varepsilon = 0.03$ and $A = 69$. The defect is positioned at $(-0.5, 0, 0)$, and we set $C = 1$, making the radius of the hole equal to $\varepsilon$ (see Figure \ref{fig:domain_initial}).

By solving the equilibrium problem, we find that the equilibrium position of the spot is $(0.2956, 0, 0)$, with a spot source strength of $S_{\varepsilon1} = 15.7073$ and a hole source strength of $S_{\varepsilon0} = 7.2927$. These values match those obtained from FlexPDE.

The leading order threshold $\tau_0$ is determined by the source strength of the spot, $S_{\varepsilon1}$. Consequently, $\tau_0$ is the same for all three directions, while the preferred direction of translational instability is determined by the first correction of $\tau$. Due to the initial placement of the spot on the $x_1$-axis, we expect two distinct thresholds: one corresponding to the $x_1$ direction and the other to the $x_{2,3}$ directions. The matrix $\mathcal{Q}_{\mu 1}$ is computed by applying central difference on the resulting regular part in (\ref{R_mu}).  As a result, $\tau_0 = 0.3230$, while one value of $\tau_1$ is $1.0051$, corresponding to the $x_1$ direction, and the other is $-0.3894$, corresponding to the $x_{2,3}$ directions. In this case, the smaller critical value of $\tau$ serves as the HB threshold, beyond which the translational oscillatory instability is triggered. The preferred oscillatory mode at onset is the one along the tangent plane and perpendicular to the $x_1$ axis. Figure \ref{fig:defect1} shows the numerical results in FlexPDE for $\tau_1 = 1$. As we expected, the oscillation on $x_3$-direction is decaying while that on $x_{2}$-, $x_3$-direction are growing, indicating the oscillation mode along $x_{2,3}$ plane is selected as a dominant mode.

\subsection{Hopf bifurcation for large eigenvalue problem}
In this section, we test the oscillatory instability of the spot's amplitude for the critical $\tau$ obtained in Section \ref{large_stability}. In the first two experiments, we use FlexPDE to find the lower bound and upper bound for the threshold in each regime. In the last experiment, as we demonstrate in Section \ref{large_stability}, there is an overlap between two regimes. We select an intermediate value of $A$ and test the system’s stability when $\tau$ lies between the two critical values. The aim of this experiment is to understand the behavior of the dominant eigenvalue and the changes in stability as $\tau$ increases.

\textbf{Experiment 4}. In this experiment, We numerically validate the threshold for the $\tau \sim \mathcal{O}(1)$ regime (regime 1) and the $\tau \sim \mathcal{O}(\varepsilon^{-3})$ regime (regime 2), respectively. Here, we define $\tau_{upper}$ and $\tau_{lower}$ as the upper and lower bounds of the true threshold, where a growing oscillation in amplitude indicates $\tau_{upper}$ and a decaying oscillation indicates $\tau_{lower}$. The 'exact' HB threshold lies between these bounds. 

\begin{figure}[h]
    \centering
    \begin{subfigure}[b]{\textwidth}
    \centering
        \includegraphics[width=0.8\textwidth]{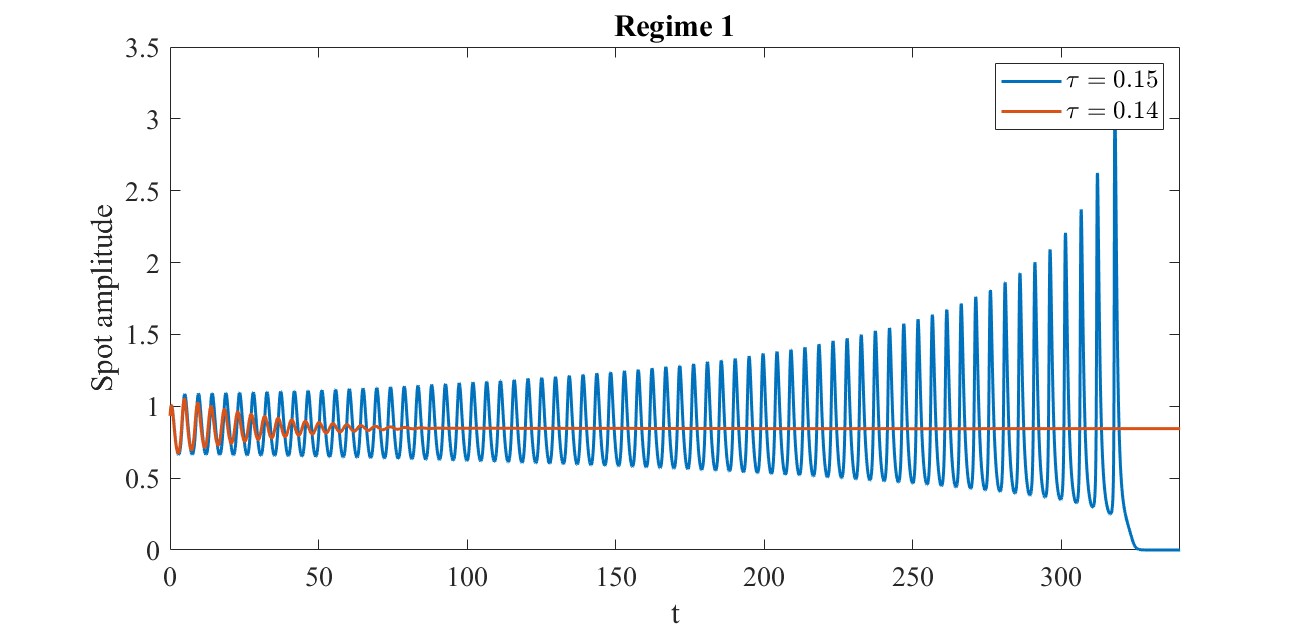} 
        \caption{}
       \label{fig:regime1_1}
    \end{subfigure}
    \begin{subfigure}[b]{0.4\textwidth}
    \centering
        \includegraphics[width=\textwidth]{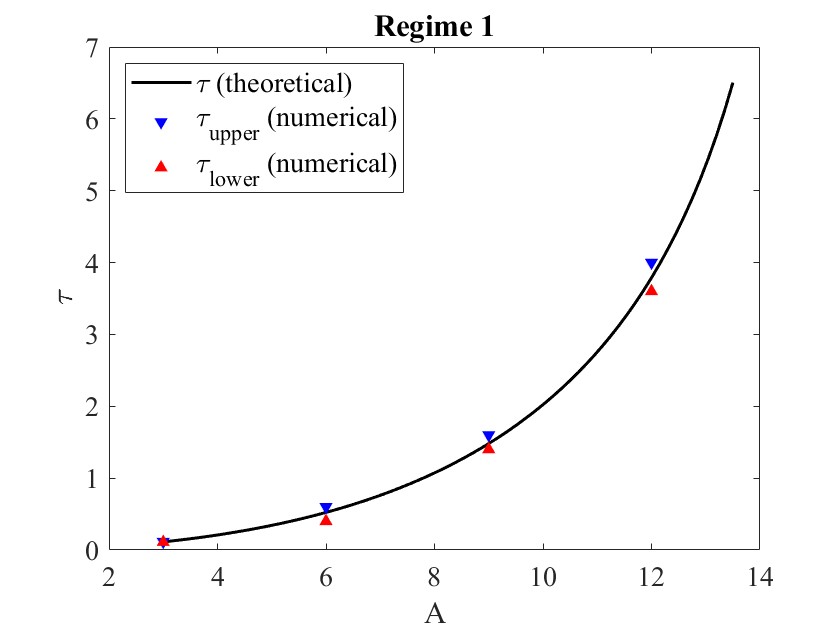} 
        \caption{}
       \label{fig:regime1_2}
    \end{subfigure}
    \begin{subfigure}[b]{0.4\textwidth}
        \centering
        \includegraphics[width=\textwidth]{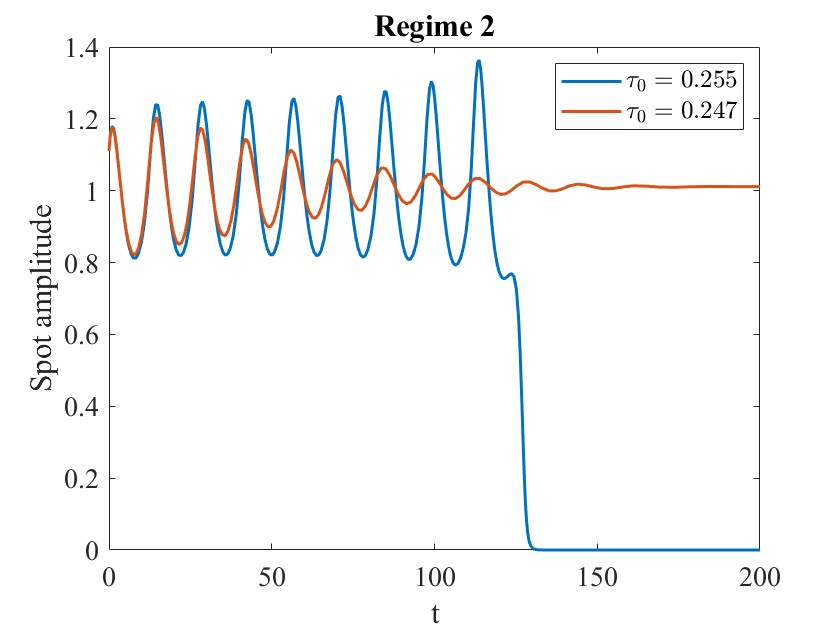} 
        \caption{}
        \label{fig:regime2}
    \end{subfigure}
    \caption{(a) A plot of the spot height from numerically solving (\ref{mod}) using FlexPDE7 in a unit ball, with $A = 3$ and $\varepsilon = 0.03$, for the indicated values of $\tau$ in regime 1. The exact threshold lies between $1.4$ and $1.5$. The theoretical threshold $1.4814$ aligns closely with the numerical result. (b) A comparison of theoretical and numerical results for various values of $A$ in regime 1, with $\varepsilon = 0.03$. We determine $\tau_{upper}$($\tau_{lower})$ is the upper(lower) bound of the true threshold when a growing(decaying) oscillation on the amplitude is observed in FlexPDE. (c) A plot validating the leading order of the HB threshold in regime 2. The theoretical value of $\tau_0$ is $0.2472$, while numerical predictions suggest it lies between $0.247$ and $0.255$. Other parameters are set to $A = 10$ and $\varepsilon = 0.02$}
    \label{fig:regime1}
\end{figure}

For regime 1, we solve (\ref{BTauLamb}) from Section \ref{regime1} in a unit ball with $A = 3$ and $\varepsilon = 0.03$, yielding a theoretical value of $\tau_{H} = 1.4814$. The numerical threshold, as shown in Figure \ref{fig:regime1_1}, is between $1.4$ and $1.5$ which coincide with our theoretical result. We then plot the numerical thresholds for various values of $A$ and compare them to our predicted results in Figure \ref{fig:regime1_2}. The theoretical thresholds align closely with the numerical results.

For regime 2, we test the system's instability by setting $\tau$ near the theoretical threshold calculated in Section \ref{regime2} for $A = 10$. The theoretical threshold is $\tau_0 = 0.2472$, where $\tau_0 = \varepsilon^3 \tau$ with $\varepsilon = 0.02$. As shown in Figure (\ref{fig:regime2}), the plot shows that the exact value of $\tau_0$ lies between $0.247$ to $0.255$. 

\begin{figure}[ht!]
    \centering
    \begin{subfigure}[b]{0.45\textwidth}
        \centering
        \includegraphics[width=\textwidth]{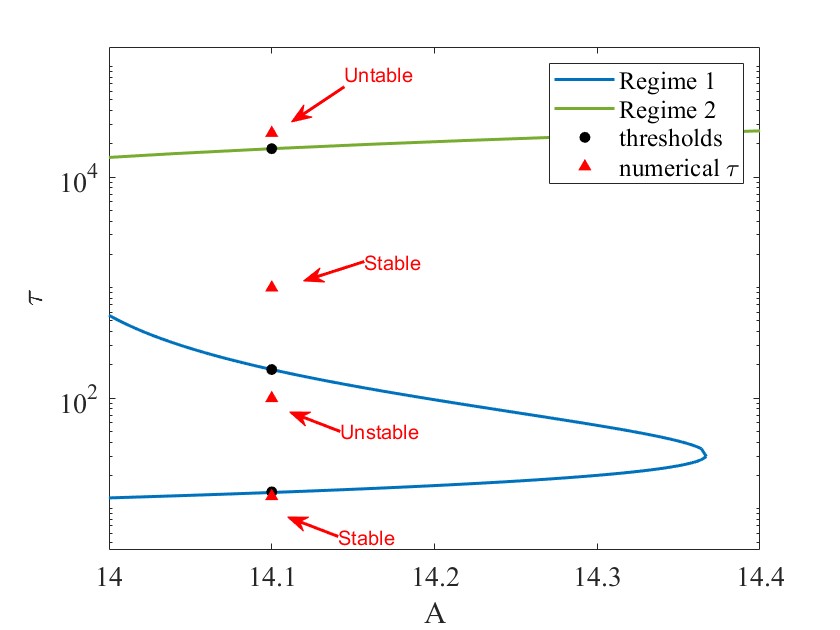} 
        \caption{}
        \label{fig:overlap1}
    \end{subfigure}
    \begin{subfigure}[b]{0.45\textwidth}
    \centering
        \includegraphics[width=\textwidth]{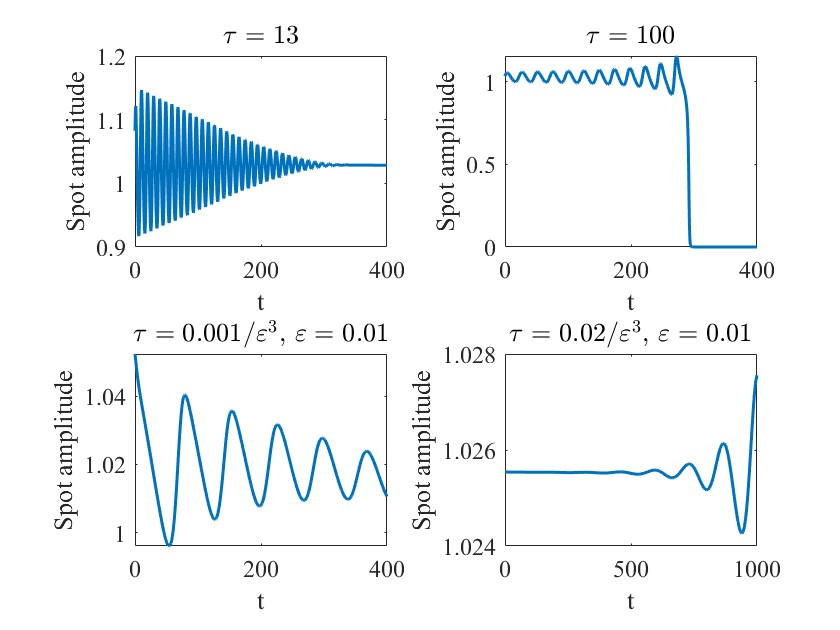} 
        \caption{}
       \label{fig:overlap2}
    \end{subfigure}
    \caption{We examine the stability in the overlapping region for $\varepsilon = 0.01$. (a) We set $A = 14.1$, and the three predicted thresholds are $\tau_H = 13.9956$, $\tau_H = 184.04$, and $\tau_H = 0.0181/\varepsilon^2$. We then select four different values of $\tau$: $13$, $100$, $0.001/\varepsilon^3$, and $0.02/\varepsilon^3$, and predict the stability as indicated. (b) The results of these four $\tau$ values in FlexPDE are shown. The left two plots indicate that the system is stable when $\tau = 13$ and $\tau = 0.001/\varepsilon^3$, while the right two plots show that it is unstable when $\tau = 100$ and $\tau = 0.02/\varepsilon^3$. The behavior of the spot's height matches our predictions.}
    \label{fig:overlap}
\end{figure}

\textbf{Experiment 5} This experiment aims to investigate amplitude instability in the overlapping region of the two regimes. As $\tau$ increases and crosses each threshold, the stability status changes, as analyzed in Section \ref{large_stability}. We set $\varepsilon = 0.01$, resulting in a saddle-node at $A_\varepsilon = 14.37$. The experiment is conducted at $A = 14.1$, where three distinct thresholds can be found. Our predicted thresholds are $\tau_H = 13.9956$, $\tau_H = 180.04$, and $\tau_H = 0.0181/\varepsilon^3$ as shown in Figure \ref{fig:overlap1}. We then select four different values of $\tau$: $\tau = 13$, $100$, $0.001/\varepsilon^3$, and $0.02/\varepsilon^3$. These values of $\tau$ are examined later in FlexPDE to confirm the stability of the system. In Figure \ref{fig:overlap2}, the left two plots show that the system is stable when $\tau = 13$ and $\tau = 0.001/\varepsilon^3$, while the right two plots indicate that it is unstable when $\tau = 100$ and $\tau = 0.02/\varepsilon^3$. The behavior of the spot's height matches our predictions.

\section{Discussion}\label{conclusion}
We have developed a $3 \times 3$ complex matrix eigenvalue problem through formal asymptotic analysis, yielding the Hopf bifurcation threshold, oscillatory frequency, and direction of oscillation at the onset of a one-spot pattern equilibrium solution in the Schnakenberg reaction-diffusion system. This result is applicable to both general and defected bounded three-dimensional domains. Additionally, we provide a detailed analysis of parameter values associated with the onset of temporal oscillations in spot amplitudes for the one-spot pattern in this system. 

In the small eigenvalue problem, we derived a complex matrix eigenvalue problem. By solving this, we obtained the leading order of the oscillation frequency $\lambda$ and the bifurcation threshold $\tau_{H}$ at least up to the $\mathcal{O}(\varepsilon)$ correction term.

We then extended these results to a perturbed sphere with radius in polar coordinates given by $r = 1 + \sigma f(\theta, \varphi)$, where $\theta \in [0, \pi)$, $\varphi \in [0, 2\pi)$, and $\varepsilon \ll \sigma \ll 1$. Here, $f(\theta, \varphi)$ is a Spherical Harmonic function of the form $P_n^m(\cos{\theta}) \cos{m \varphi}$. We observed that for $n=1$, the perturbation on the geometry results in a translation along with an $\mathcal{O}(\sigma^2)$ dilation, which does not affect preferred osculation direction. For $n \geq 3$, the perturbation does not affect the computation of thresholds. However, for $n = 2$, the geometry is stretched along a specific direction, resulting in an ellipsoid. In this configuration, the spot tends to oscillate along the major axis at onset.

We also examined the problem on a defected sphere. Here, the threshold is closely related to the defect size: a larger defect size increases the threshold, making translational oscillatory instability more difficult to trigger. Moreover, the presence of the defect breaks the high symmetry of the sphere when the defect is not centered. Consequently, the critical value of $\tau$ is no longer identical in all three directions. In this case, the smallest critical $\tau$ value is defining as the HB threshold.

In the large eigenvalue problem, we analyzed the Hopf bifurcation threshold $\tau_{H}$ in relation to the feed rate $A$. Two regimes were found in this setup: the $\tau \sim \mathcal{O}(1)$ regime (regime 1) and the $\tau \sim \mathcal{O}(\varepsilon^{-3})$ regime (regime 2). For regime 1, we applied a hybrid asymptotic-numerical method to determine the HB threshold. For regime 2, the threshold was determined based on the spectrum of the leading order eigenvalue problem in the inner region.

An overlap between these two regimes was observed for a range of $A$ values. Within this interval, as $\tau$ increases, the system undergoes a stable-unstable-stable-unstable transition with respect to amplitude oscillatory instability. An interesting finding is an anomalous scaling between the $\mathcal{O}(1)$ regime and the $\mathcal{O}(\varepsilon^{-3})$ regime. In the zoom-in box of Figure \ref{fig:full_large&numeric}, a gap between these two regimes is observed, and the scaling of the threshold transitions continuously from $\mathcal{O}(1)$ to $\mathcal{O}(\varepsilon^{-3})$ as $A$ increases. Anomalous scaling for the threshold of temporal spot-height oscillatory instability in the 2D problem has been discussed in \cite{hopf2d}, and this could serve as a potential direction for future work.

Our analysis was developed for a single-spot pattern in the Schnakenberg model. However, a similar approach could be applied to other activator-inhibitor reaction-diffusion models, such as the Gray-Scott and Brusselator models. Additionally, extending this analysis to multi-spot patterns presents a potential direction for future research. 

It would also be interesting to explore other types of heterogeneity in similar problems. The effect of a spatially uniform feed rate on spot height in a $2$D problem has been analyzed in \cite{tony-defect-2d}. Extending this discussion to the effects on translational and amplitude oscillatory instabilities in a $3$D domain would be valuable. Additionally, in our problem, the chemicals' leakage from a defect is modeled by Dirichlet boundary conditions. However, Neumann boundary conditions could be applied if no chemical leakage occurs. In this context, a different type of Green's function, as discussed in \cite{Maz'ya2009}, would be suitable for modeling this situation during the process of construction the solutions.

Lastly, the relationship between the equilibrium configurations of $N$-spot patterns in the Schnakenberg model and optimal target configurations in the narrow escape optimization problem has been well investigated in a unit sphere ($3$D) and a unit disk ($2$D). (see \cite{Bressloff2020, narrowEscap_3d, tony-defect-2d, Tzou2014MeanFP}). It would be worthwhile to explore whether this correspondence persists in more general three-dimensional domains.

\section*{Acknowledgement}
S. Deng and J.C. Tzou were supported by ARC Discovery Project DP220101808.

\appendix
\section{Equilibrium solution of a single spot pattern on a defected domain}\label{equl_defect}
The idea here follows \cite{tony-defect-2d}. We construct the equilibrium solution for a one-spot pattern problem with one defect on a $3$D domain. We assume that $|\bold{x}_i-\bold{x}_j| = \mathcal{O}(1)$ for $i\neq j$, and $\text{dist}(\bold{x}_i,\partial\Omega) = \mathcal{O}(1)$ and $\text{dist}(\bold{x}_i,\partial\Omega_\varepsilon) = \mathcal{O}(1)$, where $i = 1$ is the position of the spot and $i = 0$ is the position of the defect.

We then introduce the inner variables
\begin{equation}\label{inner_var_defect}
    \begin{aligned} 
   & \bold{y}_j = \frac{\bold{x}-\bold{x}_j}{\varepsilon} = \rho_j\mathbf{e}_j, \quad \mathbf{e}_j = \left(
    \begin{array} {cc}
    \sin\theta_j \cos\varphi_j\\
   \sin\theta_j \sin\varphi_j\\
   \cos\theta_j
    \end{array}\right),\\
&v_e \sim \sqrt{D} (V_{\varepsilon}(\rho_j)+\varepsilon^3 V_{3}(\bold{y}_j)+\dots), \quad u_e \sim \frac{1}{\sqrt{D}} (U_{\varepsilon}(\rho_j)+\varepsilon^3 U_{3}(\bold{y}_j)+\dots),   
\end{aligned} 
\end{equation}
where $j = 0$ representing the defect and $j= 1$ is the spot. $U_{\varepsilon j}$, $V_{\varepsilon j}$ are weakly depending on $\varepsilon$, and they satisfy the radially symmetric core problem

\begin{equation}\label{core_defect}
    \begin{aligned}
           \Delta_{\rho_j} V_{\varepsilon j} - V_{\varepsilon j} + U_{\varepsilon j}V_{\varepsilon j}^2 &= 0, \quad V'_{\varepsilon j}(0) =0,\\
    \Delta_{\rho_j} U_{\varepsilon j} -U_{\varepsilon j}V_{\varepsilon j}^2 &=0, \quad U'_{\varepsilon j}(0) = 0 
    \end{aligned} 
\end{equation}
with far-field behavior
\begin{equation}\label{core_farfield_defect}
    V_{\varepsilon j} \rightarrow 0,\quad U_{\varepsilon j} \sim \chi(S_{\varepsilon j})- S_{\varepsilon j}/\rho_j+..., \quad \text{as} \quad \rho \rightarrow \infty,
\end{equation}

Applying the divergence theorem on the second equation of (\ref{core_defect}), we have
\begin{equation}\label{Sj_defect}
    \begin{aligned} 
    S_{\varepsilon j} = \int_0^\infty U_{\varepsilon j}V_{\varepsilon j}^2\rho_j^2 d\rho_j.
\end{aligned} 
\end{equation}

Now we use strong localized perturbation theory to replace the effect of the defect with Dirac singularity. Let
\begin{equation}
    \bold{y}_0 = \varepsilon^{-1}(\bold{x}-\bold{x}_0) \quad \text{and} \quad U \sim \frac{U_{\varepsilon 0 }}{\sqrt{D}},
\end{equation}
where $U_{\varepsilon0}$ is $\mathcal{O}(1+\varepsilon)$, the leading order hence satisfies
\begin{equation}\label{leadingU0}
    \Delta_{\bold{y}_0} U_{\varepsilon0} = 0, \quad |\bold{y}_0|\geq C; \quad U_{\varepsilon0} = 0 \quad \text{on}\quad |\bold{y}_0| = C.
\end{equation}
We emphasize that $V_{\varepsilon 0}$ near the defect is zero because $V_{\varepsilon 0} = 0$ when $|\bold{y}_0|= C$ and the far-field of $V_{\varepsilon 0}$ is also zero.

The solution to (\ref{leadingU0}) is $U_{\varepsilon0} = S_{\varepsilon0}\left(\frac{1}{C}-\frac{1}{|\bold{y}_0|}\right)$, where $S_{\varepsilon0}$ to be determined. It gives the local behavior of the outer $u$, which is
\begin{equation}\label{farfieldU0}
    u \sim \frac{S_{\varepsilon0}}{\sqrt{D}}\left(\frac{1}{C}-\frac{\varepsilon}{|\bold{x} - \bold{x}_0|}\right), \quad \text{as} \quad \bold{x} \rightarrow \bold{x}_0.
\end{equation}
Therefore, we have the identity
\begin{equation}
    \int_{\partial\Omega_\varepsilon} -D\partial_n u|_{\partial\Omega_\varepsilon}ds \sim 4\pi S_{\varepsilon0}\sqrt{D},
\end{equation}
where $\partial_n$ is the outward normal derivative to $\bar{\Omega}$. That implies the constant $S_{\varepsilon0}$ is proportional to the diffusive flux of inhibitor through the defect.

By imposing the Dirac singularity $\frac{4\pi S_{\varepsilon0}}{\sqrt{D}}\delta(\bold{x}-\bold{x}_0)$ on the outer problem and expanding $u\sim u_0 + \varepsilon u_1$, we obtain, in $\mathcal{O}(\varepsilon)$,
\begin{equation}\label{outer_u1}
    \begin{aligned}
        &D\Delta u_1 + A = 4\pi \Big[ S_{\varepsilon1}\delta(\bold{x}-\bold{x}_1)+S_{\varepsilon0}\delta(\bold{x}-\bold{x}_0)\Big]\quad \text{in} \quad \Omega;\\
        &\partial_n u_1 = 0 \quad \text{on} \quad \partial\Omega.
    \end{aligned}
\end{equation}
The solution to (\ref{outer_u1}) is represented in terms of the Neumann Green's function as
\begin{equation}
    u_1 = -\frac{4\pi}{\sqrt{D}}\big(S_{\varepsilon1}G(\bold{x};\bold{x}_1)+S_{\varepsilon0}G(\bold{x};\bold{x}_0) \big) + \bar{u}_1,
\end{equation}
 where $G(\bold{x};\bold{x}_j)$ satisfies (\ref{Neumann_Green}). Therefore, $u$ can be expressed as
\begin{equation}\label{sol_outer_u}
    u = u_0 + \varepsilon \Big( \bar{u}_1  -\frac{4\pi}{\sqrt{D}}\big(S_{\varepsilon1}G(\bold{x};\bold{x}_1)+S_{\varepsilon0}G(\bold{x};\bold{x}_0) \big)\Big).
\end{equation}
By applying the divergence theorem to (\ref{outer_u1}) we get
\begin{equation}\label{S_defect}
    S_{\varepsilon1}+S_{\varepsilon0} = \frac{A|\Omega|}{4\pi\sqrt{D}}.
\end{equation}
We next complete the matching by analysing the local behavior at the spot and the defect, respectively. Further to determine the constant, we find the expanding form of the local behavior of the outer solution. When $\bold{x} \rightarrow \bold{x}_0$, we substitute the local behavior of the Green's function given in (\ref{Neumann_Green}) into (\ref{sol_outer_u}) and match with (\ref{farfieldU0}), yielding
\begin{equation}\label{matching_0}
    u_0+\varepsilon \bar{u}_1 -\frac{4\pi\varepsilon S_{\varepsilon0}}{\sqrt{D}}R(\bold{x}_0;\bold{x}_0)-\frac{4\pi \varepsilon S_{\varepsilon1}}{\sqrt{D}}G(\bold{x}_0;\bold{x}_1) = \frac{S_{\varepsilon0}}{C\sqrt{D}}.
\end{equation}
Recall that the far-field of $U_{\varepsilon 1}$ in the core problem is given by $U_{\varepsilon 1}\sim \chi(S_{\varepsilon1})-\frac{\varepsilon S_{\varepsilon1}}{|\bold{x} - \bold{x}_1|}$. Similarly, when $\bold{x}\rightarrow \bold{x}_1$, we have
\begin{equation}\label{matching_1}
    u_0+\varepsilon \bar{u}_1 -\frac{4\pi \varepsilon S_{\varepsilon1}}{\sqrt{D}}R(\bold{x}_1;\bold{x}_1)-\frac{4\pi \varepsilon S_{\varepsilon0}}{\sqrt{D}} G(\bold{x}_1;\bold{x}_0) = \chi(S_{\varepsilon1}).
\end{equation}
Since we treat the defect as a pined spot in our analysis, we then use the equilibrium condition given in \cite{jst-3d}, which is
\begin{equation}\label{equilibrium}
    \frac{d\bold{x}_1}{dt} = -\frac{12\pi \varepsilon^3}{k_1}\big(S_{\varepsilon1}\nabla_\bold{x}R(\bold{x}_1;\bold{x}_1)+S_{\varepsilon0}\nabla_\bold{x}G(\bold{x}_1;\bold{x}_0)\big).
\end{equation}

By solving (\ref{S_defect}), (\ref{matching_0}), (\ref{matching_1}) and (\ref{equilibrium}) simultaneously, we obtain the source strength of the spot $S_{\varepsilon1}$, the source strength of the hole $S_{\varepsilon0}$, the equilibrium position of the spot, and all other constants.

\section{The regular part of the Helmholtz Green's function}\label{Rmu}
In this section, we derive $\mathcal{Q}_{\mu1}$ for the computation in Section \ref{defected}. The approach begins with constructing a general form of the Helmholtz Green's function $G\mu$ solution using a Fourier series which satisfies (\ref{Helm}). To isolate the regular part, we use the expression $G_\mu - G_{\mu f} + R_{\mu f}$, where $G_{\mu f}$ represents the corresponding free space Helmholtz Green's function, and $R_{\mu f}$ is its regular part

We begin by expressing the Helmholtz Green's function. Using separation of variables, we solve for $G_\mu(\bold{x}; \bold{x}_0)$ in (\ref{Helm}) and represent it in polar series form, with $\bold{x} = \rho \bold{e}$ and $\bold{x}_0 = \rho_0 \bold{e}_0$, where $\bold{e}_j$ is defined in (\ref{inner_var_defect}), yielding
\begin{equation}\label{G_mu}
    \begin{aligned}
    G_\mu(\rho;\phi;\theta;\rho_0;\phi_0;\theta_0)=\sum_{n=0}^\infty A_{\mu n}(\rho,\rho_0)\sum_{m=0}^n Y_n^m(\phi,\theta,\phi_0,\theta_0),
    \end{aligned} 
\end{equation}
where 
\begin{equation}\label{spherical_harmonics}
    \begin{aligned}
       & A_{\mu n}(\rho,\rho_0) =(2n+1)\rho^{-\frac{1}{2}}\rho_0^{-\frac{1}{2}} \Bigg\{\begin{array}{c}
              \left[Q_{q}(\mu)I_q(\mu\rho_0)+K_q(\mu \rho_0)\right]I_q(\mu \rho),\qquad 0\leq\rho<\rho_0 \\
             \left[Q_{q}(\mu)I_q(\mu\rho)+K_q(\mu \rho)\right]I_q(\mu \rho_0),\qquad \rho_0 \leq \rho \leq 1
        \end{array},\\
       & Y_n^m(\phi,\theta,\phi_0,\theta_0) = c P_n^m(\cos{\theta})P_n^m(\cos{\theta_0})\cos{[m(\phi-\phi_0)]}.
    \end{aligned}
\end{equation}
Here, $I_q(z)$ and $K_q(z)$ are the first and second kind of modified Bessel function, respectively, and $q = n+\frac{1}{2}$, $Q_q(\mu) = -\frac{2\mu K'_q(\mu)-K_q(\mu)}{2\mu I'_q(\mu)-I_q(\mu)}$. The constant coefficient $c$ is given by $c = -\frac{1}{4\pi}$ when $m=0$, and $c = -\frac{1}{2\pi} \frac{(n - m)!}{(n + m)!}$ when $m \geq 1$.

In a free domain, the exact solution of the Helmholtz Green's PDE in $3$D is given by $G_{\mu f} = -\frac{e^{-\mu|\bold{x}-\bold{x}_0|}}{4\pi|\bold{x}-\bold{x}_0|}$. To make $R_\mu(\bold{x};\bold{x}_0)$ consistent, we write $G_{\mu f}$ in a polar series form. Hence we have
\begin{equation}\label{G_muf}
    \begin{aligned}
        &G_{\mu f}(\rho;\phi;\theta;\rho_0;\phi_0;\theta_0)= \sum_{n=0}^\infty A_{fn}(\rho,\rho_0)\sum_{m=0}^nY_n^m(\phi,\theta,\phi_0,\theta_0),\\
        &A_{fn} =  (2n+1)\rho^{-\frac{1}{2}}\rho_0^{-\frac{1}{2}}\Bigg\{\begin{array}{c}
             K_q(\mu \rho_0)I_q(\mu \rho),\qquad 0\leq\rho<\rho_0  \\
             K_q(\mu \rho)I_q(\mu \rho_0),\qquad \rho_0 \leq \rho \leq 1.
        \end{array}
    \end{aligned}
\end{equation}
By applying a Taylor expansion to the exact form of $G_{\mu f}$, the regular part of $G_{\mu f}$ is readily obtained as
\begin{equation}
    R_{\mu f} = \frac{\mu}{4\pi} + \frac{\mu^3}{24\pi}|\bold{x}-\bold{x}_0|^2.
\end{equation}

Therefore, the regular part of Helmholtz Green's function is
\begin{equation}\label{R_mu}
    \begin{aligned}
        R_\mu(\rho;\phi;\theta;\rho_0;\phi_0;\theta_0) = \frac{\mu}{4\pi} + \frac{\mu^3}{24\pi}|\bold{x}-\bold{x}_0|^2 +  
        \sum_{n=0}^\infty A_{n}(\rho,\rho_0)\sum_{m=0}^nY_n^m(\phi,\theta,\phi_0,\theta_0),
    \end{aligned}
\end{equation}
where 
\begin{equation}\label{Amn}
    \begin{aligned}
        A_{n}(\rho,\rho_0) =(2n+1) Q_{q}(\mu)I_q(\mu\rho_0) I_q(\mu \rho)\rho^{-\frac{1}{2}}\rho_0^{-\frac{1}{2}},
    \end{aligned}
\end{equation}
and $Y_n^m$ is defined in (\ref{spherical_harmonics}). We emphasize that only one branch remains in $R_\mu$ due to its inherent symmetry.

\bibliographystyle{plain}
\bibliography{reference}

\end{document}